\documentclass[usenatbib]{mn2e}
\usepackage{mycite}
\usepackage{amssymb}
\usepackage{amsmath}
\usepackage{graphicx}
\usepackage{natbib}

\def\Zsol{\hbox{Z$_{\odot}$}}
\def\Msol{\hbox{M$_{\odot}$}}

\usepackage{epsfig}

\newcommand{\hi}{H\,{\sc i}}
\newcommand{\hii}{H~{\sc ii}}
\newcommand{\hei}{He~{\sc i}}
\newcommand{\heii}{He~{\sc ii}}
\newcommand{\eld}{$N_{\rm e}$}

\newcommand{\elt}{$T_{\rm e}$}

\newcommand{\opp}{O$^{2+}$}

\newcommand{\np}{N$^+$}

\newcommand{\foiii}{[O~{\sc iii}]}

\newcommand{\foii}{[O~{\sc ii}]}
\newcommand{\fsii}{[S~{\sc ii}]}
\newcommand{\fsiii}{[S~{\sc iii}]}

\newcommand{\fnii}{[N~{\sc ii}]}

\newcommand{\fneiii}{[Ne~{\sc iii}]}

\newcommand{\civ}{C~{\sc iv}}

\newcommand{\fariii}{[Ar~{\sc iii}]}

\newcommand{\hp}{H$^+$}

\title[A Detailed IFS Study of UM~420 \& UM~462]
{A VLT VIMOS integral field spectroscopic study of perturbed blue compact
galaxies: UM~420 and UM~462\thanks{Based on observations collected at the
European Southern Observatory, Chile, under programmes 078.B-0353(B, E)}
\author[B. L. James et al.]
{B.~L. James$^{1}$\thanks{E-mail:
bj@star.ucl.ac.uk (BJ)}, Y.~G. Tsamis$^{2, 3}$, M.~J. Barlow$^{1}$\\
$^{1}$Department of Physics and Astronomy, University College London, Gower Street, London WC1E 6BT\\
$^{2}$Instituto de Astrof\'{i}sica de Andaluc\'{i}a (CSIC), Apartado 3004, 18080 Granada, Spain\\
$^{3}$Department of Physics and Astronomy, The Open University, Walton Hall,
Milton Keynes MK7 6AA }}

\begin{document}
\date{Accepted 2009 September 11. Received 2009 September 11; in original form 2009 July 17 }
\pagerange{\pageref{firstpage}--\pageref{lastpage}} \pubyear{2009}
\maketitle
\label{firstpage}
\begin{abstract}

We report on optical integral field spectroscopy of two unrelated blue compact
galaxies mapped with the 13 $\times$ 13 arcsec$^2$ VIMOS integral field unit at
a resolution of 0.33 $\times$ 0.33 arcsec$^2$. Continuum and background
subtracted emission line maps in the light of \foiii\ $\lambda$5007, H$\alpha$,
and \fnii\ $\lambda$6584 are presented. Both galaxies display signs of ongoing
perturbation and/or interaction. UM~420 is resolved for the first time to be a
merging system composed of two starbursting components with an `arm-like' structure
associated with the largest component. UM~462 which is a disrupted system of
irregular morphology is resolved into at least four starbursting regions. Maps
of the H$\alpha$ radial velocity and FWHM are discussed. No underlying broad
line region was detected from either galaxy as the emission lines are
well-fitted with single Gaussian profiles only. Electron temperatures and
densities as well as the abundances of helium, oxygen, nitrogen, and sulphur
were computed from spectra integrated over the whole galaxies and for each area
of recent star formation. Maps of the O/H ratio are presented: these galaxies
show oxygen abundances that are $\sim$ 20 per cent solar. No evidence of
substantial abundance variations across the galaxies that would point to
significant nitrogen or oxygen self-enrichment is found ($\lesssim$ 0.2 dex
limit). Contrary to previous observations, this analysis does not support the
classification of these BCGs as Wolf-Rayet galaxies as the characteristic broad
emission line features have not been detected in our spectra.
Baldwin-Phillips-Terlevich emission line ratio diagrams which were constructed
on a pixel by pixel basis indicate that the optical spectra of these systems
are predominantly excited by stellar photoionization.

\end{abstract}

\section{Introduction}
\label{sec:intro}

Blue compact dwarf galaxies (BCGs) offer a means of exploring star-formation in
low-mass, low-metallicity (1/50--1/3~\Zsol,\citet{Kunth:2000}) systems and, by
analogy, the chemically un-evolved systems in the high-$z$ primordial universe.
Typically, they are characterised by their strong emission lines superimposed
on a faint, blue continuum, resulting from one or more recent bursts of
star-formation. These compact, gas-rich objects are estimated to have high
star-formation rates in the order of 0.1 to 1 \Msol\ yr$^{-1}$
\citep{Fanelli:1988} with typical \hi\ masses of M$_{HI}\sim10^7-10^8$~\Msol\
 \citep{Thuan:1981}, and are ideal laboratories for many topics related to
star-formation, partly because they do not display more complicated phenomena,
such as density waves, that operate in larger galaxies \citep{Cairos:2001}.
This is a study of two BCGs, UM~420 and UM~462, first discovered by the
University of Michigan survey for extragalactic emission-line objects
\citep{Macalpine:1978}. Their individual properties may provide insights into
the role of galaxy interactions in chemical recycling and starbursting
episodes. Several studies have suggested that interactions with other systems
are a contributing factor to large-scale starbursts in dwarf galaxies
\citep{Brinks:1990,Mendez:2000,Iglesias:2001,Verdes:2002,Tran:2003}. Hence it may be no
coincidence that both galaxies studied here display signs of interaction and/or
perturbation in the form of tails, multiple nuclei, or disrupted morphology.

UM~420 was reported by \citet{Lopez-Sanchez:2008} as having an irregular and
elongated morphology, with two long regions extending in different directions
from the brightest central region. A bright spiral galaxy, UGC~1809, lies
16.5$''$ west of UM~420, but at less than half the redshift of UM~420 is not
physically associated to it and cannot cause an effect.  UM~420 was once
thought to be a large \hii\ region near the edge of UGC~1809's spiral arms, and
as a result the SIMBAD data base currently incorrectly lists its RA and DEC as
those of UCG~1809 (the correct RA and DEC for UM~420 are given in
Table~\ref{tab:galaxies}). \citet{Pustilnik:2004} describe UM~420 as being a
possible `sinking merger', i.e. harbouring a satellite galaxy that is sinking
into a larger, gas-rich companion.

UM~462 with UM~461 form part of an interacting binary pair separated by only
$\sim$50$''$ \citep{Taylor:1995,Telles:1995}.  In a \hi\ survey of \hii\
galaxies\footnote{\hii\ galaxies are galaxies with spectra similar to \hii\
regions but more metal-rich than blue compact dwarfs \citep{Kunth:2000}} with
companions \citep{Taylor:1995}, UM~461 was observed to have a tidal arm
extending in the direction of UM~462 and outer \hi\ contours that are clearly
disturbed by its nearby neighbour.  \citet{Telles:1995} interpreted UM~462 as
being part of a group of \hii\ galaxies, UM~461, UM~463 and UM~465, whose
star-bursting episodes are synchronised on time scales of less than
10$^7$~years, despite being separated by $\sim$1--2~Mpc. Following the
classification system of \citet{Loose:1985}, \citet{Cairos:2001} classified
UM~462 as being of `iE' type, i.e. showing a complex inner structure with
several star-forming regions superimposed on an extended regular envelope. This
type of structure is commonly interpreted as a sign of interaction.
Spectroscopically, both galaxies have been classified as Wolf-Rayet (WR)
galaxies (this is however disputed in the present work; see Sec. 5.3), with
broad \heii\ $\lambda$4686 being detected in their spectra
\citep[][]{Izotov:1998,Guseva:2000}. \citet{Guseva:2000} also claimed to see
broad \civ\ $\lambda$4658 and \civ\ $\lambda$5808 emission in the spectra of
UM~420, indicating the presence of WCE stars, although the detections were
doubted by \citet{Schaerer:1999} on the basis of the S/N ratio of the
\citet{Guseva:2000} spectrum.  The relative numbers of WR stars to O-type stars
were estimated by \citet{Guseva:2000} as 0.03 and 0.005 for UM~420 and UM~462,
respectively. \citet{Schaerer:1999} have defined WR galaxies as being those
whose ongoing or recent star formation has produced stars sufficiently massive
to evolve to the WR stage. However, the same authors also note that the
definition of a WR galaxy is dependent on the quality of the spectrum, and the
location and size of the aperture.

UM~420 falls within a small sub-set of BCGs reported by \citet{Pustilnik:2004}
as having anomalously high N/O ratios ($\sim$ 0.5~dex) compared to BCGs of
similar oxygen metallicity \citep[12+log(O/H)=7.93;][IT98
hereafter]{Izotov:1998}. They attributed this over-abundance of nitrogen to the
effects of N-rich winds from large numbers of Wolf-Rayet (WNL-type) stars. A
previous IFU study by \citet{James:2009} (J09 hereafter) of another BCG in this
`high N/O' group, Mrk~996, derived a significantly higher O/H ratio than
previously and an N/O ratio for its narrow-line component which is now typical
for its metallicity. Mrk 996, however, was shown by J09 to also contain an
extremely dense, extended broad-line component displaying a N/O ratio up to 20
times higher than the galaxy's narrow-line component; this large N/H and N/O
enrichment was attributed to the cumulative effects of a large population of
$\sim$2600 WNL-type stars whose number was estimated from fitting the broad
emission features at 4640 and 4650-\AA\ with theoretical WR spectral templates.
By conducting a similar study of UM~420 we aimed to investigate its N/O ratio
status, and to study potential effects to the gas abundance patterns due to WR
stars. On the other hand, UM~462 provides a useful comparison, as it has been
reported to have a rather normal N/O ratio close to the average observed for
many BCGs at its published metallicity of 12 $+$ log(O/H) $=$ 7.95 (IT98).

In order to understand interaction-induced starbursts, it is essential to gain
spatial information regarding the properties (e.g. physical conditions,
chemical abundances) of the starburst regions and any spatial correlations that
may hold.  This type of information has been limited in previous long-slit
spectroscopic studies of \hii\ galaxies. In this paper we present high
resolution optical observations obtained with the VIMOS integral field unit
(IFU) spectrograph on the ESO 8.2m Very Large Telescope UT3/Melipal.  The data
afford us new spatiokinematic `3-D' views of UM~420 and UM~462.  The spatial
and spectral resolutions achieved (Section~\ref{sec:observations}) allow us to
undertake a chemical and kinematical analysis of both systems, providing a
clearer picture of the ionised gas within their star-forming regions. We adopt
distances of 238~Mpc and 14.4~Mpc for UM~420 and UM~462, respectively,
corresponding to their redshifted velocities measured here of $+$17,454  and
$+$1057 km~s$^{-1}$ ($z$ $=$ 0.060604 and 0.003527, as given in
Table~\ref{tab:galaxies}), for a Hubble constant of H$_{\rm o}$ = 73.5
km~s$^{-1}$~Mpc$^{-1}$ \citep{Bernardis:2008}.

\begin{table*}
\begin{center}
\begin{small}
\caption{General Parameters of UM~420 and UM~462}
\begin{tabular}{lccccl}
\hline
 & \multicolumn{2}{c}{Coordinates (J2000)} & \multicolumn{3}{c}{}\\
Name &  $\alpha$ & $\delta$ & $z$$^a$ & Distance (Mpc) & Other Names\\
\hline
UM~420 & 02 21 36.2 & 00 29 34.3 & 0.060604 & 238 & \\
UM~462 & 11 53 18.5 & -02 32 20.6 & 0.003527 &  14.4 & Mrk~1307, UGC~06850\\
\hline \label{tab:galaxies}
\end{tabular}
\begin{description}
\item[$^a$]Derived from the present observations
\end{description}
\end{small}
\end{center}
\end{table*}

\section{VIMOS IFU Observations and Data Reduction}
\subsection{Observations}
\label{sec:observations}

Two data sets for each galaxy were obtained with the Visible Multi-Object
Spectrograph (VIMOS) IFU mounted on the 8.2m VLT at the Paranal Observatory in
Chile. All data sets were taken with the high-resolution and high-magnification
settings of the IFU, which resulted in a field-of-view (FoV) of 13$''
\times$13$''$ covered by 1600 spaxels (spatial pixels) at a sampling of
0.33$''$ per spaxel. The data consist of high-resolution blue grism spectra
(HRblue, $\sim$0.51\,\AA/pixel, 2.01$ \pm$ 0.23\,\AA\ FWHM resolution) covering
4150--6200~\AA\ and high-resolution orange grism spectra (HRorange,
$\sim$0.6\,\AA/pixel, 1.92 $\pm$ 0.09\,\AA\ FWHM resolution) covering
5250--7400~\AA.  Spectra from both grisms for each galaxy are shown in
Figure~\ref{fig:fullSpectra}. The observing log can be found in
Table~\ref{tab:obslog}. Four exposures were taken per grism, per observation
ID.  The third exposure within each set was dithered by $+$0.25$''$ in RA and
by $+$0.61$''$ in DEC for both galaxies (corresponding to a 2 spaxel offset in
the $X$-direction) in order to remove any broken fibres when averaging
exposures.  All observations were taken at a position angle of 20$^\circ$ for
UM~420 and 22$^\circ$ for UM~462.

\begin{table*}
\begin{center}
\begin{small}
\caption{VIMOS IFU observing log}
\begin{tabular}{|cccccc|}
\hline
Observation ID & Date &Grism    &Exp. time (s) & Airmass range & FWHM seeing (arcsec)\\
\hline
\multicolumn{6}{c}{{\bf UM~420}}\\
250536-9 & 14/11/2006 & HR blue & 4 ${\times}$ 402 & 1.11--1.40  & 0.85 \\
250534,250535 & 14/11/2006 & HR orange & 4 ${\times}$ 372 & 1.13--1.36 & 0.79\\
\multicolumn{6}{c}{{\bf UM~462}}\\
250563-6 & 22-23/03/2007 & HR blue & 4 ${\times}$ 402 & 1.60 -- 1.10 & 1.02\\
250559,250562 & 22-23/03/2007& HR orange & 4 ${\times}$ 372 & 2.16 -- 1.73 & 0.74\\
\hline \label{tab:obslog}
\end{tabular}
\end{small}
\end{center}
\end{table*}

\subsection{Data Reduction}

Data reduction was carried out using the ESO pipeline via the {\sc
gasgano}{\footnote{http://www.eso.org/sci/data-processing/software/gasgano}
software package and followed the sequence outlined by J09.  The reduction
involves three main tasks: {\it vmbias}, {\it vmifucalib} and {\it
vmifustandard}. The products of these were fed into {\it vmifuscience} which
extracted the bias-subtracted, wavelength- and flux-calibrated, and relative
fibre transmission-corrected science spectra.  The final cube construction from
the quadrant-specific science spectra output by the pipeline utilised an IFU
table that lists the one to one correspondence between fibre positions on the
IFU head and stacked spectra on the IFU \citep{Bastian:2006}. A final data cube
for each of the HRblue and HRorange grisms was then created by averaging over
the final flux-calibrated science cubes created from each observation block.
For a more detailed description of the data reduction and processing from
science spectra to spatially reconstructed cube, see J09.  A schematic
representation of the data reduction processes performed within {\sc gasgano}
is provided by \citet{Zanichelli:2005}. Sky subtraction was performed by
locating a background region within each quadrant, summing the spectra over
each of the spaxels in the reconstructed cube and subtracting the median sky
spectrum of the region from its corresponding quadrant.  Median combining was
needed to ensure that any residual contamination from faint objects was
removed. In the case of UM~420, the western edge of the IFU frame sampled the
outermost faint regions of the unrelated foreground spiral galaxy UGC~1809
(e.g. \citet{Lopez-Sanchez:2008}), and this background subtraction procedure
enabled the removal of any contaminating signal. Before averaging to produce a
final data cube for each grism, each cube was corrected for differential
atmospheric refraction (DAR). This correction accounts for the refraction of an
object's spectrum along the parallactic angle of the observation as it is
observed through the atmosphere. The cubes were corrected using an {\sc
iraf}-based programme written by J. R. Walsh (based on an algorithm described
in \citet{Walsh:1990}). The procedure calculates fractional spaxel shifts for
each monochromatic slice of the cube relative to a fiducial wavelength (i.e. a
strong emission line), shifts each slice with respect to the orientation of the
slit on the sky and the parallactic angle and recombines the DAR-corrected data
cube. For both galaxies a high S/N ratio fiducial emission line (\hei\
$\lambda$5876) was available within the wavelength overlap of the HRblue and
HRorange cubes. Although not essential, correcting multiple spectral cubes to a
common emission line allows one to check that the alignment between them is
correct.

\subsection{Emission Line Profile Fitting}

Taking into account that there are 1600 spectra across the FoV of each data
cube, we utilised an automated fitting process called PAN \citep[Peak ANalysis;
][]{Dimeo:2005}. This is an IDL-based general curve-fitting package, adapted by
\citet{Westmoquette:2007} for use with FITS data. The user can interactively
specify the initial parameters of a spectral line fit (continuum level, line
peak flux, centroid and width) and allow PAN to sequentially process each
spectrum, fitting Gaussian profiles accordingly.  The output consists of the
fit parameters for the continuum and each spectral line's profile and the
$\chi^2$ value for the fit. It was found that all emission lines in the spectra
of both UM~420 and UM~462 could be optimally fitted with a single narrow
Gaussian.  For high S/N ratio cases, such as Balmer lines and \foiii\
$\lambda$5007, attempts were made to fit an additional broad Gaussian profile
underneath the narrow component. These fits were assessed using the statistical
F-test procedure, a process that determines the optimum number of Gaussians
required to fit each line (see J09 for a more detailed discussion). We found
that it was not statistically significant to fit anything more than a single
Gaussian to the emission lines in the spectra of either object. Thus single
Gaussian profiles were fitted to each emission line profile, restricting the
minimum FWHM to be the instrumental width. Suitable wavelength limits were
defined for each emission line and continuum level fit. Further constraints
were applied when fitting the \fsii\ doublet: the wavelength difference between
the two lines was taken to be equal to the redshifted laboratory value when
fitting the velocity component, and their FWHM were set equal to one another.
The fitting errors reported by PAN underestimate the true uncertainties. We
thus follow the error estimation procedure outlined by J09, which involves the
visual re-inspection of the line profile plus fit after checking which solution
was selected by our tests, and taking into account the S/N ratio of the
spectrum. By comparing PAN fits to those performed by other fitting techniques
(e.g. {\sc iraf}'s {\sc splot} task) on line profiles with an established
configuration (i.e. after performing the F-test) we find that estimated
uncertainties of $\sim$5--10 per cent are associated with the single Gaussian
fits. A listing of the measured flux and FWHM uncertainties appears in Table
\ref{tab:fluxes}, where errors are quoted for individual component fits to
emission lines detected on the integrated spectra from across all of UM~420 and
UM~462.

\section{Mapping Line Fluxes}
\subsection{Line fluxes and reddening corrections}
\label{sec:redCorrect}
\begin{figure*}
\begin{center}
\includegraphics[scale=0.70, angle=0]{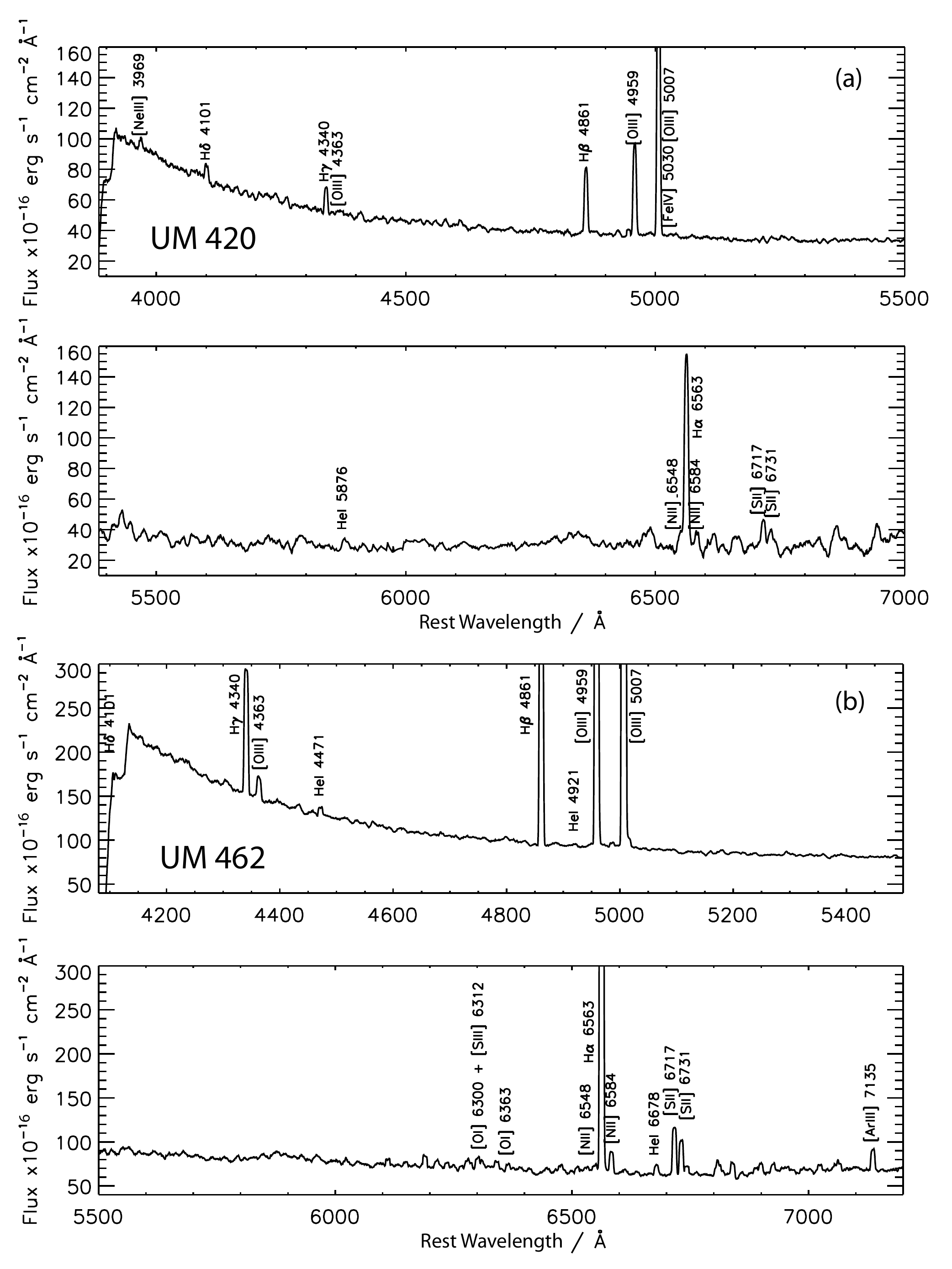}
\caption{VIMOS IFU summed spectra of UM~420 after smoothing with a 5 pixel
boxcar function: (a) UM~420, top panel: high resolution blue spectrum, bottom
panel: high resolution orange spectrum.  The spectra are integrated over an
area of 7.6$''\times$6.3$''$ (8.8$\times$7.2 kpc$^2$) and correspond to
exposure times of 4$\times$402~s  and 4$\times$372~s for the HRblue and
HRorange grisms, respectively; (b) Same as (a), for UM~462.  Spectra are
integrated over an area of 10.2$''\times$10.2$''$ (714.1$\times$714.1 pc$^2$).}
\label{fig:fullSpectra}
\end{center}
\end{figure*}

Full HRblue and HRorange spectra for UM~420 and UM~462 are shown in
Figure~\ref{fig:fullSpectra}(a) and (b), respectively. Table \ref{tab:fluxes}
lists the measured FWHMs and observed and de-reddened fluxes of the detected
emission lines within UM~420 and UM~462, respectively. The listed fluxes are
from IFU spectra summed over each galaxy and are quoted relative to
$F$(H$\beta$) $=$ 100.0. Foreground Milky Way reddening values of $E$($B-V$)
$=$ 0.04 and 0.02 were adopted, from the maps of \citet{Schlegel:1998},
corresponding to $c$(H$\beta$) $=$ 0.05 and 0.03 in the directions of UM~420
and UM~462, respectively. The line fluxes were then corrected for extinction
using the Galactic reddening law of \citet{Howarth:1983} with $R_V$ $=$ 3.1
using $c$(H$\beta$) values derived from each galaxy's H$\alpha$/H$\beta$ and
H$\gamma$/H$\beta$ emission line ratios, weighted in a 3 : 1 ratio,
respectively, after comparison with the theoretical Case~B ratios from
\citet{Hummer:1987} of $j_{\rm H\alpha}/j_{\rm H\beta}$ $=$ 2.863 and $j_{\rm
H\gamma}/j_{\rm H\beta}$ $=$ 0.468 (at \elt\ $=$ 10$^4$~K and \eld\ $=$
100~cm$^{-3}$). The same method was used to create $c$(H$\beta$) maps for each
galaxy, using H$\alpha$/H$\beta$ and H$\gamma$/H$\beta$ ratio maps, which were
employed to deredden other lines on a spaxel by spaxel basis prior to creating
electron temperature, electron density and abundance maps.  As stated in Table
\ref{tab:fluxes}, overall $c$(H$\beta$) values of 0.25$\pm$0.08 and
0.19$\pm$0.04 were found to be applicable to the integrated emission from
UM~420 and UM~462, respectively.

\begin{table*}
\caption{Emission line measurements for summed spectra of UM~420 and UM~462. Observed and de-reddened fluxes are relative to F(H$\beta$)=I(H$\beta$)=100.  Line fluxes were extinction-corrected using the c(H$\beta$) values shown at the bottom of the table, calculated from the relative H$\alpha$, H$\beta$ and H$\gamma$ fluxes. FWHMs have been corrected for the instrumental FWHM resolution (Section \ref{sec:observations})}
\begin{center}
\begin{small}
\begin{tabular}{lcccccc}
\hline
  & \multicolumn{3}{c}{UM~420} & \multicolumn{3}{c}{UM~462}\\
 Line ID  & FHWM (km s$^{-1}$) & F($\lambda$) & I($\lambda$)& FHWM (km s$^{-1}$) & F($\lambda$) & I($\lambda$)\\
\hline
 3969  [Ne~III] &   431.0$\pm$   52.2     &     27.1$\pm$     2.5 &     30.5$\pm$     2.8 & --- & --- & --- \\

 4101  H$\delta$ &   205.7$\pm$   12.1    &     22.8$\pm$     1.2 &     25.3$\pm$     1.3 & --- & --- & --- \\

 4340 H$\gamma$ &   183.3$\pm$    5.3     &     42.0$\pm$     1.1 &     45.1$\pm$     1.1 &   143.1$\pm$    1.8     &     43.3$\pm$     0.5 &     45.7$\pm$     0.5 \\
	    
 4363 [O~III] &   239.1$\pm$   64.0    &      6.3$\pm$     1.3 &      6.8$\pm$     1.4 &   156.6$\pm$   13.8     &      7.1$\pm$     0.5 &      7.4$\pm$     0.5 \\
	    
 4471 He~I & 143.5$\pm$26.3     &      3.5$\pm$0.5 &      3.7$\pm$0.5  &  113.7$\pm$   12.4     &      3.1$\pm$     0.3 &      3.2$\pm$     0.3\\
	    
 4686  He~II & --- & --- & --- &   175.4$\pm$    5.1     &      1.7$\pm$     0.1 &      1.7$\pm$     0.1 \\
	    
 4861 H$\beta$ &   187.9$\pm$    2.0     &    100.0$\pm$     1.3 &    100.0$\pm$     0.9 &   123.4$\pm$    0.7     &    100.0$\pm$     0.7 &    100.0$\pm$     0.5 \\
	    
 4921 He~I & --- & --- & --- &   122.5$\pm$    7.3     &      0.9$\pm$     0.2 &      0.9$\pm$     0.2 \\
	    
 4959 [O~III] &   182.3$\pm$    1.8     &    139.2$\pm$     1.7 &    137.3$\pm$     1.1 &   118.6$\pm$    1.2     &    164.6$\pm$     1.7 &    162.9$\pm$     1.5 \\
	    
 5007 [O~III] &   181.4$\pm$    1.0     &    430.5$\pm$     4.5 &    421.9$\pm$     1.9 &   116.4$\pm$    0.7     &    477.4$\pm$     3.5 &    470.2$\pm$     2.5 \\
	    
 5015 He~I or [FeIV] & --- & --- & --- &   104.2$\pm$    6.6     &      1.8$\pm$     0.1 &      1.8$\pm$     0.1 \\

5030  [Fe~IV] &    79.9$\pm$   15.2     &      3.5$\pm$     0.5 &      3.4$\pm$     0.5 & --- & --- & --- \\

5876  He~I &   184.0$\pm$   14.4     &     13.2$\pm$     0.9 &     11.7$\pm$     0.7 & 115.8$\pm$  11.6     &     12.1$\pm$     1.0 &     11.0$\pm$     0.9 \\

6051  [Fe~IV] &    36.7$\pm$    7.2     &      6.8$\pm$     0.4 &      5.9$\pm$     0.3 & --- & --- & --- \\
	    
 6300 [O~I] & --- & --- & --- &   116.8$\pm$   17.2     &      5.0$\pm$     0.6 &      4.4$\pm$     0.5 \\
	    
 6312 [S~III]  & --- & --- & --- &   123.6$\pm$   21.9     &      1.9$\pm$     0.4 &      1.7$\pm$     0.3  \\
	    
 6363 [O~I]  & --- & --- & --- &   150.9$\pm$   28.8     &      3.5$\pm$     0.7 &      3.1$\pm$     0.6 \\
	    
 6548 [N~II] &   238.3$\pm$   56.6     &     10.9$\pm$     2.1 &      9.1$\pm$     1.7 &   127.3$\pm$   13.5     &      3.1$\pm$     0.3 &      2.7$\pm$     0.2 \\
	    
 6563 H$\alpha$ &   178.8$\pm$    1.1     &    334.4$\pm$     3.6 &    279.1$\pm$     1.4 &   103.1$\pm$    0.6     &    323.4$\pm$     2.5 &    282.2$\pm$     1.6 \\
	    
 6584 [N~II] &   189.7$\pm$   14.4     &     29.5$\pm$     1.8 &     24.6$\pm$     1.5 &   109.8$\pm$    4.2     &      8.5$\pm$     0.3 &      7.4$\pm$     0.2 \\
	    
 6678 He~I &   135.7$\pm$105.2    &      4.7$\pm$1.6 &      3.8$\pm$1.3 &   127.6$\pm$    6.3     &      4.2$\pm$     0.4 &      3.7$\pm$     0.3 \\
	    
 6716 [S~II] &   196.5$\pm$    9.3     &     33.9$\pm$     1.4 &     27.9$\pm$     1.1 &   101.0$\pm$    5.2     &     19.0$\pm$     0.8 &     16.4$\pm$     0.7 \\
	    
 6731 [S~II] &   201.2$\pm$   12.6     &     26.9$\pm$     1.4 &     22.1$\pm$     1.1 &   105.9$\pm$    7.4     &     14.4$\pm$     0.8 &     12.4$\pm$     0.7 \\
	    
 7136 [Ar~III]  & --- & --- & --- &   104.6$\pm$    8.9     &      9.2$\pm$     0.6 &      7.8$\pm$     0.5 \\
 
 \multicolumn{7}{c}{}  \\

c(H$\beta$) dex &\multicolumn{3}{c}{0.25$\pm$0.08} &\multicolumn{3}{c}{0.19$\pm$0.04}\\

 F(H$\beta$)$\times$10$^{-16}$erg s$^{-1}$ cm$^{-2}$ & \multicolumn{3}{c}{359.1$\pm$6.1} & \multicolumn{3}{c}{2660$\pm$14}\\

\hline\label{tab:fluxes}
\end{tabular}
\end{small}
\end{center}
\end{table*}

\subsection{H$\alpha$ line properties}

Here we discuss the appearance of the targets in the light of the highest S/N
ratio \hi\ emission line, H$\alpha$. Figure \ref{fig:HaMaps} shows flux, radial
velocity and FWHM maps for H$\alpha$ from UM~420 and UM~462, respectively.

\begin{figure*}
\begin{center}
\includegraphics[scale=0.65, angle=0]{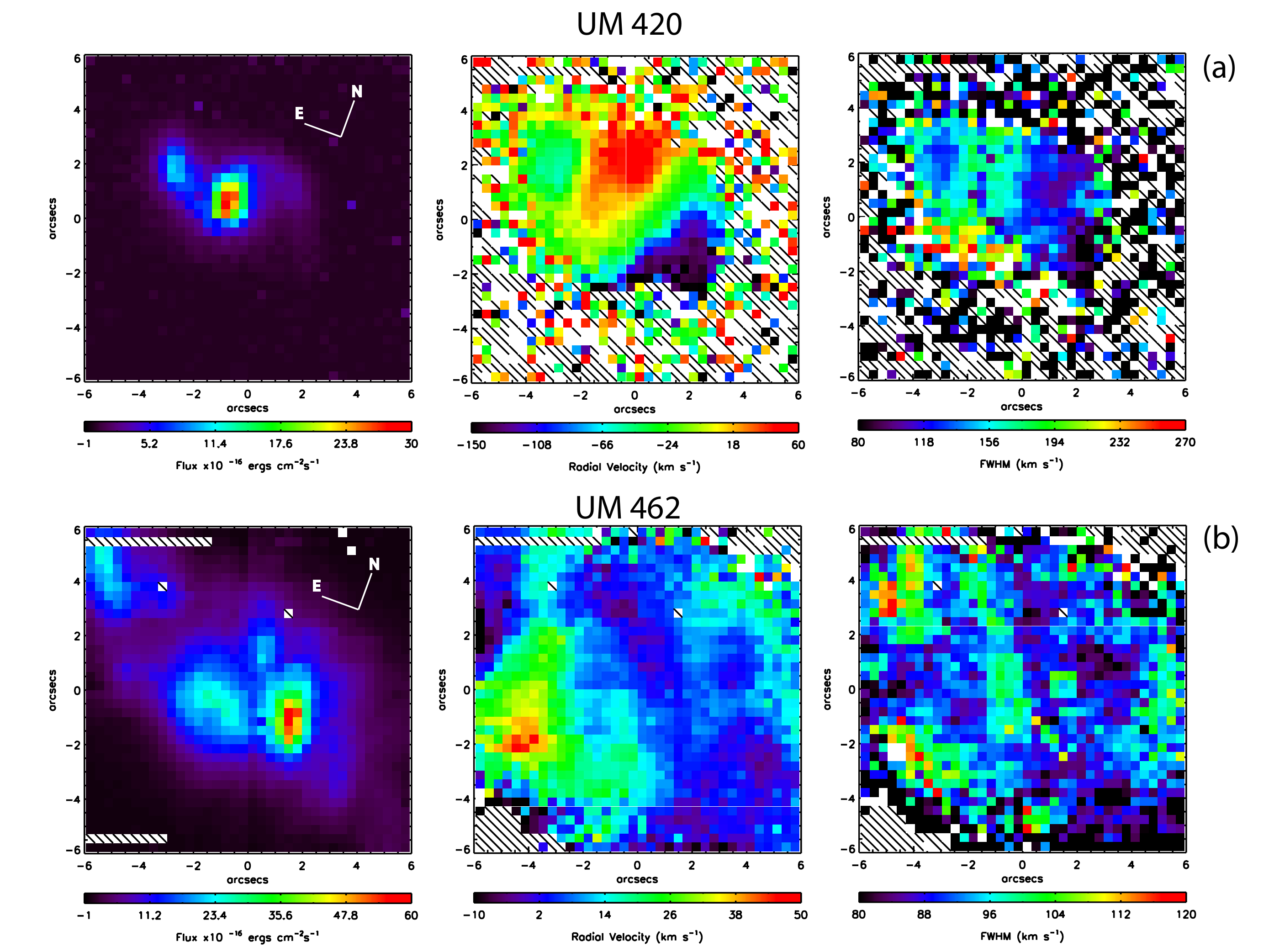}
\caption{Maps of UM~420 (a) and UM~462 (b) in H$\alpha$; left column: flux per
0.33 arcsec$^2$ spaxel; centre column: radial velocity (relative to
heliocentric systemic velocity of $+$17,545 and $+$1057 km~s$^{-1}$ for UM~420
and UM~462, respectively; right column: H$\alpha$ FWHM corrected for the
instrumental point spread function. See text for details.} \label{fig:HaMaps}
\end{center}
\end{figure*}

\subsubsection{UM~420}
\label{sec:HaUM420}

The H$\alpha$ flux map shown in the left-hand panel of
Figure~\ref{fig:HaMaps}(a) displays a double peak; the first peak is centrally
located at the RA and DEC listed in Table \ref{tab:galaxies}, with a surface
brightness of 2.75$\times$10$^{-14}$ erg cm$^{-2}$ s$^{-1}$ arcsec$^{-2}$ and
the second peak is located $\sim$2.2$''$ east of that central position, with a
peak surface brightness of 9.72$\times$10$^{-15}$ erg cm$^{-2}$ s$^{-1}$
arcsec$^{-2}$. These two H$\alpha$ peaks provide the basis for defining two
main star-forming regions, as displayed in Figure~\ref{fig:UM420Regions}. An
arm-like structure can be seen north-west of the central peak, extending in a
south-westerly direction; it could be the result of an ongoing merger or
interaction between the two identified regions. A more luminous spiral galaxy,
UGC 1809, lies 16.5$''$ westward of the galaxy, as noted by
\citet{Takase:1986}, but at less than half the redshift of UM~420 it cannot be
identified as an interacting companion. The VIMOS IFU H$\alpha$ flux is a
factor of 1.1 larger than the one measured by IT98 in a 3$''$ $\times$ 200$''$
slit and a factor of 1.8 times larger than that estimated by
\citet{Lopez-Sanchez:2008} from broad band H$\alpha$ $+$ \fnii\ images. The
H$\alpha$ radial velocity map shown in the central panel of
Figure~\ref{fig:HaMaps} shows two separate velocity gradients. Firstly, there
is a balanced velocity gradient at position angle (PA) 50$^\circ$ east of north
(labelled as `Position 1' on the central panel of Figure~\ref{fig:PVum420}).
This linear velocity gradient may be indicative of solid body rotation between
the main emission peak (Region 1) and the `arm-like' structure north-west of
Region 1. Secondly, the main body of the galaxy, i.e. the double peaked
emission structure, shows a velocity gradient at a PA of 110$^\circ$ east of
north (labelled as `Position 2' on the central panel of
Figure~\ref{fig:PVum420}).


Figure~\ref{fig:PVum420} shows position-velocity (P-V) diagrams along both
these axes. These diagrams were created by placing a spaxel-wide (i.e. with a
width of 0.33$''$) pseudo-slit across the radial velocity map, and plotting the
radial velocity as a function of distance from the slit's centre. The smooth
and symmetrically structured velocity gradient along Position 1 suggests that
the axis orthogonal to it could be adopted as a rotational axis for UM~420
(projected on the plane of the sky).
From the top panel of Figure~\ref{fig:PVum420} we can then make a new estimate
of the heliocentric systemic velocity of UM~420; a radial velocity of
$+$17,545$\pm$30 km~s$^{-1}$ is needed to normalise the distribution along the
proposed axis of rotation to zero (cf. the $+$17,514$\pm$12~km~s$^{-1}$
estimation of IT98). Thus a velocity gradient ranging from $\sim$-100
km~s$^{-1}$ to $\sim$+100 km~s$^{-1}$ is seen in the direction normal to the
rotation axis, with the negative radial velocity of $-$100 km s$^{-1}$ lying
within the south-west arm of the galaxy. Also, a velocity gradient exists
between Region 1 and 2, ranging from $\sim$ $-$5 km~s$^{-1}$ to $\sim$+100
km~s$^{-1}$. This velocity difference of $\sim$100 km~s$^{-1}$ between the two
peaks in H$\alpha$ flux may be evidence that Region 2 is a satellite galaxy
falling into or merging with a larger companion, Region 1.

Another scenario could also be considered.  Based on the H$\alpha$ emission line map (Figure~\ref{fig:UM420Regions}), UM~420 could also be identified as a spiral galaxy with weak spiral arms (of which we detect only one arm - the south-west arm), a nuclear starburst (Region 1) and a faint, outer-lying starburst or superstar cluster situated at the edge of the galaxy (Region 2).  However, for a distance of 238\,Mpc (Section~\ref{sec:intro}), UM~420 has an outer radius of only $\sim$3.5\,kpc (at isophotes at 3\% of the central peak flux), which would be very small for a spiral galaxy.  Further to this, we find a spiral galaxy scenario improbable when considering the radial velocity map (central panel of Figure~\ref{fig:HaMaps}a) which clearly shows that Region 2 is kinematically decoupled from the rest of the galaxy.


\begin{figure}
\begin{center}
\includegraphics[scale=0.35, angle=0]{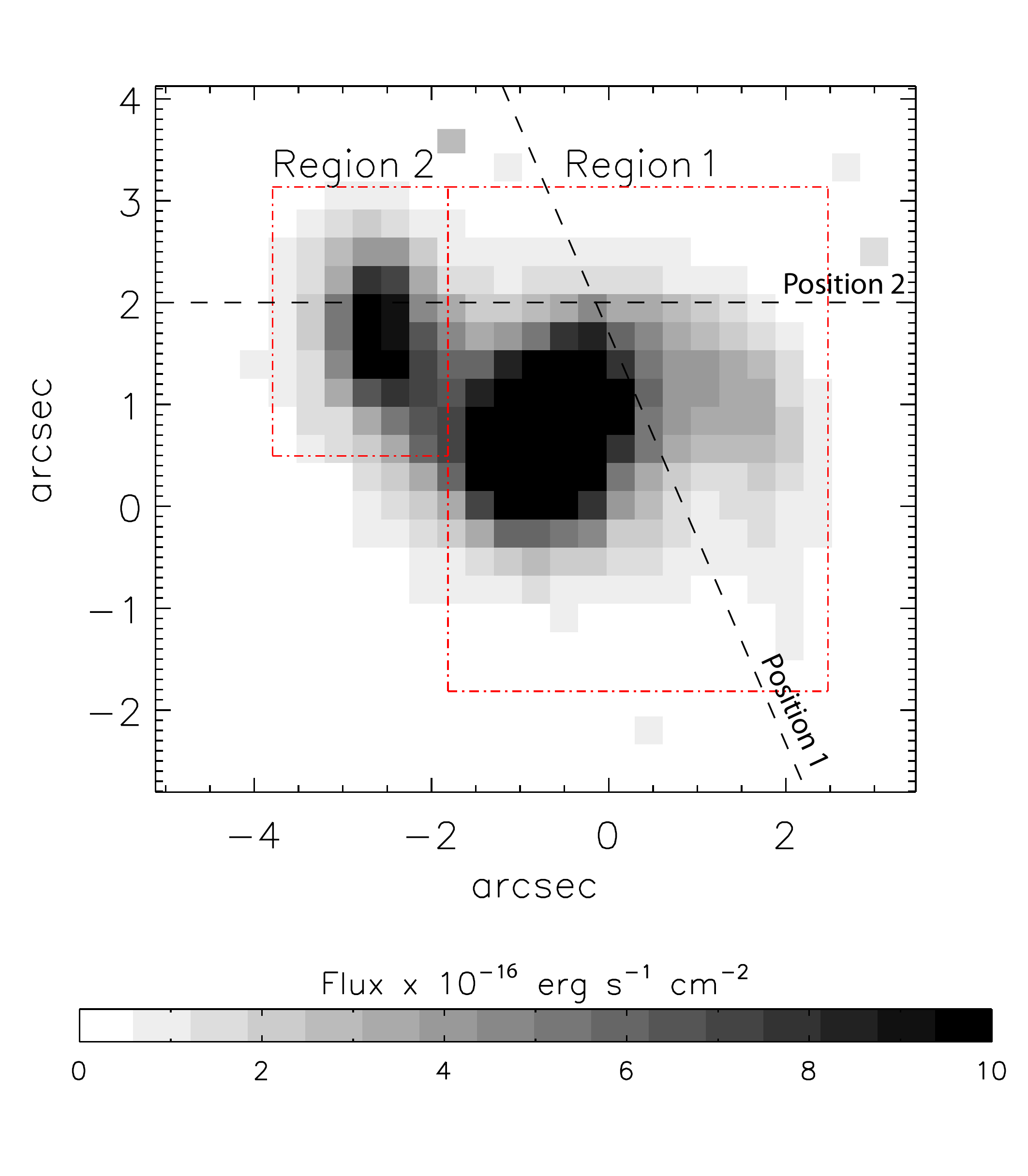}
\caption{UM~420: H$\alpha$ emission per 0.33$\times$0.33 arcsec$^2$ spaxel. The
two main areas of H$\alpha$ emission have been used to define two main
star-forming regions.  Also overlaid on the map are the locations of position-velocity cuts used for the P-V diagrams shown in Figure~\ref{fig:PVum420}.} \label{fig:UM420Regions}
\end{center}
\end{figure}

A FWHM map for H$\alpha$ is shown in the right-hand panel of
Figure~\ref{fig:HaMaps}(a). The highest measured FWHM is $\sim$270 km s$^{-1}$,
located southwards of the central emission peak, along the southern edge of the
galaxy.  The central flux peak is aligned with a peak of $\sim$160 km s$^{-1}$
whereas the second, easterly, flux peak is located within an area of decreased
FWHM, $\sim$130 km s$^{-1}$, surrounded by a slightly higher FWHM area of
$\sim$150 km s$^{-1}$. The south-western arm harbours the lowest line FWHM of
$\sim$80--120 km s$^{-1}$.

\begin{figure}
\begin{center}
\includegraphics[scale=0.5, angle=0]{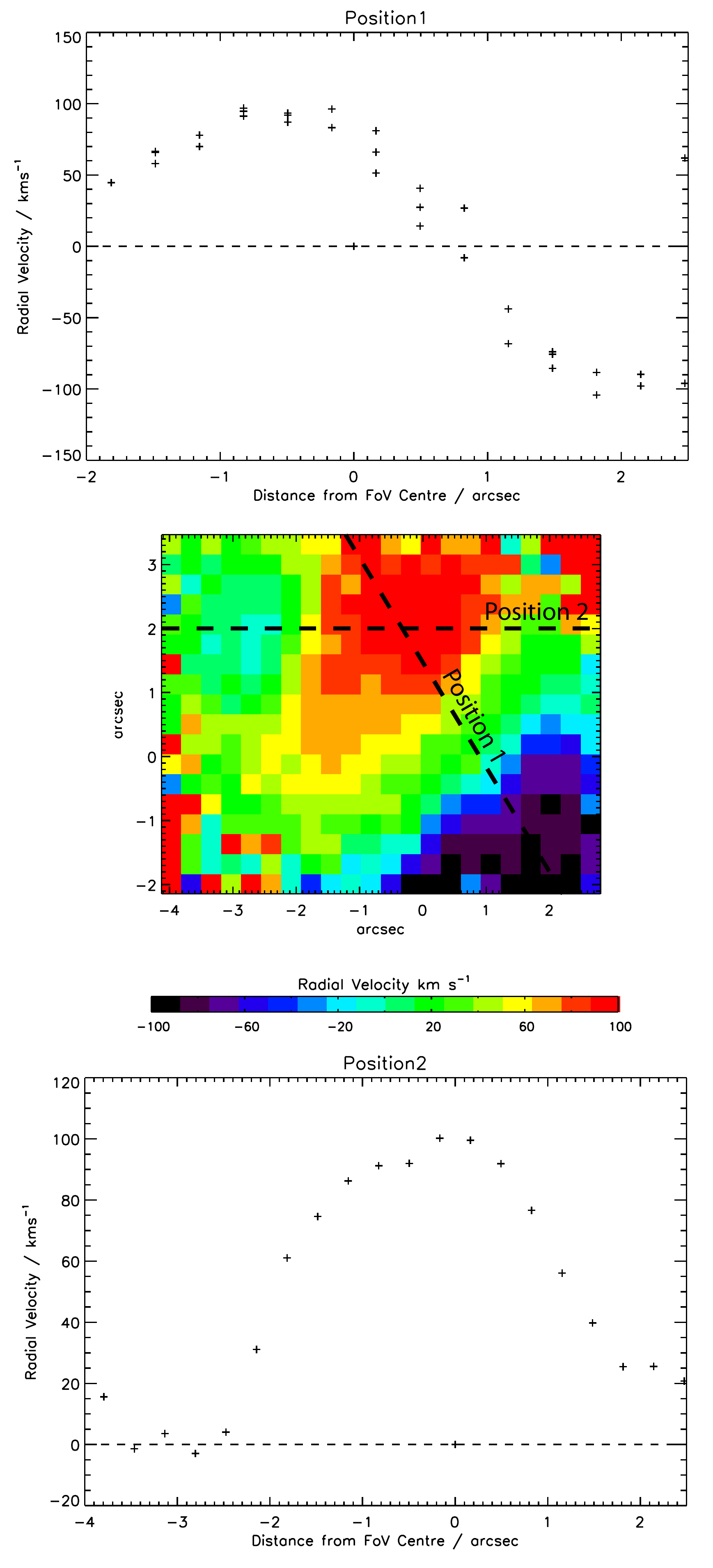}
\caption{Position velocity (P-V) diagrams for UM~420 in H$\alpha$: central
panel: radial velocity map of UM~420 (see Figure~\ref{fig:HaMaps}(a) for
details) where the black dotted line overlaid on the map represents cuts
orthogonal to the two proposed axes of rotation, as used for the P-V diagrams
in the top and bottom panels; top panel: P-V diagram for H$\alpha$ along
Position 1, defined as 50$^\circ$ east of north; bottom-panel: P-V diagram for
H$\alpha$ along Position 2, at 110$^\circ$ east of north. Zero-point velocities
are relative to a recession velocity of $+$17,514 km~s$^{-1}$. Typical relative
radial velocity errors are estimated as 30~km s$^{-1}$, with an estimated
spatial uncertainty of 0.5 spaxels ($\pm$0.17$''$).} \label{fig:PVum420}
\end{center}
\end{figure}

\subsubsection{UM~462}

UM~462 exhibits a far more disrupted morphology than UM~420. The left-hand
panel of Figure~\ref{fig:HaMaps}(b) shows a peak in H$\alpha$ flux located at
$\alpha = 11^{\rm h}\, 53^{\rm m}\, 20.25\pm0.1^{\rm s}$, $\delta =
-02^\circ\,32'\,19.62\pm0.1''$ (J2000), with a surface brightness of
5.34$\times$10$^{-14}$ erg cm$^{-2}$ s$^{-1}$ arcsec$^{-2}$.  Three additional
flux peaks can be seen, located 2.5$''$ northeast, 3.05$''$ east and 8.9$''$
east of this central peak. These H$\alpha$ peaks provide the basis for defining
four main star-forming regions, as displayed in Figure~\ref{fig:UM462Regions}.
Three of the regions have a similar peak surface brightness of
2.0--2.4$\times$10$^{-13}$ erg cm$^{-2}$ s$^{-1}$ arcsec$^{-2}$. Region 4 lies
farthest away from the brightest SF region (Region 1), at almost 9$''$; a large
low surface brightness envelope ($\sim$5$\times$10$^{-14}$ erg cm$^{-2}$
s$^{-1}$ arcsec$^{-2}$) separates it from Regions 1--3.  It could be the result
of interaction with the nearby galaxy UM~461 \citep{Taylor:1995}.  Previous
$R$-band and \hi\ imaging by \citet{Taylor:1995} and VRI surface photometry by
\citet{Telles:1995thesis} revealed only two star-forming regions within UM~462
which correspond to Regions~1 and 2 in Figure~\ref{fig:UM462Regions}.  The
H$\alpha$ flux measured by IT98 amounts to 35\% of the corresponding VIMOS IFU
flux, in accordance with the fraction of the galaxy's emitting area intercepted
by IT98's slit. The radial velocity distribution of H$\alpha$, displayed in the
central panel of Figure~\ref{fig:HaMaps}(b), shows no overall velocity
structure through the galaxy. UM~462, being part of an interacting binary pair
of galaxies, has a highly disturbed velocity distribution which shows no
spatial correlation with flux (Figure~\ref{fig:HaMaps}(b): left-hand panel).
The brightest region (Region 1) has a radial velocity of $\sim$0--10 km
s$^{-1}$, as does the majority of the emitting gas. The largest radial velocity
lies directly south-east of this region, with a magnitude of $+$50 km s$^{-1}$,
in an area of low surface brightness emission. This velocity space extends into
the envelope separating Region 4 from Regions 1--3. Region 4 itself lies within
a space of negative radial velocity, where again the peaks in velocity and flux
are spatially uncorrelated.

The H$\alpha$ FWHM map of UM~462 (right-hand panel of
Figure~\ref{fig:HaMaps}(b)) shows no spatial correlation with the flux
distribution and only a minimal correlation with radial velocity
(Figure~\ref{fig:HaMaps}(b):central panel). A peak in the FWHM of H$\alpha$ is
seen $\sim$1$''$ south of the emission peak of Region 4, at 130 km s$^{-1}$,
which aligns with the negative radial velocity peak.

\subsection{\foiii\ and \fnii}

Emission line maps of both galaxies in the light of \foiii\ $\lambda$5007 and
\fnii\ $\lambda$6584 lines are shown in Figs\,~\ref{fig:O3N2UM420} and
\ref{fig:O3N2UM462}. For UM~420 the morphology of \foiii\ is similar to that of
H$\alpha$ exhibiting two main starbursting regions along with the arc-like
 `arm' protruding from the brighter region. In UM~462, which is the
nearest of the two objects, the \foiii\ map shows more structure that the
corresponding H$\alpha$ map with an additional area of compact emission to the
south-west of Region 1 of Fig\,~\ref{fig:UM462Regions}. According to the
morphological criteria of Cairos et al. (2001) these maps support the
designation of UM~462 as being of `iI,C' type (irregular; cometary appearance)
rather than of `iE' type; UM~420 can be classified as `iI, M' (merging). For
both BCGs, the \fnii\ morphology is more diffuse and extended that that of
\foiii. The wider distribution of \fnii\ is probably due to the effect of lower
energy photons surviving farther away from the massive ionising clusters and
allowed to produce \np\ beyond the \opp\ zone.


\begin{figure}
\begin{center}
\includegraphics[scale=0.45, angle=0]{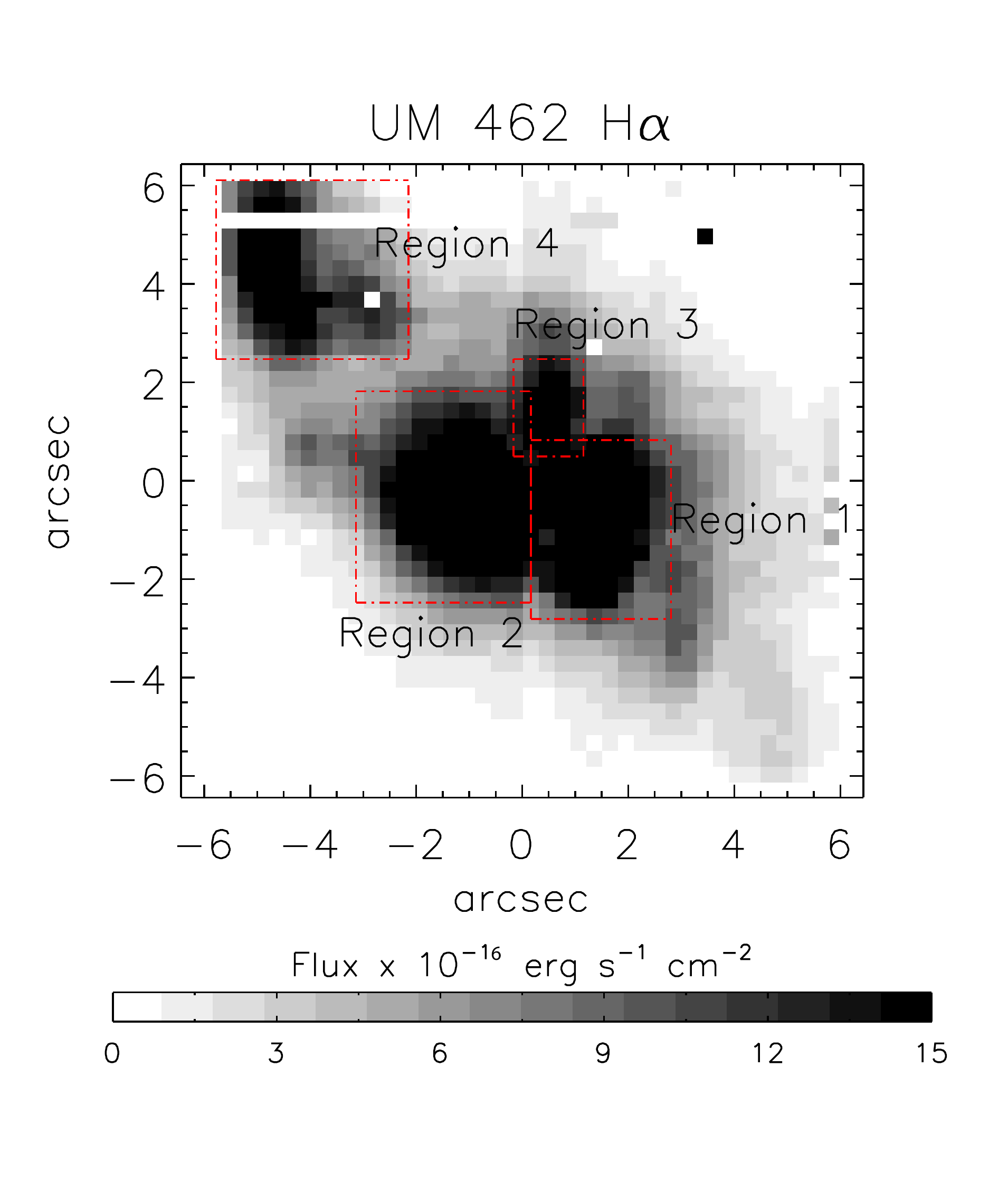}
\caption{UM~462: H$\alpha$ emission per 0.33$\times$0.33 arcsec$^2$ spaxel. The
four main H$\alpha$ emission areas have been labelled Regions 1--4.}
\label{fig:UM462Regions}
\end{center}
\end{figure}

\section{Electron Temperature and Density Diagnostics}

The dereddened \foiii\ ($\lambda$5007 $+$ $\lambda$4959)/$\lambda$4363
intensity ratios were used to determine electron temperatures within each
galaxy.  The \elt\ values were computed by inputting \foiii\ intensity ratios
and an adopted electron density into \textsc{iraf's}\footnote{IRAF is
distributed by the National Optical Astronomy Observatory, which is operated by
the Association of Universities for Research in Astronomy} {\tt TEMDEN} task in
the {\tt NEBULA} package. Atomic transition probabilities and collisional
strengths for O$^{2+}$ were taken from \citet{Wiese:1996} and
\citet{Lennon:1994}, respectively. From the summed spectra over each galaxy
(Table~\ref{tab:fluxes}) and adopting the electron densities quoted below we
derived mean electron temperatures of $\sim$14,000~K and $\sim$13,700~K for
UM~420 and UM~462, respectively (Table~\ref{tab:abundances}). Using the same
method, \elt\ maps were derived using \foiii\ ($\lambda$5007 $+$
$\lambda$4959)/$\lambda$4363 emission map ratios. Average \elt\ values for the
regions corresponding to peaks in H$\alpha$ emission for UM~420 and UM~462 are
given in Tables \ref{tab:um420Abs} and \ref{tab:um462Abs}, respectively, using
the regionally integrated relative line intensities
(Tables~\ref{tab:um420RegFlux},~\ref{tab:um462RegFlux}). The \elt\ values from
the summed spectra were adopted when computing \eld\ values from the \fsii\
doublet ratio $\lambda$6717/$\lambda$6731: we derived average \eld\ values of
$\sim$170~cm$^{-3}$ and $\sim$90~cm$^{-3}$ for UM~420 and UM~462, respectively
(Table~\ref{tab:abundances}). Similarly, the \elt\ maps were used to compute
\eld\ maps from $\lambda$6717/$\lambda$6731 ratio maps, which were then adopted
for the computation of ionic abundance maps discussed in the following Section.
Average \eld\ values for the regions corresponding to peaks in H$\alpha$
emission for UM~420 and UM~462 are given in Tables~\ref{tab:um420Abs} and
\ref{tab:um462Abs}, respectively.

\section{Chemical Abundances}
\label{sec:abundances}

\begin{figure*}
\begin{center}
\includegraphics[scale=0.5, angle=90]{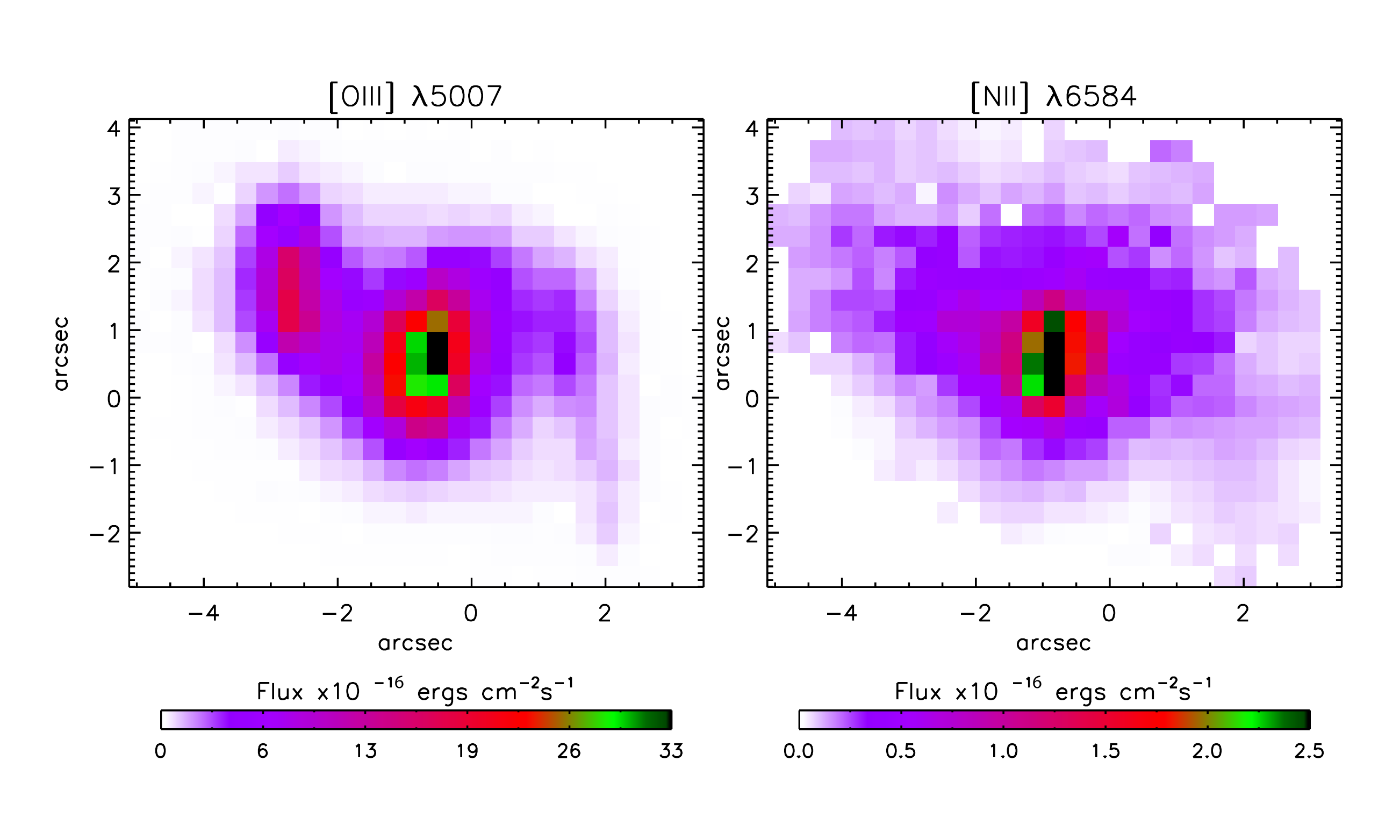}
\caption{UM~420: The distribution of flux per 0.33$\times$0.33 arsec$^2$
spaxels of \foiii\ $\lambda$5007 (left panel) and \fnii\ $\lambda$6584 (right
panel)} \label{fig:O3N2UM420}
\end{center}
\end{figure*}

\begin{figure*}
\begin{center}
\includegraphics[scale=0.5, angle=90]{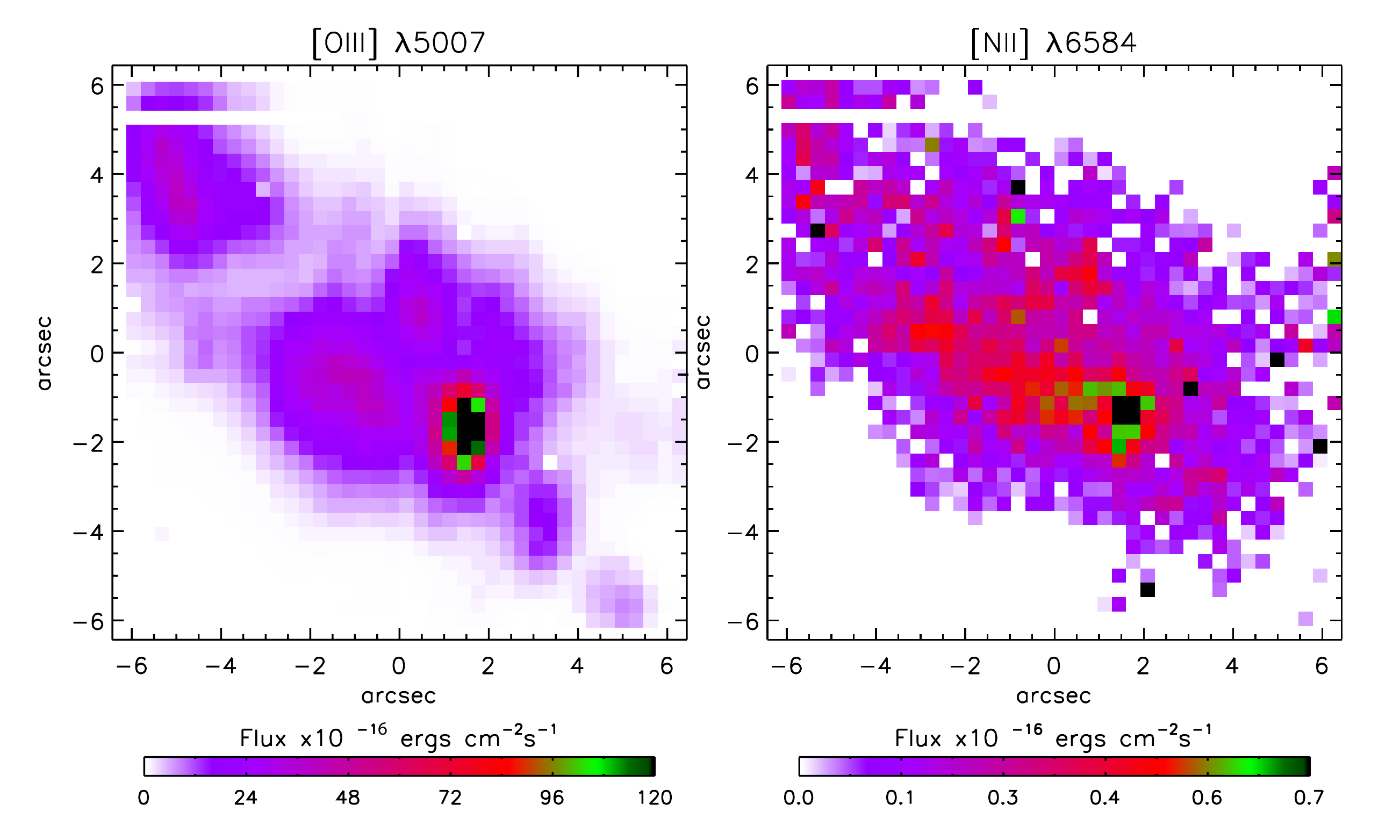}
\caption{UM~462: The distribution of flux per 0.33$\times$0.33 arsec$^2$
spaxels of \foiii\ $\lambda$5007 (left panel) and \fnii\ $\lambda$6584 (right
panel)} \label{fig:O3N2UM462}
\end{center}
\end{figure*}

\begin{table}
\caption{Ionic and elemental abundances for UM~420 and UM~462 derived, from summed IFU spectra over each galaxy.  All ionisation correction factors (ICFs) are taken from Kingsburgh \& Barlow, (1994).}
\begin{center}
\begin{small}
\begin{tabular}{lcc}
Property  & UM~420 & UM~462\\
\hline
T$_e$ (O~III)/ K &   14000$\pm$   1500 &   13700$\pm$    350\\
N$_e$ (S~II)/ cm$^{-3}$ &  170$\pm$  80 &   90$\pm$  75\\
\vspace{1.8mm}
c(H$\beta$) & 0.25$\pm$0.08 & 0.19$\pm$0.04 \\
He$^+$/H$^+$ ($\lambda$4471) $\times10^2$ & 7.65$\pm$1.03 & 6.68 $\pm$0.62  \\
He$^+$/H$^+$ ($\lambda$5876) $\times10^2$ & 9.07$\pm$0.54 & 8.63 $\pm$0.71  \\
He$^+$/H$^+$ ($\lambda$6678) $\times10^2$ & 10.6$\pm$3.6 & 10.4 $\pm$ 0.72 \\
He$^+$/H$^+$ mean $\times10^2$ & 9.09$\pm$1.20 & 8.58 $\pm$0.72  \\
He$^{2+}$/H$^+$ ($\lambda$4686)$\times10^3$ & --- & 1.45 $\pm$ 0.08 \\
\vspace{1.5mm}
He/H $\times10^2$ & 9.09$\pm$1.20 & 8.73 $\pm$0.72  \\
O$^+$/H$^+$$^a$ $\times10^5$ &   5.51$\pm^{  3.35}_{  1.56}$ & 
  4.32$\pm^{  0.43}_{  0.36}$\\
O$^{++}$/H$^+$ $\times10^5$ &   5.24$\pm^{  1.96}_{  1.13}$ & 
  6.30$\pm^{  0.50}_{  0.44}$\\
O/H $\times10^4$ &  1.08$\pm^{  0.53}_{  0.27}$ & 
  1.06$\pm^{  0.09}_{  0.08}$\\
\vspace{1.5mm}
12+log(O/H) &  8.03$\pm^{  0.17}_{  0.13}$ & 
  8.03$\pm^{  0.04}_{  0.07}$\\
N$^+$/H$^+$ $\times10^6$ &  1.94$\pm^{  0.68}_{  0.44}$ & 
  0.61$\pm^{  0.05}_{  0.05}$\\
ICF(N) &  1.95$\pm^{  2.12}_{  1.04}$ & 
  2.46$\pm^{  0.46}_{  0.39}$\\
N/H  $\times10^6$ &   3.79$\pm^{  6.86}_{  2.42}$ & 
  1.50$\pm^{  0.43}_{  0.33}$\\
12+log(N/H) &  6.58$\pm^{  0.45}_{  0.44}$ & 
  6.18$\pm^{  0.11}_{  0.11}$\\
\vspace{1.5mm}
log(N/O) & $-$1.45$\pm^{  0.57}_{  0.62}$ & 
 $-$1.85$\pm^{  0.14}_{  0.15}$\\
Ne$^{++}$/H$^+$ $\times10^5$ &  3.33$\pm^{  2.24}_{
  1.09}$ & ---\\
ICF(Ne) &  2.05$\pm^{  1.86}_{  0.93}$ & --- \\
Ne/H $\times10^5$&  6.84$\pm^{ 14.98}_{
  4.33}$ & --- \\
12+log(Ne/H) &  7.83$\pm^{  0.50}_{  0.43}$ & --- \\
\vspace{1.5mm}
log(Ne/O) & $-$0.20$\pm^{  0.63}_{  0.61}$ & --- \\
S$^+$/H$^+$ $\times10^7$ &  5.58$\pm^{  1.66}_{  1.12}$ & 
  3.30$\pm^{  0.33}_{  0.30}$\\
S$^{++}$/H$^+$ $\times10^6$&  2.55$^b\pm^{  1.84}_{  0.85}$ & 
  1.20$\pm^{  0.34}_{  0.29}$\\
ICF(S) &  0.81$\pm^{  0.19}_{  0.49}$ &   0.62$\pm^{
  0.14}_{  0.13}$\\
S/H  $\times10^6$ &  2.53$\pm^{  2.58}_{  1.84}$ & 
  0.96$\pm^{  0.51}_{  0.35}$\\
12+log(S/H) &  6.40$\pm^{  0.31}_{  0.56}$ &   5.98$\pm^{
  0.18}_{  0.20}$\\
\vspace{1.5mm}
log(S/O) & $-$1.63$\pm^{  0.43}_{  0.74}$ &  $-$2.05$\pm^{
  0.22}_{  0.24}$\\
  Ar$^{++}$/H$^+$ $\times10^7$& --- & 3.71$\pm^{0.19}_{0.17}$s \\
  ICF(Ar) & --- & 1.07$\pm^{0.05}_{0.03}$ \\
  Ar/H $\times10^7$& --- & 3.98$\pm^{0.38}_{0.31}$ \\
  log(Ar/O) & --- & $-$2.43$\pm^{0.08}_{0.07}$ \\
\hline
\end{tabular}\label{tab:abundances}
\begin{description}
\item[$^a$]Derived using [O~II]~$\lambda$3727 flux from IT98
\item[$^b$]Derived using the relationship between S$^{2+}$ and S$^+$ (Equation A38) of Kinsgburgh \& Barlow (1994)
\end{description}
\end{small}
\end{center}
\end{table}

Abundance maps relative to H$^+$ were created for the N$^{+}$, O$^{2+}$,
S$^{+}$ and S$^{2+}$ ions, using the $\lambda\lambda$6584, 5007, 6717$+$6731
and 6312 lines, respectively, when the appropriate line was detected.  Examples
of flux maps used in the derivations of these abundance maps can be seen in
Figures \ref{fig:O3N2UM420} and \ref{fig:O3N2UM462}, showing the distribution
of flux in the \foiii\ $\lambda$5007 and \fnii\ $\lambda$6584 emission lines
for UM~420 and UM~462, respectively.  Since the \fsiii\ $\lambda$6312 line was
not detected from UM~420, its S$^{2+}$/H$^+$ abundance was estimated using an
empirical relationship between the S$^{2+}$ and S$^{+}$ ionic fractions
(equation A38) from \citet{Kingsburgh:1994}. Ionic abundances were calculated
using \textsc{iraf's} {\tt IONIC} task, using each galaxy's respective \elt\
and \eld\ maps described above, with each VIMOS spaxel treated as a distinct
`nebular zone' with its own set of physical conditions.  Ionic abundance ratios
were also derived from the summed spectra fluxes, whose \elt's and \eld's are
listed in Table \ref{tab:abundances}, using the program {\sc equib06}
(originally written by I. D. Howarth and S. Adams). Table \ref{tab:abundances}
also includes a Ne$^{2+}$/H$^+$ abundance ratio for UM~420, as the \fneiii\
$\lambda$3969 line was redshifted into the HRblue spectral range, and an
Ar$^{2+}$/H$^+$ ratio for UM~462. Abundance maps were however not created for
either ion due to the low S/N ratio \fneiii\ $\lambda$3969 and \fariii\
$\lambda$7135 emission line maps, and the N/H and Ar/H listed in Table
\ref{tab:abundances} have been computed from integrated line ratios across the
IFU. Differences between abundances derived using the different atomic data
tables utilised by {\tt IONIC} and {\sc equib06} were investigated by also
using {\tt IONIC} to compute abundances from the summed spectra. Typical
differences were found to be $\sim$2\% or less. Since \foii\ $\lambda$3727 is
not redshifted into the VIMOS spectral range for either galaxy we have to rely
on published fluxes. We therefore adopted $F$($\lambda$3727)/$F$(H$\beta$)
values of 2.02$\pm$0.04 and 1.46$\pm$0.01 for UM~420 and UM~462, respectively
(from IT98), and corrected them for reddening using the mean $c$(H$\beta$)
values derived here (see Table~\ref{tab:fluxes}). The spectra of IT98 were
obtained using the 2.1-m Kitt Peak Observatory GoldCam spectrograph's
3$''\times$200$''$ slit aligned to collect the maximum amount of flux from the
targets. For UM~420, whose GoldCam and VIMOS H$\alpha$ fluxes agree within 11
per cent, the adoption of \foii\ fluxes from IT98 should introduce only minor
uncertainties in our analysis; for UM~462 whose H$\alpha$ flux measured by IT98
is 35\% of the VIMOS flux, larger uncertainties might be expected. It is
encouraging however that the {\it relative} line fluxes and intensities listed
in Table~\ref{tab:fluxes} differ from those of IT98 by an average factor of
only 0.96 for UM~420 and 1.09 for UM~462. Hence for the derivation of the
O$^{+}$ abundance we adopted a constant $I$($\lambda$3727)/$I$(H$\beta$) ratio
for each galaxy. As a result we cannot, for instance, comment on the spatial
variance of the N/O abundance ratio across the galaxies. An alternative method
for deriving an O$^{+}$ abundance map was investigated in which global
O$^{+}$/N$^{+}$ ratios of 16.49 and 37.64 from IT98 for UM~420 and UM~462,
respectively, were multiplied by the corresponding N$^{+}$/H$^{+}$ IFU maps.
Overall it was found that maps derived using the `global factor' method yielded
elemental oxygen abundances $\sim$40\% higher and nitrogen abundances
$\sim$55\% lower than those derived from a constant
$I$($\lambda$3727)/$I$(H$\beta$) ratio for each galaxy. However, by deriving an
O$^{+}$ abundance map in this way we would be prevented from creating a
meaningful N/H map since ICF(N) $=$ O/O$^+$. We therefore opted to use the
O$^+$ abundance maps derived from the constant $\lambda$3727 fluxes when
creating ICF(N), ICF(S) and O/H abundance maps.

Ionic nitrogen, neon and sulphur abundances were converted into N/H, Ne/H and
S/H abundances using ionisation correction factors (ICFs) from
\citet{Kingsburgh:1994}. Average abundances from integrated galaxy spectra are
listed in Table \ref{tab:abundances}. Elemental O/H abundance maps for UM~420
and UM~462 are shown in Figure \ref{fig:OElAb}, while average ionic and
elemental abundances derived from spectra summed over each individual
star-forming region are listed in Tables \ref{tab:um420Abs} and
\ref{tab:um462Abs} for UM~420 and UM~462, respectively.\footnote{Regional
abundances derived directly by taking averages over the abundance maps were
found to be inaccurate due to occasionally large \elt\ uncertainties resulting
from poor S/N ratio in some spaxels of the \foiii\ $\lambda$4363 maps. In
contrast, summed regional spectra increased the S/N ratio for $\lambda$4363,
allowing more reliable average line ratios and hence temperatures and
abundances for each region to be derived.}

\subsection{UM~420}

\begin{table*}
\caption{UM~420 regional fluxes and intensities used for regional T$_e$ and N$_e$ diagnostics and regional ionic abundance calculations.}
\begin{center}
\begin{small}
\begin{tabular}{lcccc}
\hline
  & \multicolumn{2}{c}{Region 1} & \multicolumn{2}{c}{Region 2} \\
\cline{2-5}
 Line ID  & F($\lambda$) & I($\lambda$) & F($\lambda$) & I($\lambda$)\\
\hline
4340 H$\gamma$ 
 & 40.43$\pm$  0.49
 & 45.05$\pm$  0.54
 & 32.44$\pm$  1.15
 & 35.86$\pm$  1.27
 \\
4363 \foiii\ 
 &  4.51$\pm$  0.31
 &  5.00$\pm$  0.35
 &  5.34$\pm$  0.78
 &  5.88$\pm$  0.86
 \\
4471 \hei\ 
 &  4.14$\pm$  0.33
 &  4.49$\pm$  0.36
 &  4.92$\pm$  0.59
 &  5.30$\pm$  0.64
 \\
4861 H$\beta$ 
 &100.00$\pm$  1.04
 &100.00$\pm$  1.04
 &100.00$\pm$  1.80
 &100.00$\pm$  1.80
 \\
4959 \foiii\ 
 &140.57$\pm$  1.47
 &137.72$\pm$  1.44
 &168.37$\pm$  2.69
 &165.20$\pm$  2.64
 \\
5007 \foiii\ 
 &441.01$\pm$  4.40
 &427.64$\pm$  4.27
 &514.34$\pm$  8.13
 &499.91$\pm$  7.91
 \\
5876 \hei\ 
 & 13.04$\pm$  0.31
 & 10.86$\pm$  0.26
 & 11.29$\pm$  0.79
 &  9.53$\pm$  0.67
 \\
6563 H$\alpha$ 
 &366.84$\pm$  5.21
 &278.98$\pm$  3.96
 &296.63$\pm$  3.88
 &230.33$\pm$  3.01
 \\
6584 \fnii\ 
 & 29.66$\pm$  0.31
 & 22.50$\pm$  0.24
 & 16.80$\pm$  0.98
 & 13.02$\pm$  0.76
 \\
6678 \hei\ 
 &  2.24$\pm$  0.32
 &  1.68$\pm$  0.24
 & --- & ---
 \\
6716 \fsii\ 
 & 30.15$\pm$  0.51
 & 22.52$\pm$  0.38
 & 24.11$\pm$  0.89
 & 18.40$\pm$  0.68
 \\
6731 \fsii\ 
 & 23.31$\pm$  0.51
 & 17.37$\pm$  0.38
 & 20.81$\pm$  0.94
 & 15.86$\pm$  0.72
 \\
F(H$\beta$)$^a$ & \multicolumn{2}{c}{300.4$\pm$  3.330} & \multicolumn{2}{c}{ 48.414$\pm$  0.874}\\
\hline
\end{tabular}\label{tab:um420RegFlux}
\begin{description}
\item{$^a$} in units of $10^{-16}$erg s$^{-1}$ cm$^{-2}$
\end{description}
\end{small}
\end{center}
\end{table*}

The line fluxes for the summed spectra over Regions 1 and 2 of UM~420 are listed in Table~\ref{tab:um420RegFlux} and the average physical properties from which they are derived are listed in Table \ref{tab:um420Abs}.  A graphical representation of the variations in elemental and ionic abundances
between Regions 1 and 2 is shown in Figure~\ref{fig:abugrid}.

\begin{table}
\caption{Ionic and elemental abundances for UM~420, derived from summed IFU spectra over each region.  All ionisation correction factors (ICFs) are taken from Kingsburgh \& Barlow, (1994).}
\begin{center}
\begin{small}
\begin{tabular}{lcc}
\hline
 Property & Region 1 & Region 2 \\
\hline
T$_e$ (O~III)/ K & 12300$\pm$330 & 12900$\pm$680\\
N$_e$ (S~II)/ cm$^{-3}$ & 120$\pm$55 & 300$\pm$160  \\
\vspace{1.8mm}
c(H$\beta$) & 0.37$\pm$0.04 & 0.34$\pm$0.08 \\
O$^+$/H$^+$$^a$ $\times10^4$ &   0.95$\pm$  0.12&  0.89$\pm$  0.25\\
O$^{++}$/H$^+$ $\times10^4$ &   0.79$\pm$  0.07&  0.92$\pm$  0.18\\
O/H $\times10^4$ &  1.74$\pm$  0.19&  1.81$\pm$  0.43\\
\vspace{1.5mm}
12+log(O/H) &  8.24$\pm$  0.05&  8.26$\pm$  0.09\\
N$^+$/H$^+$ $\times10^6$ &   2.67$\pm$  0.18&  1.55$\pm$  0.22\\
N/H $\times10^6$ &  4.89$\pm$  1.60&  3.15$\pm$  2.39\\
12+log(N/H) &  6.69$\pm$  0.12&  6.50$\pm$  0.25\\
\vspace{1.5mm}
log(N/O) & $-$1.55$\pm$  0.17& $-$1.76$\pm$  0.33\\
S$^+$/H$^+$ $\times10^7$ &  13.38$\pm$  0.92& 11.26$\pm$  1.65\\
S$^{++}$/H$^+$$^b$ $\times10^6$ &   7.47$\pm$  0.49&  6.39$\pm$  0.89\\
S/H $\times10^6$ &  7.58$\pm$  1.51&  5.87$\pm$  2.62\\
12+log(S/H) &  6.88$\pm$  0.08&  6.77$\pm$  0.16\\
log(S/O) & $-$1.36$\pm$  0.12& $-$1.49$\pm$  0.24\\
\hline
\end{tabular}\label{tab:um420Abs}
\begin{description}
\item[$^a$]Derived using [O~II]~$\lambda$3727 flux from IT98
\item[$^b$]Derived using the relationship between S$^{2+}$ and S$^+$(Equation A38) of Kinsgburgh \& Barlow (1994)
\end{description}
\end{small}
\end{center}
\end{table}


The elemental oxygen abundance map shown in the left-hand panel of
Figure~\ref{fig:OElAb} displays two peaks whose locations correlate with the
two H$\alpha$ peaks of UM~420 (marked in Figure~\ref{fig:UM420Regions}).
Minimal abundance variations are seen between Regions 1 and 2, whose average
metallicity is 12 $+$ log(O/H) $\sim$ 8.25 $\pm$ 0.07. Adopting a solar oxygen
abundance of 8.71$\pm$0.10 relative to hydrogen \citep{Scott:2009}, this
corresponds to an oxygen abundance of $\sim$0.35 solar for Regions 1 and 2. This is $\sim$0.2~dex higher than the value derived from the integrated spectrum of the whole galaxy.  This could point to some limited oxygen enrichment in the areas corresponding to the peak H$\alpha$ emission, that is the clusters associated with Regions 1 and 2.  However, since all the oxygen abundances overlap within their 1$\sigma$ uncertainties, we cannot conclude that this variation in oxygen abundance is significant.  The nitrogen abundances of the two regions of star-formation are consistent within
their uncertainties (Table~\ref{tab:um420Abs} ). The N/O ratios for both
regions are consistent with those of other metal-poor emission line galaxies of
similar oxygen metallicity \citep{Izotov:2006a}, and do not show the nitrogen
excess reported by \citet{Pustilnik:2004}. The S/O ratio for Region 1 is
slightly higher than expected for BCGs, although Region 2 is closer to the
average reported range of log(S/O) $\sim$ $-$1.4 -- $-$1.7
\citep{Izotov:2006a}.

\subsection{UM~462}
\begin{table*}
\caption{UM~462 regional fluxes and intensities used for regional T$_e$ and N$_e$ diagnostics and regional ionic abundance calculations.}
\begin{center}
\begin{small}
\begin{tabular}{lcccccccc}
\hline
  & \multicolumn{2}{c}{Region 1} & \multicolumn{2}{c}{Region 2}& \multicolumn{2}{c}{Region 3} & \multicolumn{2}{c}{Region 4} \\
\cline{2-9}
 Line ID  & F($\lambda$) & I($\lambda$) & F($\lambda$) & I($\lambda$) & F($\lambda$) & I($\lambda$) & F($\lambda$) & I($\lambda$)\\
\hline
4340 H$\gamma$ &
      40.92$\pm$     0.72 &
      46.49$\pm$     0.49 &
      45.97$\pm$     0.41 &
      43.39$\pm$     0.54 &
      45.98$\pm$  0.41
 & 53.86$\pm$  0.48
 & 43.39$\pm$  0.54
 & 44.45$\pm$  0.56
 \\
4363 \foiii\ 
 &  8.61$\pm$  0.18
 &  9.26$\pm$  0.19
 &  7.27$\pm$  0.26
 &  7.55$\pm$  0.27
 &  9.17$\pm$  0.70
 & 10.67$\pm$  0.82
 &  6.62$\pm$  0.53
 &  6.77$\pm$  0.55
 \\
4471 \hei\ 
 &  3.01$\pm$  0.14
 &  3.19$\pm$  0.15
 &  3.51$\pm$  0.16
 &  3.61$\pm$  0.16
 &  3.66$\pm$  0.34
 &  4.12$\pm$  0.38
 &  3.63$\pm$  0.40
 &  3.70$\pm$  0.40
 \\
4686 \heii\ 
 & --- & ---
 &  1.21$\pm$  0.12
 &  1.22$\pm$  0.12
 & --- & ---
 & --- & ---
 \\
4861 H$\beta$ 
 &100.00$\pm$  0.65
 &100.00$\pm$  0.65
 &100.00$\pm$  0.65
 &100.00$\pm$  0.65
 &100.00$\pm$  0.65
 &100.00$\pm$  0.65
 &100.00$\pm$  0.76
 &100.00$\pm$  0.76
 \\
4959 \foiii\ 
 &218.90$\pm$  1.90
 &215.76$\pm$  1.87
 &169.17$\pm$  1.30
 &167.95$\pm$  1.29
 &182.95$\pm$  1.23
 &177.54$\pm$  1.19
 &174.63$\pm$  1.52
 &173.83$\pm$  1.51
 \\
5007 \foiii\ 
 &630.70$\pm$  5.95
 &617.18$\pm$  5.82
 &503.02$\pm$  2.64
 &497.58$\pm$  2.61
 &564.84$\pm$  2.32
 &539.98$\pm$  2.22
 &507.96$\pm$  4.43
 &504.49$\pm$  4.40
 \\
5876 \hei\ 
 & 14.47$\pm$  0.16
 & 12.72$\pm$  0.14
 &  9.77$\pm$  0.59
 &  9.16$\pm$  0.55
 & 15.02$\pm$  0.64
 & 11.48$\pm$  0.49
 & ---
 &  ---
 \\
6563 H$\alpha$ 
 &332.75$\pm$  0.98
 &274.41$\pm$  0.81
 &325.84$\pm$  2.44
 &295.81$\pm$  2.22
 &483.79$\pm$  3.16
 &324.20$\pm$  2.12
 &293.19$\pm$  4.92
 &275.84$\pm$  4.63
 \\
6584 \fnii\ 
 &  4.25$\pm$  0.11
 &  3.50$\pm$  0.09
 &  6.88$\pm$  0.23
 &  6.24$\pm$  0.21
 &  9.88$\pm$  0.48
 &  6.60$\pm$  0.32
 &  5.04$\pm$  0.42
 &  4.74$\pm$  0.40
 \\
6678 \hei\ 
 &  2.34$\pm$  0.10
 &  1.91$\pm$  0.08
 &  3.21$\pm$  0.13
 &  2.90$\pm$  0.12
 &  5.19$\pm$  0.44
 &  3.41$\pm$  0.29
 &  ---
 &  ---
 \\
6716 \fsii\ 
 &  7.53$\pm$  0.10
 &  6.13$\pm$  0.08
 & 16.11$\pm$  0.20
 & 14.53$\pm$  0.18
 & 28.09$\pm$  0.72
 & 18.32$\pm$  0.47
 & 11.99$\pm$  0.38
 & 11.23$\pm$  0.35
 \\
6731 \fsii\ 
 &  5.56$\pm$  0.10
 &  4.52$\pm$  0.08
 & 11.91$\pm$  0.20
 & 10.74$\pm$  0.18
 & 19.90$\pm$  0.73
 & 12.95$\pm$  0.47
 &  8.79$\pm$  0.40
 &  8.23$\pm$  0.37
 \\
F(H$\beta$)$^a$ & \multicolumn{2}{c}{1065.01$\pm$  6.32} & \multicolumn{2}{c}{594.32$\pm$6.03} & \multicolumn{2}{c}{ 96.24$\pm$2.33} & \multicolumn{2}{c}{ 275.04$\pm$ 15.62}\\
\hline
\end{tabular}\label{tab:um462RegFlux}
\begin{description}
\item{$^a$} in units of $10^{-16}$erg s$^{-1}$ cm$^{-2}$
\end{description}
\end{small}
\end{center}
\end{table*}

Table~\ref{tab:um462RegFlux} lists the fluxes from summed spectra over Regions 1--4 of UM~462. The average ionic and elemental abundances, along with the average \elt\ and \eld\ values, derived from the summed spectra over the respective regions are listed in Table~\ref{tab:um462Abs}.  A graphical representation of the variation in elemental and ionic abundances across Regions~1--4 is shown in Figure~\ref{fig:abugrid}.
\begin{table*}
\caption{Ionic and elemental abundances for UM~462, derived from summed IFU spectra over each region.  All ionisation correction factors (ICFs) are taken from Kingsburgh \& Barlow, (1994).}
\begin{center}
\begin{small}
\begin{tabular}{lcccc}
\hline
 Property & Region 1 & Region 2 & Region 3 & Region 4 \\
\hline
T$_e$ (O~III)/ K & 13400$\pm$150 & 13500$\pm$230 & 15200$\pm$500 & 12800$\pm$440 \\
N$_e$ (S~II)/ cm$^{-3}$ & 70$\pm$40 & 70 $\pm$40 & 20$\pm^{70}_{10}$ & 60$\pm$50 \\
\vspace{1.8mm}
c(H$\beta$) & 0.26$\pm$0.03 & 0.13$\pm$0.03 & 0.54$\pm$0.03 & 0.08$\pm$0.08 \\
O$^+$/H$^+$$^a$ $\times10^5$ &   4.85$\pm$  0.28&  4.38$\pm$  0.33&
  3.91$\pm$  0.44&  5.09$\pm$  0.81\\
O$^{++}$/H$^+$ $\times10^5$ &   8.86$\pm$  0.30&  6.99$\pm$  0.34&
  5.58$\pm$  0.49&  8.17$\pm$  0.88\\
O/H $\times10^4$ &  1.37$\pm$  0.06&  1.14$\pm$  0.07&
  0.95$\pm$  0.09&  1.33$\pm$  1.49\\
\vspace{1.5mm}
12+log(O/H) &  8.14$\pm$  0.02&  8.06$\pm$  0.02&
  7.98$\pm$  0.04&  8.12$\pm$  0.05\\
N$^+$/H$^+$$\times10^6$ &   0.34$\pm$  0.01&  0.60$\pm$  0.02&
  0.50$\pm$  0.03&  0.51$\pm$  0.04\\
N/H $\times10^6$ &  0.97$\pm$  0.13&  1.56$\pm$  0.28&
  1.22$\pm$  0.41&  1.33$\pm$  0.56\\
12+log(N/H) &  5.99$\pm$  0.05&  6.19$\pm$  0.07&
  6.08$\pm$  0.13&  6.12$\pm$  0.15\\
\vspace{1.5mm}
log(N/O) & $-$2.15$\pm$  0.07& $-$1.86$\pm$  0.10&
 $-$1.89$\pm$  0.17& $-$2.00$\pm$  0.21\\
S$^+$/H$^+$ $\times10^7$ &   3.01$\pm$  0.09&  7.04$\pm$  0.28&
  7.06$\pm$  0.45&  6.01$\pm$  0.51\\
S$^{++}$/H$^+$ $\times10^6$ &   0.96$\pm$  0.04&  1.27$\pm$  0.08&
  1.37$\pm$  0.15&  1.97$\pm$  0.26\\
S/H $\times10^6$ &  0.66$\pm$  0.12&  1.15$\pm$  0.27&
  1.32$\pm$  0.51&  1.49$\pm$  0.80\\
12+log(S/H) &  5.82$\pm$  0.07&  6.06$\pm$  0.09&
  6.12$\pm$  0.14&  6.17$\pm$  0.19\\
log(S/O) & $-$2.32$\pm$  0.09& $-$2.00$\pm$  0.12&
 $-$1.86$\pm$  0.19& $-$1.95$\pm$  0.24\\
\hline
\end{tabular}\label{tab:um462Abs}
\begin{description}
\item[$^a$]Derived using [O~II]~$\lambda$3727 flux from IT98
\end{description}
\end{small}
\end{center}
\end{table*}


The right-hand panel of Figure~\ref{fig:OElAb}(a) shows the O/H abundance ratio
map for UM~462. The oxygen abundance varies spatially across the different
star-forming regions of UM~462 (as marked in Figure~\ref{fig:UM462Regions}).
Four maxima in oxygen abundance are seen across the map, all aligning spatially
with peaks in H$\alpha$ emission (shown as overlaid contours), with the
exception of Region 1, where the maximum in oxygen abundance appears to lie
$\sim$1.5$''$ south of the H$\alpha$ peak of Region 1.  A decrease in the
oxygen abundance can be seen $\sim$1.0$''$ north-east of the peak in Region 1.
Oxygen abundances for Regions 1--4 are listed in Table \ref{tab:um462Abs}. A
maximum variation of 40\% is seen between Region 1 (displaying the highest
metallicity) and Region 3 (displaying the lowest). The mean oxygen abundance of
the four identified star-forming regions in UM~462 is 12 $+$ log(O/H) $=$ 8.08
$\pm$ 0.94, that is, $\sim$0.25 solar. This is in good agreement with the value derived from the integrated spectrum of the whole galaxy. Regional N/H and S/H abundance ratios
are given in Table \ref{tab:um462Abs}. Regions 2, 3 and 4 have the same
nitrogen abundance within the uncertainties. The maximum difference (0.2 dex)
is between Regions 2 and 1; the latter shows the lowest nitrogen abundance. In
comparison with metal-poor emission line galaxies of similar metallicity
\citep{Izotov:2006a}, UM~462 appears to be overall slightly nitrogen poor. The
sulphur abundance across the galaxy is variable with Region~1 having the
lowest, a factor of $\sim$ 2 below the other three regions. In comparison with
other emission line galaxies, UM~462 displays a lower than average log(S/O) of
$-$2.03 [typical values range from $\sim$ $-$1.4 to $-$1.7;
\citet{Izotov:2006a}].

\begin{figure*}
\begin{center}
\includegraphics[scale=0.6, angle=0]{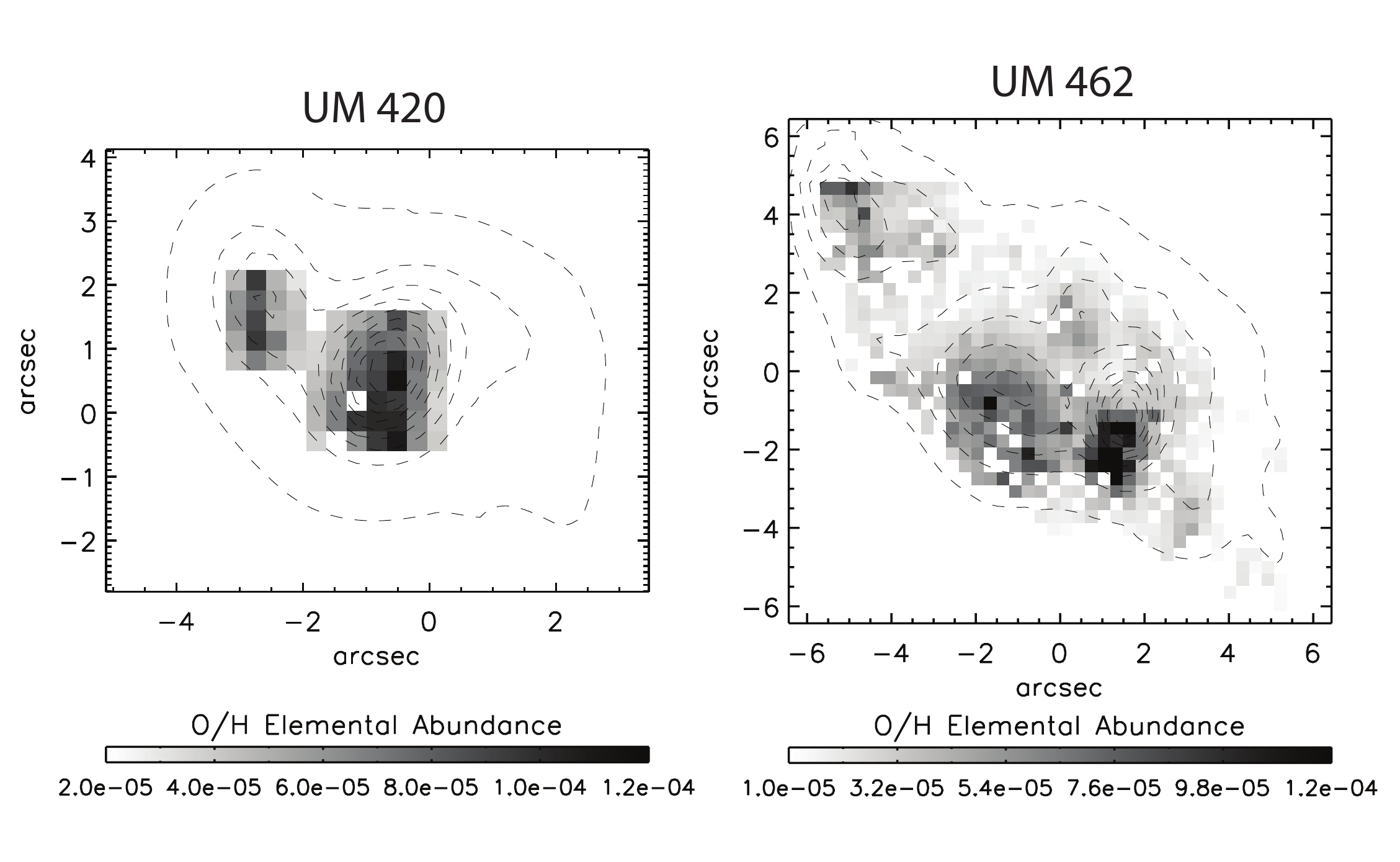}
\caption{O/H abundance ratio maps of UM~420 (left panel) and UM~462 (right
panel) derived from their respective O$^{2+}$ and O$^+$ ionic abundance maps
and \elt\ and \eld\ maps, as described in Section \ref{sec:abundances}.
Overlaid are the H$\alpha$ flux contours from Figure~\ref{fig:HaMaps}(a).}
\label{fig:OElAb}
\end{center}
\end{figure*}

\begin{figure}
\begin{center}
\includegraphics[scale=0.45, angle=0]{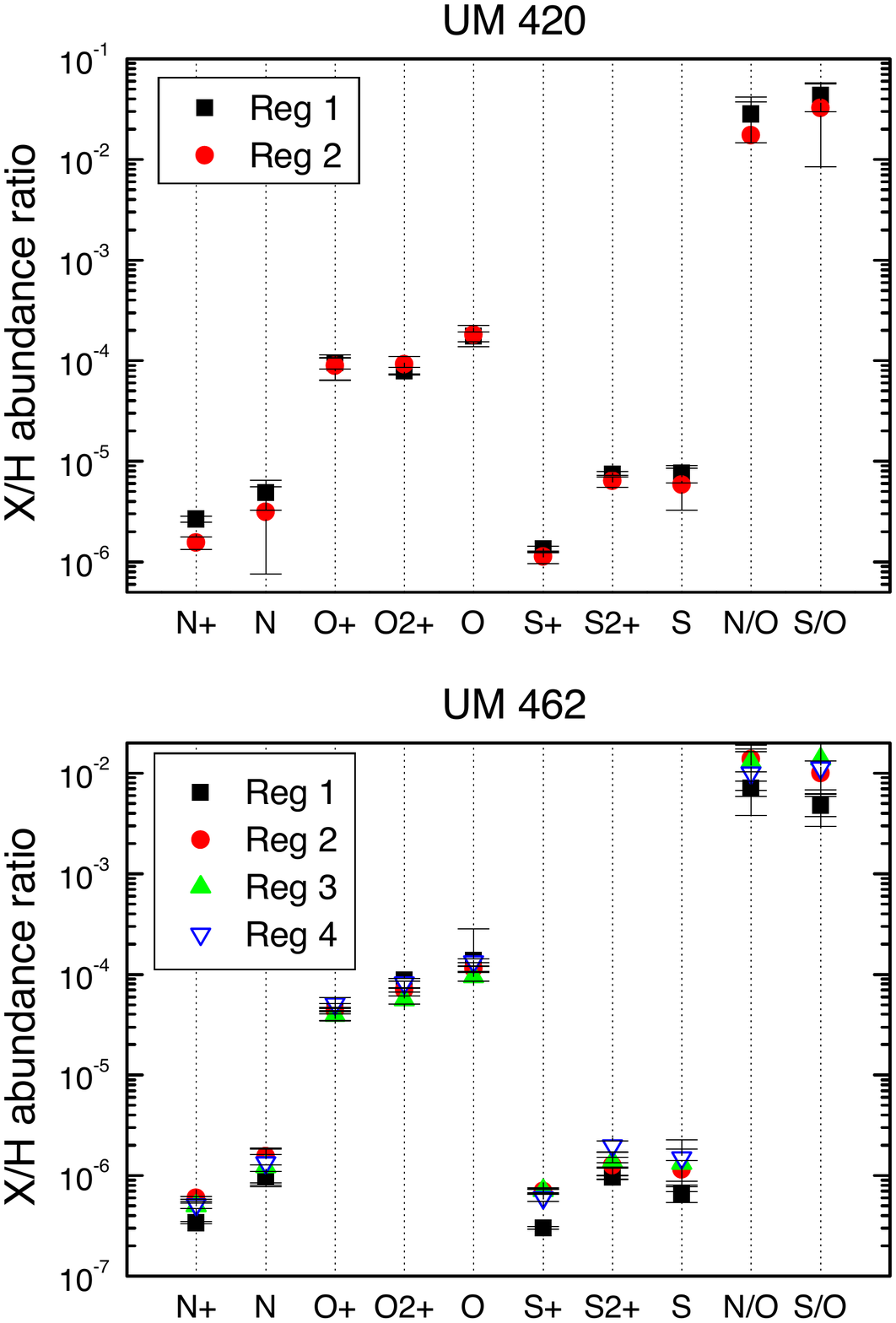}
\caption{Graphical representation of the variations in ionic and elemental
abundances across UM~420 and UM~462, as given in Tables~\ref{tab:um420Abs} and
\ref{tab:um462Abs}, respectively.} \label{fig:abugrid}
\end{center}
\end{figure}

\subsubsection{Helium abundances}

Following the method outlined by \citet{Tsamis:2003} and
\citet{Wesson:2008}, abundances for helium were derived using atomic data from
\citet{Smits:1996}, accounting for the effects of collisional excitation using
the formulae in \citet{Benjamin:1999}. Ionic and total helium abundances
relative to hydrogen derived from helium recombination lines measured from
summed spectra over the entire galaxies are given in Table~\ref{tab:abundances}
(derived using the global \elt\ and \eld\ values from the same Table). Given
that the reddening varies across the galaxies (predominantly in the case of
UM~462), and in order to minimise the uncertainties, we have also computed
helium abundances per each individual star forming region; these are given in
Tables~\ref{tab:HeAbundsum420} and \ref{tab:HeAbundsum462}, for UM~420 and
UM~462, respectively. These values are derived from summed spectra over each
star-forming region of the galaxy dereddened with the corresponding
$c$(H$\beta$). Our adopted mean values for He$^+$/H$^+$ were derived from the
$\lambda$4471, $\lambda$5876 and $\lambda$6678 lines, averaged with weights
1:3:1.  The helium abundances were calculated for the average temperatures and
densities given in Tables~\ref{tab:um420Abs} and \ref{tab:um462Abs}. Regional
helium abundances were not calculated for Region~4 of UM~462 because only one
\hei\ line was detected (see Table~\ref{tab:um462RegFlux}). Helium abundances
derived from summed spectra are found to agree with those derived in IT98,
within the uncertainties.

\begin{table}
\begin{center}
\begin{small}
\caption{Ionic and elemental helium abundances for UM~420 derived from summed
spectra over each region.$^a$}
\begin{tabular}{lcc}
\hline
  & Region 1 & Region 2 \\
\hline
He$^+$/H$^+$ ($\lambda$4471) $\times10^2$ & 9.23$\pm$0.73 & 10.77$\pm$1.23\\
He$^+$/H$^+$ ($\lambda$5876) $\times10^2$ &  8.31$\pm$0.20& 7.15$\pm$0.52\\
He$^+$/H$^+$ ($\lambda$6678) $\times10^2$ & ~4.56$\pm$0.66 : & --- \\
He$^+$/H$^+$ mean $\times10^2$            & 8.54$\pm$0.33 & 8.05$\pm$0.70\\
He$^{2+}$/H$^+$ ($\lambda$4686)$\times10^3$ & --- & --- \\
He/H $\times10^2$ &  8.54$\pm$0.33 & 8.05$\pm$0.70\\
\hline \label{tab:HeAbundsum420}
\end{tabular}
\begin{description}
\item[$^a$] Entries followed by `:' were excluded from the final average.

\end{description}

\end{small}
\end{center}
\end{table}

\begin{table*}
\begin{center}
\begin{small}
\caption{Ionic and elemental helium abundances for UM~462 derived from summed
spectra over each Region (with the exclusion of Region~4, where only one \hei\
line was detected).$^a$}
\begin{tabular}{lccc}
\hline
  & Region 1 & Region 2 & Region 3\\
\hline
He$^+$/H$^+$ ($\lambda$4471) $\times10^2$ &6.66$\pm$0.32  & 7.54$\pm$0.34  &   8.75$\pm$0.81  \\
He$^+$/H$^+$ ($\lambda$5876) $\times10^2$ & 9.97$\pm$0.10  & 7.20$\pm$0.43  & 9.33$\pm$ 0.40 \\
He$^+$/H$^+$ ($\lambda$6678) $\times10^2$ & ~5.34$\pm$0.23 :  & 8.11$\pm$ 0.33 & 9.83$\pm$0.84  \\
He$^+$/H$^+$ mean $\times10^2$ &9.14$\pm$0.16 &  7.45$\pm$0.39  &   9.31$\pm$0.57  \\
He$^{2+}$/H$^+$ ($\lambda$4686)$\times10^3$ & --- & 1.04$\pm$0.11 &   ---\\
He/H $\times10^2$ &   9.14$\pm$0.16 &  7.55$\pm$0.40 &  9.31$\pm$0.57 \\
\hline \label{tab:HeAbundsum462}
\end{tabular}
\begin{description}
\item[$^a$] Entries followed by `:' were excluded from the final average.

\end{description}

\end{small}
\end{center}
\end{table*}

\subsection{Stellar Properties}
\subsubsection{Wolf-Rayet features?}

Contrary to previous long slit studies
\citep{Izotov:1998,Schaerer:1999,Guseva:2000}, which had limited spatial
coverage, the VIMOS IFU spectra do not reveal any evidence for broad WR
emission features at either $\sim$4686~\AA\ or at $\sim$5808, 5812~\AA. This is
true for single spaxel spectra, summed spectra over Region 1 (the highest
surface brightness SF region in both galaxies) or in the summed spectra over
the whole of each galaxy, as shown in Figure~\ref{fig:WRspec}. Hence this
analysis does not support the classification of UM~420 and UM~462 as Wolf-Rayet
galaxies. When searching for WR features one must consider the width of the
extraction aperture, which can sometimes be too large and thus dilute weak WR
features by the continuum flux \citep{Lopez-Sanchez:2008}. We therefore also
examined the summed spectra over each individual star-forming region, with
extraction apertures equal to those displayed in Figures~\ref{fig:UM420Regions}
and ~\ref{fig:UM462Regions} aiming to reduce the dilution of any WR features.
However, those remained undetected. This supports the statement by
\citet{Schaerer:1999} regarding the dependence of a object's classification as
a `WR galaxy' on the quality of the spectrum, and location and size of the
aperture. Given that our spectroscopy was obtained with an 8.2~m telescope,
with exposure times of $\sim$1500~s for each spectral region, we conclude that
the presence of WR stars in UM~420 and UM~462 is not yet proven.

\begin{figure*}
\begin{center}
\includegraphics[scale=0.75, angle=0]{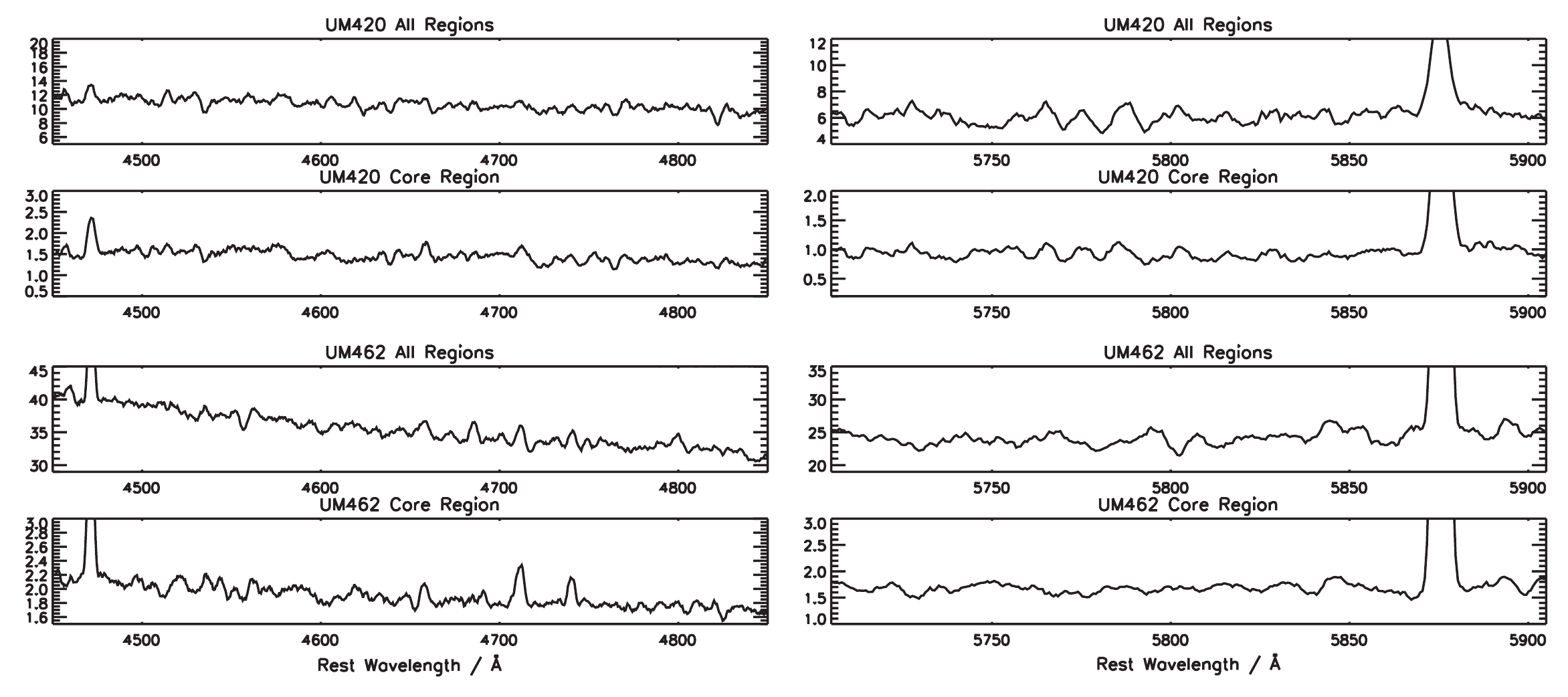}
\caption{Sections of VIMOS IFU spectra, in flux units of
$\times10^{-16}$\,erg~s$^{-1}$~cm$^{-2}$, where the WR `blue bump' feature
(left column) and \civ\ `red' feature (right column) would be located within
summed spectra over the core region (labelled Region~1 in
Figures~\ref{fig:UM420Regions} and \ref{fig:UM462Regions} for UM~420 and
UM~462, respectively) and over the entire galaxy. There are no detections of WR
emission features in any of the spectra.} \label{fig:WRspec}
\end{center}
\end{figure*}

\subsubsection{Starburst ages and star formation rates}
\label{sec:stellarAge}

Luminosities of hydrogen recombination lines, particularly H$\beta$, can
provide estimates of the ionising flux present, assuming a radiation-bounded
nebula \citep{Schaerer:1998}. Thus, the equivalent width ($EW$) of H$\beta$ is
commonly used as an age indicator of the ionising stellar population at a given
metallicity. Maps of $EW$(H$\beta$) for UM~420 and UM~462 can be seen in
Figure~\ref{fig:EWHbeta}(a) \& (b), respectively. We can use these maps, in
conjunction with the metallicity maps described in Section
\ref{sec:abundances}, to estimate the age of the latest star forming episodes
throughout UM~420 and UM~462 by comparing the regional observed average
$EW$(H$\beta$) values with those predicted by the spectral synthesis code
STARBURST99 \citep{Leitherer:1999}. For the models we chose metallicities of
0.4 and 0.2\,\Zsol, representative of the average metallicities of UM~420 and
UM~462 (derived from regional summed spectra), together with assumptions of an
instantaneous burst with a Salpeter initial mass function (IMF), a total mass
of 1$\times10^6$~\Msol\ (the default mass chosen by \citet{Leitherer:1999} to
produce properties that are typical for actual star-forming galaxies), and a
100\Msol\ upper stellar mass limit (which approximates the classical
\citet{Salpeter:1955} IMF). Models were run for Geneva tracks with ``high''
mass loss rates (tracks recommended by the Geneva group) and Padova tracks with
thermally pulsing AGB stars included. For each of these evolutionary tracks,
two types of model atmosphere were used; firstly the Pauldrach-Hillier (PH, the
recommended atmosphere) and secondly Lejeune-Schmutz (LS). The latter was
chosen because it incorporates stars with strong winds, which would be
representative of the WR population within each galaxy (if present).  The
stellar ages predicted by each model and the observed average $EW$(H$\beta$)
within each peak emission region in UM~420 and UM~462 were derived from
Figure~\ref{fig:EWSB99} and are listed in Table~\ref{tab:EWs}. The difference
between the ages predicted by the Geneva and Padova stellar evolutionary tracks
is relatively small, with the Geneva tracks predicting lower ages by up to
20\%. The difference in ages predicted by the PH and LS atmospheres are smaller
still, with LS predicting lower ages by $\sim$6\%.  We cannot comment on which
atmosphere model is more appropriate given that the existence of WR stars
within each galaxy is questioned by the current study, hence we adopt average
stellar ages from the four model combinations. The current (instantaneous) star
formation rates (SFRs) based on the H$\alpha$ luminosities were calculated
following \citet{Kennicutt:1998} and are given in Table~\ref{tab:EWs}. SFRs
corrected for the sub-solar metallicities of these targets are also given,
derived following the methods outlined in \citet{Lee:2002}.

\begin{table*}
\begin{center}
\begin{small}
\caption{Age of the latest star formation episode and current star formation
rates.}
\begin{tabular}{lcc|cccc}
\hline
 & \multicolumn{6}{c}{Starburst ages (Myr)}\\
 \cline{2-7}
 & \multicolumn{2}{c}{UM~420} & \multicolumn{4}{c}{UM~462} \\
 \hline
 Model & Reg.~1 & Reg.~2 & Reg.~1 & Reg.~2 & Reg.~3 & Reg.~4 \\
Padova-AGB PH & 4.51$\pm$0.21 & 5.26$\pm$0.19 & 4.85$\pm$0.32 & 5.45$\pm$0.13 & 5.85$\pm$0.13  & 5.90$\pm$0.27 \\
Padova-AGB LS & 4.24$\pm$0.34 & 5.16$\pm$0.19 & 4.66$\pm$0.36 & 5.31$\pm$0.10 & 5.66$\pm$0.07  & 5.76$\pm$0.24 \\
Geneva-High PH & 3.76$\pm$0.36 & 4.91$\pm$0.27 & 3.91$\pm$0.56 & 4.86$\pm$0.13 & 5.21$\pm$0.10  & 5.51$\pm$0.39 \\
Geneva-High LS & 3.50$\pm$0.34 & 4.71$\pm$0.22 & 3.76$\pm$0.48 & 4.61$\pm$0.10 & 5.01$\pm$0.10  & 5.21$\pm$0.33 \\

\\
& \multicolumn{6}{c}{Star formation rates (M$_\odot$ yr$^{-1}$)}\\
\cline{2-7}
& \multicolumn{2}{c}{UM~420} & \multicolumn{4}{c}{UM~462} \\
\hline
SFR(H$\alpha$)$^a$              & 10.5 & 1.31 &  0.104 &  0.047  &  0.021 & 0.018 \\
SFR(H$\alpha$)$^b$              & 7.0  &0.87  & 0.069  & 0.027 & 0.012 & 0.013 \\
\hline \label{tab:EWs}
\end{tabular}
\begin{description}
\item[$^a$] Derived using the relationship between SFR and $L$(H$\alpha$) from
\citet{Kennicutt:1998}.

\item[$^b$] Corrected for sub-solar metallicities following \citet{Lee:2002}.

\end{description}
\end{small}
\end{center}
\end{table*}

{\it UM~420}: Regions 1 and 2 contain ionising stellar populations with weighted average ages of 3.94$\pm$0.31 and 4.99$\pm$0.21~Myr, respectively
(Table~\ref{tab:EWs}).  Interestingly, the ages returned by all 4 models consistently show Region~2 to be older than Region~1, which could imply that these were in
fact once isolated bodies, and that star formation within Region 1 was
triggered or is still a product of an ongoing merger with Region 2. This also agrees well with the in-falling merger scenario derived from radial-velocity maps (discussed in Section~\ref{sec:HaUM420}). However, the significance of the derived difference in ages between the
two regions is only at the 2$\sigma$ level. Also in
support of this are their respective SFRs which differ by almost an order of
magnitude. Global SFRs derived by \citet{Lopez-Sanchez:2008} are 3.7 $\pm$
0.2~\Msol~yr$^{-1}$ (based on an H$\alpha$ flux estimated from broad-band
photometry) and 1.9 $\pm$ 0.9~\Msol~yr$^{-1}$ (based on the 1.4~GHz flux), and
are lower than our measurements which are based on the monochromatic H$\alpha$
flux.

{\it UM~462}: Table~\ref{tab:EWs} show average stellar population ages of
4.19$\pm$0.42, 5.07$\pm$0.11, 5.45$\pm$0.10 and 5.56$\pm$0.31~Myr for Regions
1--4, respectively. Since Region 1 shows the highest ionising flux, we would
expect it to contain the youngest stellar population. It is notable that the
age of the stellar population increases through Regions 2--4 as the distance of
separation from Region 1 increases and at the same time the SFR decreases. Our
SFRs are comparable to previously published values of 0.13 (based on FIR
fluxes), 0.00081 (blue) and 0.5 \Msol\ yr$^{-1}$ (narrow-band H$\alpha$) from
\citet{Sage:1992} and 0.36 (60~$\mu$m) and 0.13 \Msol\ yr$^{-1}$ (1.4~GHz) from
\citet{Hopkins:2002}. Evidence suggests that Regions~1--3 are linked in their
SF properties, with decreasing metallicity and SFR and increasing stellar
population ages as the distance from Region 1 increases. Region~4 breaks this
pattern in metallicity harbouring a starburst which is coeval with that of
Region~3.


\begin{figure}
\begin{center}
\includegraphics[scale=0.60, angle=0]{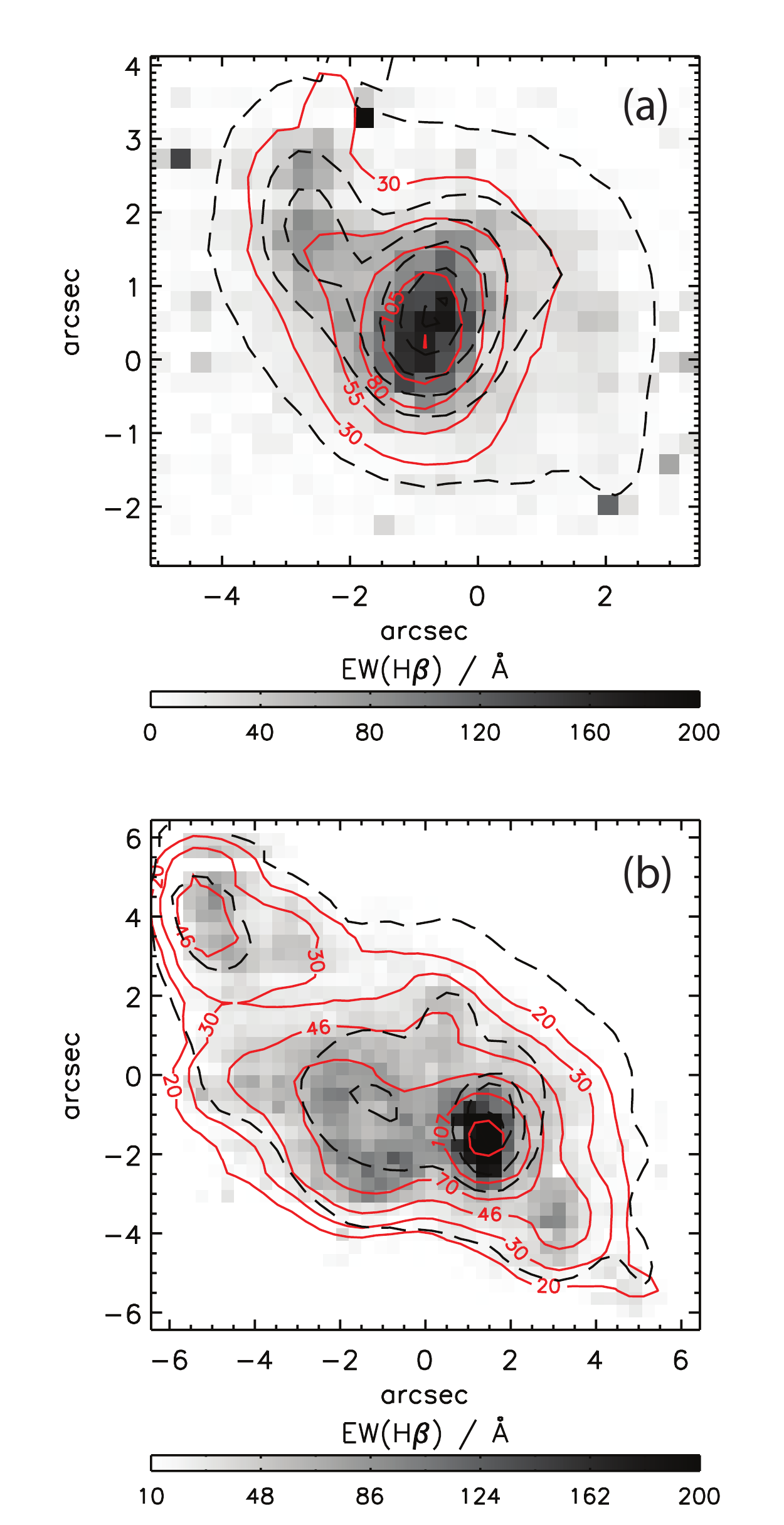}
\caption{Maps of the equivalent width of H$\beta$ across: (a) UM~420; (b)
UM~462. Overlaid red solid lines are contours of $EW$(H$\beta$); black-dashed
lines are H$\alpha$ flux contours from Figure~\ref{fig:HaMaps}. }
\label{fig:EWHbeta}
\end{center}
\end{figure}
\begin{figure}
\begin{center}
\includegraphics[scale=0.80, angle=0]{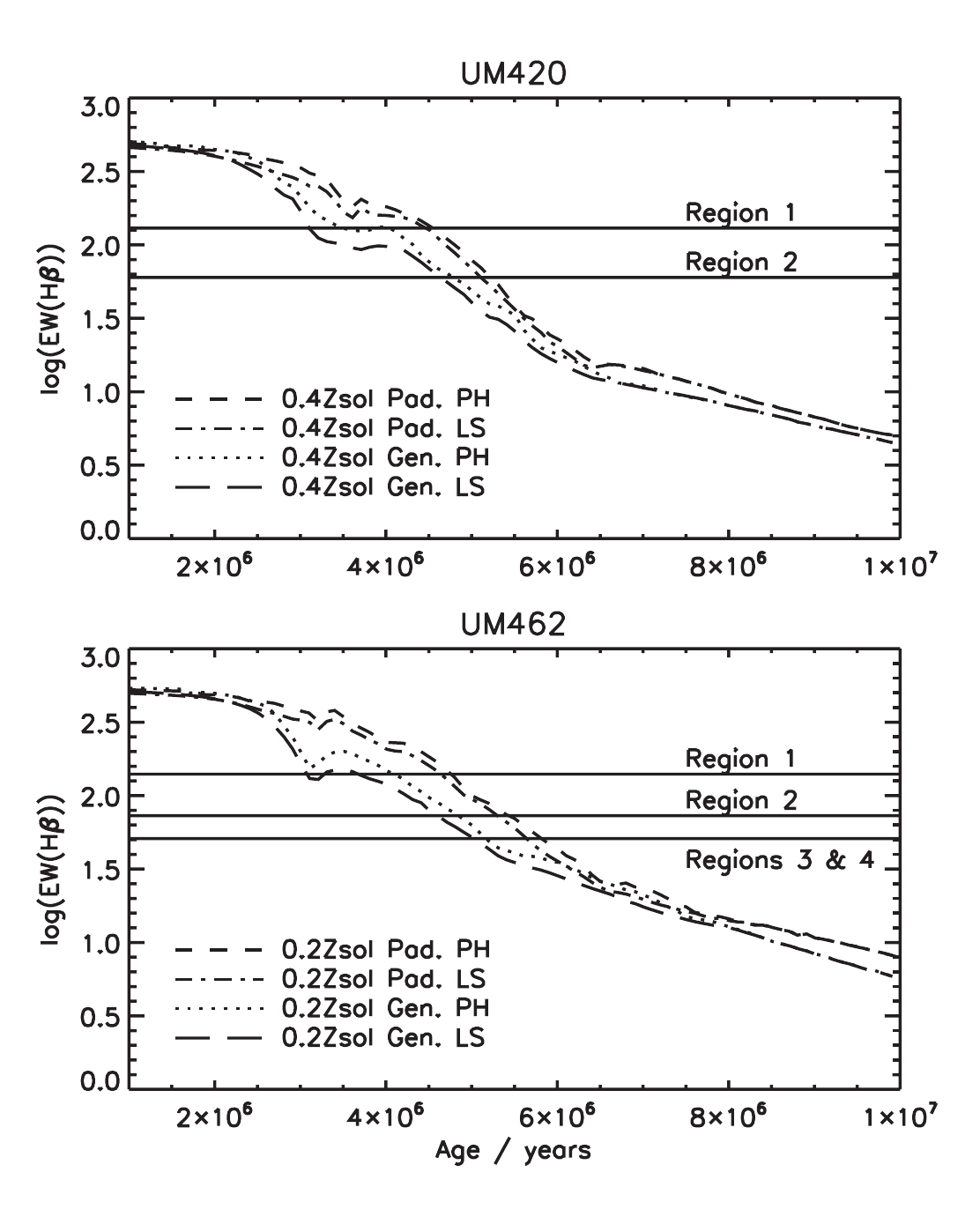}
\caption{$EW$(H$\beta$) as a function of age, as predicted by the STARBURST99
code for metallicities of 0.4\Zsol\ (UM~420) and 0.2\Zsol\ (UM~462) using a
combination of Geneva or Padova stellar evolutionary tracks and Lejeune-Schmutz
(LS) or Pauldrach-Hillier (PH) model atmospheres.  The observed average
$EW$(H$\beta$)'s for each star forming region in UM~420 (top panel) and UM~462
(bottom panel) are overlaid (solid line).} \label{fig:EWSB99}
\end{center}
\end{figure}

\section{Emission Line Galaxy Classification}

It is useful to attempt to diagnose the emission line excitation mechanisms in
BCGs using the classic diagnostic diagrams of \citet{Baldwin:1981} (the BPT
diagrams). These are used to classify galaxies according to the dominant
excitation mechanism of their emission lines, i.e. either photoionisation by
massive stars within \hii\ regions or photoionisation by non-thermal continua
from AGN.  The diagrams consist of excitation-dependent and
extinction-independent line ratios: log(\foiii\ $\lambda$5007/H$\beta$) versus
either log(\fsii\ $\lambda$6716 $+$ $\lambda$6731)/H$\alpha$) or log(\fnii\
$\lambda$6584/H$\alpha$). Star-forming galaxies fall into the lower left region
of the diagrams, AGN host galaxies fall into the upper right region and
Low-Ionisation Emission Line Regions (LINERs) fall on the lower right. Here we
adopt the `maximum starburst line' derived by \citet{Kewley:2001} from
starburst grids defined by two parameters; metallicities $Z$ = $\sim$
0.05--3.0~\Zsol\ and ionization parameters $q =
5\times10^{6}$~--~$3\times10^8$~cm~s$^{-1}$, where $q$ is the maximum velocity
of an ionisation front that can be driven by the local radiation field and is
related to the non-dimensional ionisation parameter $U=q/c$,
\citep{Dopita:2001}. This line defines the maximum flux ratio an object can
have to be successfully fitted by photoionization models alone. Ratios lying
above this boundary are inferred to require additional sources of excitation,
such as shocks or AGNs. It is thus instructive to see what area of the BPT
diagram UM~420 and UM~462 occupy as a whole, as well as their resolved
star-forming regions which were defined in Figures \ref{fig:UM420Regions} and
\ref{fig:UM462Regions}.

{\it UM~420}: Figure~\ref{fig:UM420_bpt}(a) and (b) shows the BPT diagram
locations of the two star-forming regions within UM~420.  Although a large
spread can be seen for both regions, Region 2 occupies the higher end of the
\foiii\ $\lambda$5007/H$\beta$ emission line ratio in both diagrams, and does
not exceed the \foiii\ $\lambda$5007/H$\beta$ values of Region 1. Some spaxels
do cross the `maximum starburst line', but are insufficient in frequency to be
considered as evidence for substantial non-thermal line excitation. The
\fsii/H$\alpha$ ratios are generally lower than those predicted by standard
predictions of shock models from the literature. There is a larger spread in
\fnii\ $\lambda$6584/H$\alpha$ as compared to \fsii/H$\alpha$ which is probably
due to the fact that the N$^+$/\hp\ abundance ratio differential between
Regions 1 and 2 is larger than the corresponding S$^+$/\hp\ differential by 45
per cent. In relation to the grid of ($Z$, $q$) models of \citet{Kewley:2001},
UM~420 lies within a range of $Z =$ 0.4--0.5~\Zsol and $q \ge 3\times10^8$.
This is consistent with the average metallicity of 0.35 $\pm$ 0.06~\Zsol\ for
Regions 1 and 2 (Table \ref{tab:um420Abs}). We conclude that photoionisation by
stellar populations is the dominant line excitation mechanism within UM~420.


{\it UM~462}: The BPT diagnostic diagrams shown in
Figure~\ref{fig:UM462_bpt}(a) and (b) show the emission line ratios for each
spaxel within the four star-forming regions of UM~462 (as defined in
Figure~\ref{fig:UM462Regions}). As with UM~420 the \fsii/H$\alpha$ diagnostic
ratios tend to straddle the theoretical upper limit for photoionisation.
Although their locations appear highly clustered, a slight distinction can be
seen for Region 4 which lies slightly to the left of the diagram. The large
spread in values can be seen more clearly in Figure~\ref{fig:UM462_bpt}(b),
where the spread along the $X$-axis, i.e. \fnii\ $\lambda$6584/H$\alpha$,
extends on either side of the other three regions. This range in \fnii\ and
\fsii\ excitation conditions is perhaps to be expected given the 9$''$ distance
separating Region 4 from the central regions. Several spaxels' emission line
ratios do spread into the locus of non-thermal excitation but their low
occurrence renders them insignificant. In relation to the models of
\citet{Kewley:2001}, UM~462 lies within a range of $Z=$0.2--0.5~\Zsol and
$q\ge1.5\times10^8$.  This is consistent with the average metallicity of
$\sim$0.24\Zsol\ for Regions 1--4 (Table \ref{tab:um462Abs}). We conclude that
photoionisation from stellar sources is the dominant excitation mechanism
within UM~462 as well. Comparing the two galaxies, the data points of UM~462
are more highly clustered than those of UM~420, despite the former system's
disrupted morphology and are further to the left (smaller values) on the
$X$-axis of both diagrams. The explanation probably lies in the higher S/H and
N/H abundances of UM~420 compared to those of UM~462 (by corresponding factors
of $\sim$ 2.5). Both galaxies display a similar level of \foiii\
$\lambda$5007/H$\beta$ excitation.

\begin{figure}
\begin{center}
\includegraphics[scale=0.50, angle=0]{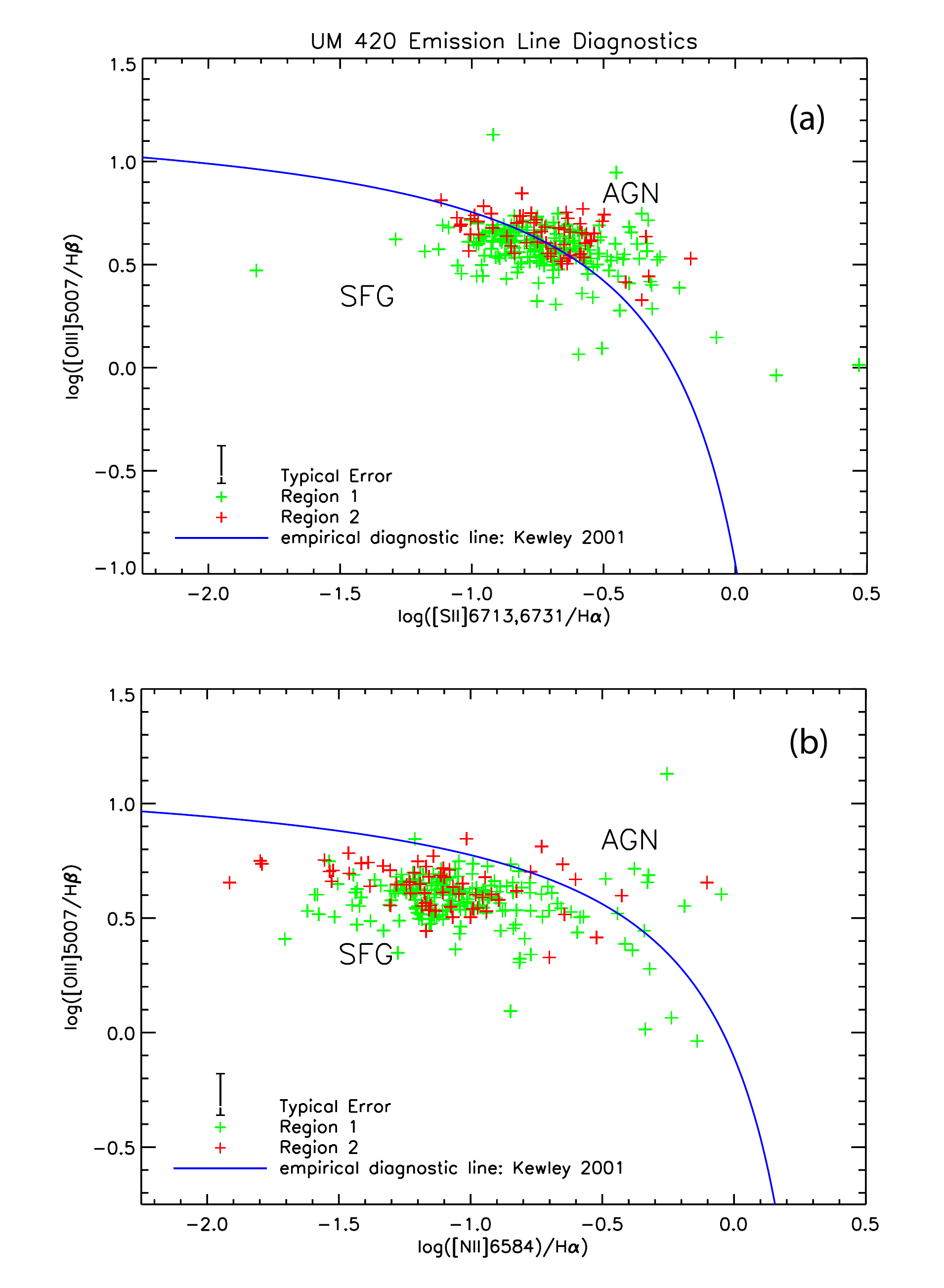}
\caption{Emission line diagnostic diagrams for UM~420.  Each data point
represents a spaxel in the ratioed dereddened flux maps corresponding to
\foiii\ $\lambda$5007/H$\beta$ versus (a) \fsii\ $\lambda$6716 $+$
$\lambda$6731/H$\alpha$ and (b) \fnii\ $\lambda$6584/H$\alpha$. Spaxels
corresponding to the two star-forming regions defined in
Figure~\ref{fig:UM420Regions} are separated by colour. The empirical diagnostic
`maximum starburst' line from \citet{Kewley:2001} is also shown, as are the
positions of emission line ratios for star-forming galaxies(SFG) and active
galaxies (AGN).} \label{fig:UM420_bpt}
\end{center}
\end{figure}

\begin{figure}
\begin{center}
\includegraphics[scale=0.50, angle=0]{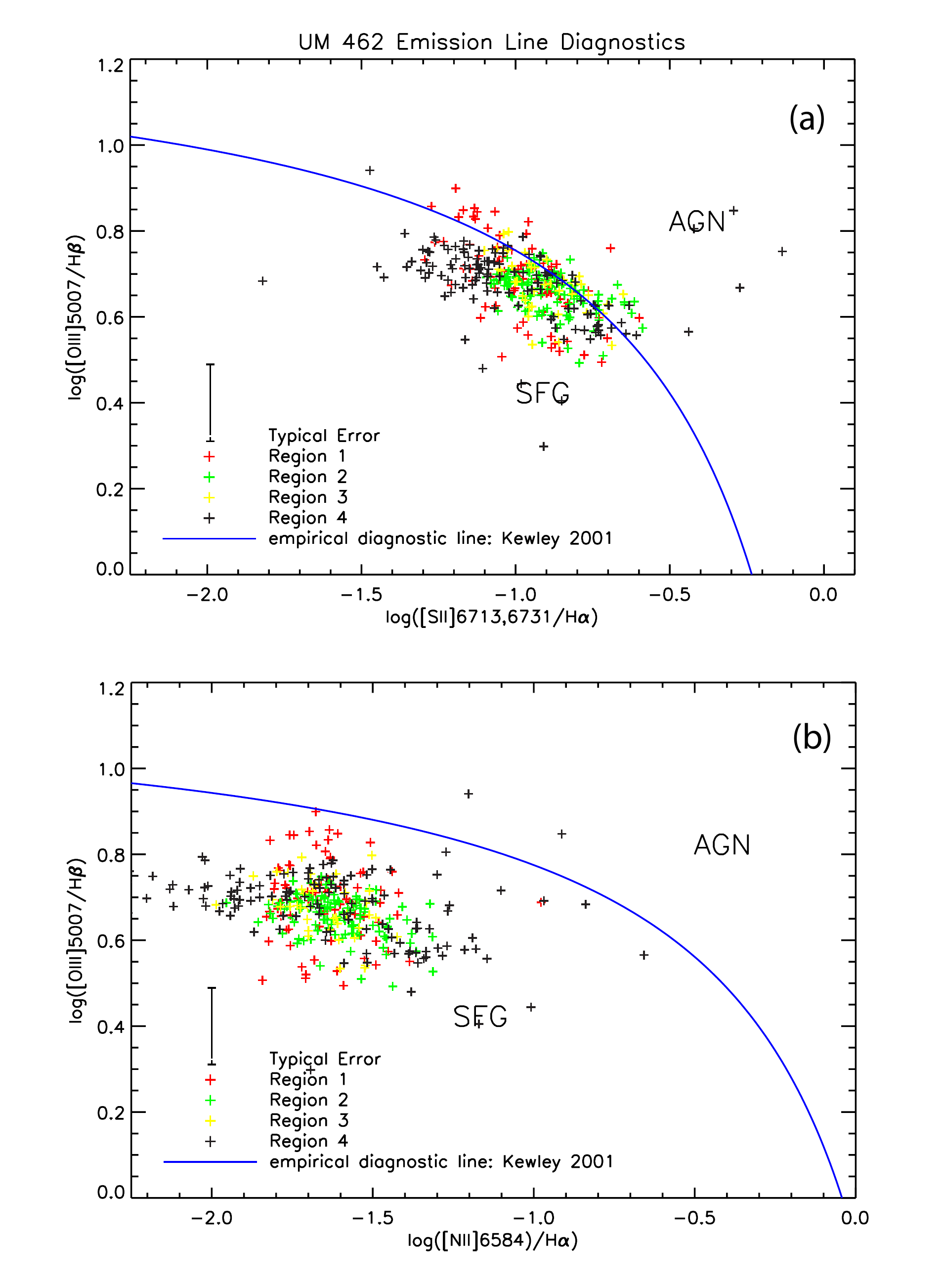}
\caption{Same as for Figure~\ref{fig:UM420_bpt}, but for UM~462.  Regions 1--4
correspond to the separate star-forming regions defined in
Figure~\ref{fig:UM462Regions}.} \label{fig:UM462_bpt}
\end{center}
\end{figure}

\section{Conclusions}

We have analyzed VIMOS IFU integral field spectroscopy of the unrelated BCGs
UM~420 and UM~462 and studied their morphology by creating monochromatic
emission line maps. Both systems show signs of interaction and/or perturbation
with the former galaxy currently undergoing a merger (type `iI,M') and the
latter displaying a highly disrupted irregular or cometary appearance (type
`iI,C'; probably due to interaction with a BCG which was not part of this
study, UM~461). The spatially resolved emission line maps in the light of
\foiii\ and H$\alpha$ have revealed two main areas of massive star formation in
UM~420 along with an `arm-like' structure, and at least four such areas in UM~462. Current
star formation rates were computed from the H$\alpha$ line luminosities for
each main starbursting region and the ages of the last major star formation
episode were estimated by fitting the observed Balmer line equivalent widths
with STARBURST99 models. The two merging components of UM~420 have SFRs that
differ by a factor of $\sim$ 8 and starburst episodes separated by 1\,Myr. The
latest major star formation event took place $\sim$ 4\,Myr ago in the largest
merging component, and has been producing stars at a rate of 10\,\Msol\
yr$^{-1}$. In UM~462 the last star forming episode was $\sim$ 4 -- 5\,Myr ago
and has been producing stars at a rate which varies across the galaxy between
$\sim$ 0.01 -- 0.10\,\Msol\ yr$^{-1}$, indicative of propagating or triggered
star formation.

For both targets the abundances of He, N, O, and S were measured and O/H
abundance ratio maps were created based on the direct method of estimating
electron temperatures from the \foiii\ $\lambda$4363/$\lambda$5007 line ratio.
The measured oxygen abundances are 12 $+$ log(O/H) $=$ 8.03 $\pm$ 0.20 for
UM~420 and 8.03 $\pm$ 0.10 for UM~462 (20 per cent solar). We find no evidence
for significant nitrogen or oxygen variations across the galaxies (at the
0.2\,dex level), which would point to self-enrichment from nucleosynthetic
products associated with the recent massive star formation activity. Regarding
the abundance of nitrogen, and the N/O ratio, this result is in qualitative
agreement with the finding that these BCGs cannot be classified as Wolf-Rayet
galaxies as the characteristic broad-line stellar features have not been
detected by VIMOS; the existence of WR stars in them is an open issue.

\section{Acknowledgments}

We would like to thank the VIMOS staff at Paranal and Garching for scheduling
and taking these service mode observations [programmes 078.B-0353(B, E); PI:
Y.G.~Tsamis]. We appreciate discussions with Carlo Izzo about the VIMOS
instrument and the GASGANO tool. Also, our thanks go to Roger Wesson and Jeremy
Walsh for help regarding the derivation of helium abundances. This research
made use of the NASA ADS and NED data bases. BLJ acknowledges support from a
STFC studentship. YGT acknowledges support from grants AYA2007-67965-C03-02 and
CSD2006-00070 CONSOLIDER-2010 ``First science with the GTC" of the Spanish
Ministry of Science and Innovation.

\bibliography{}








@article{Bernardis:2008,
	Adsnote = {Provided by the SAO/NASA Astrophysics Data System},
	Adsurl = {http://adsabs.harvard.edu/abs/2008JCAP...03..020D},
	Archiveprefix = {arXiv},
	Author = {{DeBernardis}, F. and {Melchiorri}, A. and {Verde}, L. and {Jimenez}, R.},
	Date-Added = {2008-11-17 22:03:11 +0000},
	Date-Modified = {2008-11-17 22:03:23 +0000},
	Doi = {10.1088/1475-7516/2008/03/020},
	Eprint = {0707.4170},
	Journal = {Journal of Cosmology and Astro-Particle Physics},
	Month = mar,
	Pages = {20-+},
	Title = {{The cosmic neutrino background and the age of the Universe}},
	Volume = 3,
	Year = 2008}

@article{Schlegel:1998,
	Adsnote = {Provided by the SAO/NASA Astrophysics Data System},
	Adsurl = {http://adsabs.harvard.edu/abs/1998ApJ...500..525S},
	Author = {{Schlegel}, D.~J. and {Finkbeiner}, D.~P. and {Davis}, M.},
	Date-Added = {2008-11-13 11:53:32 +0000},
	Date-Modified = {2008-11-13 11:53:46 +0000},
	Doi = {10.1086/305772},
	Eprint = {arXiv:astro-ph/9710327},
	Journal = {\apj},
	Keywords = {COSMOLOGY: DIFFUSE RADIATION, COSMOLOGY: COSMIC MICROWAVE BACKGROUND, ISM: DUST, EXTINCTION, INTERPLANETARY MEDIUM, INFRARED: ISM: CONTINUUM},
	Month = jun,
	Pages = {525-+},
	Title = {{Maps of Dust Infrared Emission for Use in Estimation of Reddening and Cosmic Microwave Background Radiation Foregrounds}},
	Volume = 500,
	Year = 1998}

@article{Hummer:1987,
	Adsnote = {Provided by the SAO/NASA Astrophysics Data System},
	Adsurl = {http://adsabs.harvard.edu/abs/1987MNRAS.224..801H},
	Author = {{Hummer}, D.~G. and {Storey}, P.~J.},
	Date-Added = {2008-11-13 11:49:17 +0000},
	Date-Modified = {2008-11-13 11:49:31 +0000},
	Journal = {\mnras},
	Keywords = {HELIUM, HYDROGEN IONS, INFRARED SPECTROSCOPY, ION RECOMBINATION, LINE SPECTRA, STELLAR SPECTRA, COLLISION RATES, DENSITY DISTRIBUTION, ELECTRON DISTRIBUTION, ION TEMPERATURE, IONIC COLLISIONS, QUANTUM NUMBERS},
	Month = feb,
	Pages = {801-820},
	Title = {{Recombination-line intensities for hydrogenic ions. I - Case B calculations for H I and He II}},
	Volume = 224,
	Year = 1987}

@article{Howarth:1983,
	Adsnote = {Provided by the SAO/NASA Astrophysics Data System},
	Adsurl = {http://adsabs.harvard.edu/abs/1983MNRAS.203..301H},
	Author = {{Howarth}, I.~D.},
	Date-Added = {2008-11-13 11:41:16 +0000},
	Date-Modified = {2008-11-13 11:41:28 +0000},
	Journal = {\mnras},
	Keywords = {GALACTIC RADIATION, INTERSTELLAR EXTINCTION, MAGELLANIC CLOUDS, SPACEBORNE ASTRONOMY, ULTRAVIOLET SPECTRA, DATA ACQUISITION, DATA REDUCTION, IUE, MILKY WAY GALAXY, PARAMETERIZATION, SATELLITE OBSERVATION},
	Month = apr,
	Pages = {301-304},
	Title = {{LMC and galactic extinction}},
	Volume = 203,
	Year = 1983}

@article{cite-key,
	Date-Added = {2008-10-24 17:57:03 +0100},
	Date-Modified = {2008-10-24 17:57:03 +0100}}

@article{Dopita:1978,
	Adsnote = {Provided by the SAO/NASA Astrophysics Data System},
	Adsurl = {http://adsabs.harvard.edu/abs/1978ApJS...37..117D},
	Author = {{Dopita}, M.~A.},
	Date-Added = {2008-10-24 14:28:18 +0100},
	Date-Modified = {2008-10-24 14:28:27 +0100},
	Doi = {10.1086/190521},
	Journal = {\apjs},
	Keywords = {EMISSION SPECTRA, HERBIG-HARO OBJECTS, OPTICAL EMISSION SPECTROSCOPY, PROTOSTARS, SHOCK WAVES, STELLAR EVOLUTION, STELLAR SPECTROPHOTOMETRY, INTERSTELLAR GAS, LINE SPECTRA, RADIAL VELOCITY, STELLAR LUMINOSITY, STELLAR MASS EJECTION, T TAURI STARS},
	Month = may,
	Pages = {117-144},
	Title = {{Optical emission from shocks. IV - The Herbig-Haro objects}},
	Volume = 37,
	Year = 1978}

@article{Shull:1979,
	Adsnote = {Provided by the SAO/NASA Astrophysics Data System},
	Adsurl = {http://adsabs.harvard.edu/abs/1979ApJ...227..131S},
	Author = {{Shull}, J.~M. and {McKee}, C.~F.},
	Date-Added = {2008-10-24 14:26:51 +0100},
	Date-Modified = {2008-10-24 14:27:01 +0100},
	Doi = {10.1086/156712},
	Journal = {\apj},
	Keywords = {ASTRONOMICAL MODELS, INTERSTELLAR CHEMISTRY, IONIZING RADIATION, RADIATIVE TRANSFER, SHOCK WAVE PROPAGATION, ULTRAVIOLET RADIATION, ATOMIC EXCITATIONS, EMISSION SPECTRA, GAS IONIZATION, IONIZATION CROSS SECTIONS, METAL IONS, RADIANT FLUX DENSITY, SHOCK FRONTS, SUPERNOVA REMNANTS},
	Month = jan,
	Pages = {131-149},
	Title = {{Theoretical models of interstellar shocks. I - Radiative transfer and UV precursors}},
	Volume = 227,
	Year = 1979}

@article{Calzetti:2004,
	Adsnote = {Provided by the SAO/NASA Astrophysics Data System},
	Adsurl = {http://adsabs.harvard.edu/abs/2004AJ....127.1405C},
	Author = {{Calzetti}, D. and {Harris}, J. and {Gallagher}, III, J.~S. and {Smith}, D.~A. and {Conselice}, C.~J. and {Homeier}, N. and {Kewley}, L.},
	Date-Added = {2008-10-24 14:23:42 +0100},
	Date-Modified = {2008-10-24 14:23:52 +0100},
	Eprint = {arXiv:astro-ph/0312385},
	Journal = {\aj},
	Keywords = {Galaxies: ISM, Galaxies: Interactions, Galaxies: Starburst, ISM: Structure},
	Month = mar,
	Pages = {1405-1430},
	Title = {{The Ionized Gas in Local Starburst Galaxies: Global and Small-Scale Feedback from Star Formation}},
	Volume = 127,
	Year = 2004}

@inproceedings{Dopita:2001,
	Adsnote = {Provided by the SAO/NASA Astrophysics Data System},
	Adsurl = {http://adsabs.harvard.edu/abs/2001sgnf.conf..225D},
	Author = {{Dopita}, M.~A. and {Kewley}, L.~J. and {Sutherland}, R.~S. and {Heisler}, C.~A.},
	Booktitle = {Starburst Galaxies: Near and Far},
	Date-Added = {2008-10-24 14:22:16 +0100},
	Date-Modified = {2008-10-24 14:22:46 +0100},
	Editor = {{Tacconi}, L. and {Lutz}, D.},
	Pages = {225-+},
	Title = {{Chemical Abundances and Evolution in Nearby Starbursts}},
	Year = 2001}

@article{Kewley:2001,
	Adsnote = {Provided by the SAO/NASA Astrophysics Data System},
	Adsurl = {http://adsabs.harvard.edu/abs/2001ApJ...556..121K},
	Author = {{Kewley}, L.~J. and {Dopita}, M.~A. and {Sutherland}, R.~S. and {Heisler}, C.~A. and {Trevena}, J.},
	Date-Added = {2008-10-24 14:21:13 +0100},
	Date-Modified = {2008-10-24 14:21:21 +0100},
	Doi = {10.1086/321545},
	Eprint = {arXiv:astro-ph/0106324},
	Journal = {\apj},
	Keywords = {Galaxies: Starburst, Radiation Mechanisms: Thermal},
	Month = jul,
	Pages = {121-140},
	Title = {{Theoretical Modeling of Starburst Galaxies}},
	Volume = 556,
	Year = 2001}

@article{Baldwin:1981,
	Adsnote = {Provided by the SAO/NASA Astrophysics Data System},
	Adsurl = {http://adsabs.harvard.edu/abs/1981PASP...93....5B},
	Author = {{Baldwin}, J.~A. and {Phillips}, M.~M. and {Terlevich}, R.},
	Date-Added = {2008-10-24 14:20:04 +0100},
	Date-Modified = {2008-10-24 14:20:12 +0100},
	Journal = {\pasp},
	Keywords = {ASTRONOMICAL SPECTROSCOPY, EMISSION SPECTRA, LINE SPECTRA, QUASARS, SEYFERT GALAXIES, CLASSIFICATIONS, H II REGIONS, PHOTOIONIZATION, PLANETARY NEBULAE, SHOCK HEATING},
	Month = feb,
	Pages = {5-19},
	Title = {{Classification parameters for the emission-line spectra of extragalactic objects}},
	Volume = 93,
	Year = 1981}

@article{Stasinska:2006,
	Adsnote = {Provided by the SAO/NASA Astrophysics Data System},
	Adsurl = {http://adsabs.harvard.edu/abs/2006MNRAS.371..972S},
	Author = {{Stasi{\'n}ska}, G. and {Cid Fernandes}, R. and {Mateus}, A. and {Sodr{\'e}}, L. and {Asari}, N.~V.},
	Date-Added = {2008-10-24 14:18:27 +0100},
	Date-Modified = {2008-10-24 14:18:42 +0100},
	Doi = {10.1111/j.1365-2966.2006.10732.x},
	Eprint = {arXiv:astro-ph/0606724},
	Journal = {\mnras},
	Keywords = {galaxies: active: galaxies: starburst},
	Month = sep,
	Pages = {972-982},
	Title = {{Semi-empirical analysis of Sloan Digital Sky Survey galaxies - III. How to distinguish AGN hosts}},
	Volume = 371,
	Year = 2006}

@article{Thuan:1995,
	Adsnote = {Provided by the SAO/NASA Astrophysics Data System},
	Adsurl = {http://adsabs.harvard.edu/abs/1995ApJ...445..108T},
	Author = {{Thuan}, T.~X. and {Izotov}, Y.~I. and {Lipovetsky}, V.~A.},
	Date-Added = {2008-10-22 15:28:57 +0100},
	Date-Modified = {2008-10-22 15:29:05 +0100},
	Doi = {10.1086/175676},
	Journal = {\apj},
	Keywords = {ABUNDANCE, COMPACT GALAXIES, H II REGIONS, HEAVY ELEMENTS, LINE SPECTRA, MASSIVE STARS, NUCLEAR FUSION, SKY SURVEYS (ASTRONOMY), ARGON, FLUX DENSITY, NEON, NITROGEN, OXYGEN, SPECTRUM ANALYSIS, STELLAR SPECTROPHOTOMETRY, SULFUR},
	Month = may,
	Pages = {108-123},
	Title = {{Heavy element abundances in a new sample of low-metallicity blue compact galaxies}},
	Volume = 445,
	Year = 1995}

@article{Thuan:2008,
	Adsnote = {Provided by the SAO/NASA Astrophysics Data System},
	Adsurl = {http://adsabs.harvard.edu/abs/2008arXiv0809.3082T},
	Archiveprefix = {arXiv},
	Author = {{Thuan}, T.~X. and {Hunt}, L.~K. and {Izotov}, Y.~I.},
	Date-Added = {2008-10-07 12:43:35 +0100},
	Date-Modified = {2008-10-07 12:44:42 +0100},
	Eprint = {0809.3082},
	Journal = {ArXiv e-prints},
	Keywords = {Astrophysics},
	Month = sep,
	Title = {{The Spitzer View of Low-Metallicity Star Formation: II. Mrk 996, a Blue Compact Dwarf Galaxy with an Extremely Dense Nucleus}},
	Year = 2008}

@article{Searle:1972,
	Adsnote = {Provided by the SAO/NASA Astrophysics Data System},
	Adsurl = {http://adsabs.harvard.edu/abs/1972ApJ...173...25S},
	Author = {{Searle}, L. and {Sargent}, W.~L.~W.},
	Date-Added = {2008-10-07 12:00:54 +0100},
	Date-Modified = {2008-10-07 12:01:06 +0100},
	Journal = {\apj},
	Month = apr,
	Pages = {25-+},
	Title = {{Inferences from the Composition of Two Dwarf Blue Galaxies}},
	Volume = 173,
	Year = 1972}

@article{Thuan:2004,
	Adsnote = {Provided by the SAO/NASA Astrophysics Data System},
	Adsurl = {http://adsabs.harvard.edu/abs/2004AJ....128..617T},
	Author = {{Thuan}, T.~X. and {Hibbard}, J.~E. and {L{\'e}vrier}, F.},
	Date-Added = {2008-10-07 11:48:43 +0100},
	Date-Modified = {2008-10-07 11:48:54 +0100},
	Doi = {10.1086/422431},
	Journal = {\aj},
	Keywords = {Galaxies: Individual: NGC Number: NGC 2366, Galaxies: Individual: NGC Number: NGC 4861, Galaxies: Individual: Alphanumeric: VII Zw 403, Galaxies: Individual: Name: Haro 2, Galaxies: Kinematics and Dynamics, Galaxies: Starburst},
	Month = aug,
	Pages = {617-643},
	Title = {{The H I Kinematics and Distribution of Four Blue Compact Dwarf Galaxies}},
	Volume = 128,
	Year = 2004}

@article{VanZee:2001,
	Adsnote = {Provided by the SAO/NASA Astrophysics Data System},
	Adsurl = {http://adsabs.harvard.edu/abs/2001AJ....122..121V},
	Author = {{van Zee}, L. and {Salzer}, J.~J. and {Skillman}, E.~D.},
	Date-Added = {2008-10-07 11:46:49 +0100},
	Date-Modified = {2008-10-07 11:47:03 +0100},
	Doi = {10.1086/321108},
	Eprint = {arXiv:astro-ph/0103454},
	Journal = {\aj},
	Keywords = {Galaxies: Dwarf, Galaxies: Evolution, Galaxies: Kinematics and Dynamics},
	Month = jul,
	Pages = {121-139},
	Title = {{Kinematic Constraints on Evolutionary Scenarios for Blue Compact Dwarf Galaxies. I. Neutral Gas Dynamics}},
	Volume = 122,
	Year = 2001}

@article{Pustilnik:2004,
	Adsnote = {Provided by the SAO/NASA Astrophysics Data System},
	Adsurl = {http://adsabs.harvard.edu/abs/2004A
	Author = {{Pustilnik}, S. and {Kniazev}, A. and {Pramskij}, A. and {Izotov}, Y. and {Foltz}, C. and {Brosch}, N. and {Martin}, J.-M. and {Ugryumov}, A.},
	Date-Added = {2008-08-26 17:11:23 +0100},
	Date-Modified = {2008-08-26 17:11:35 +0100},
	Doi = {10.1051/0004-6361:20035646},
	Eprint = {arXiv:astro-ph/0404171},
	Journal= {\aap},
	Keywords = {galaxies: starburst, galaxies: abundances, galaxies: interactions, galaxies: evolution, stars: Wolf-Rayet, galaxies: individual: HS 0837+4717},
	Month = may,
	Pages = {469-484},
	Title = {{HS 0837+4717 - a metal-deficient blue compact galaxy with large nitrogen excess}},
	Volume = 419,
	Year = 2004}

@article{Kewley:2001a,
	Adsnote = {Provided by the SAO/NASA Astrophysics Data System},
	Adsurl = {http://adsabs.harvard.edu/abs/2001ApJ...556..121K},
	Author = {{Kewley}, L.~J. and {Dopita}, M.~A. and {Sutherland}, R.~S. and {Heisler}, C.~A. and {Trevena}, J.},
	Date-Added = {2008-08-26 14:17:15 +0100},
	Date-Modified = {2008-08-26 14:17:40 +0100},
	Doi = {10.1086/321545},
	Eprint = {arXiv:astro-ph/0106324},
	Journal = {\apj},
	Keywords = {Galaxies: Starburst, Radiation Mechanisms: Thermal},
	Month = jul,
	Pages = {121-140},
	Title = {{Theoretical Modeling of Starburst Galaxies}},
	Volume = 556,
	Year = 2001}

@article{Veilleux:1987,
	Adsnote = {Provided by the SAO/NASA Astrophysics Data System},
	Adsurl = {http://adsabs.harvard.edu/abs/1987ApJS...63..295V},
	Author = {{Veilleux}, S. and {Osterbrock}, D.~E.},
	Date-Added = {2008-08-26 14:14:51 +0100},
	Date-Modified = {2008-08-26 14:15:05 +0100},
	Doi = {10.1086/191166},
	Journal = {\apjs},
	Keywords = {EMISSION SPECTRA, SEYFERT GALAXIES, SPECTRAL LINE WIDTH, SPECTRUM ANALYSIS, ACTIVE GALAXIES, ASTRONOMICAL MODELS, ASTRONOMICAL SPECTROSCOPY, H II REGIONS, HOT STARS, INTERSTELLAR EXTINCTION, PHOTOIONIZATION, STAR FORMATION},
	Month = feb,
	Pages = {295-310},
	Title = {{Spectral classification of emission-line galaxies}},
	Volume = 63,
	Year = 1987}

@article{Prieto:2001,
	Adsnote = {Provided by the SAO/NASA Astrophysics Data System},
	Adsurl = {http://adsabs.harvard.edu/abs/2001ApJ...556L..63A},
	Author = {{Allende Prieto}, C. and {Lambert}, D.~L. and {Asplund}, M.},
	Date-Added = {2008-08-14 17:48:05 +0100},
	Date-Modified = {2008-08-14 17:48:19 +0100},
	Doi = {10.1086/322874},
	Eprint = {arXiv:astro-ph/0106360},
	Journal = {\apjl},
	Keywords = {Sun: Abundances, Sun: Photosphere},
	Month = jul,
	Pages = {L63-L66},
	Title = {{The Forbidden Abundance of Oxygen in the Sun}},
	Volume = 556,
	Year = 2001}

@article{Schaerer:1998,
	Adsnote = {Provided by the SAO/NASA Astrophysics Data System},
	Adsurl = {http://adsabs.harvard.edu/abs/1998ApJ...497..618S},
	Author = {{Schaerer}, D. and {Vacca}, W.~D.},
	Date-Added = {2008-08-08 10:37:47 +0100},
	Date-Modified = {2008-08-08 10:38:02 +0100},
	Doi = {10.1086/305487},
	Eprint = {arXiv:astro-ph/9711140},
	Journal = {\apj},
	Keywords = {GALAXIES: STARBURST, GALAXIES: STELLAR CONTENT, ISM: H II REGIONS, STARS: WOLF-RAYET},
	Month = apr,
	Pages = {618-+},
	Title = {{New Models for Wolf-Rayet and O Star Populations in Young Starbursts}},
	Volume = 497,
	Year = 1998}

@article{bergvall:2002,
	Adsnote = {Provided by the SAO/NASA Astrophysics Data System},
	Adsurl = {http://adsabs.harvard.edu/abs/2002A
	Author = {{Bergvall}, N. and {{\"O}stlin}, G.},
	Date-Added = {2008-08-06 18:04:14 +0100},
	Date-Modified = {2008-08-06 18:04:26 +0100},
	Doi = {10.1051/0004-6361:20020759},
	Journal = {\aap},
	Keywords = {galaxies: evolution, galaxies: formation, galaxies: starburst, galaxies: dwarfs},
	Month = aug,
	Pages = {891-915},
	Title = {{Massive (?) starburst hosts of blue compact galaxies (BCGs). Optical/near-IR observations of 4 BCGs and their companions}},
	Volume = 390,
	Year = 2002}

@article{Garnett:1995,
	Adsnote = {Provided by the SAO/NASA Astrophysics Data System},
	Adsurl = {http://adsabs.harvard.edu/abs/1995ApJ...443...64G},
	Author = {{Garnett}, D.~R. and {Skillman}, E.~D. and {Dufour}, R.~J. and 
	{Peimbert}, M. and {Torres-Peimbert}, S. and {Terlevich}, R. and 
	{Terlevich}, E. and {Shields}, G.~A.},
	Date-Added = {2008-08-06 17:49:05 +0100},
	Date-Modified = {2008-08-06 17:49:17 +0100},
	Doi = {10.1086/175503},
	Eprint = {arXiv:astro-ph/9411011},
	Journal = {\apj},
	Keywords = {ABUNDANCE, CARBON, DWARF GALAXIES, EVOLUTION (DEVELOPMENT), H II REGIONS, IRREGULAR GALAXIES, MAGELLANIC CLOUDS, OXYGEN, ULTRAVIOLET ASTRONOMY, GALACTIC EVOLUTION, HUBBLE SPACE TELESCOPE, INTERSTELLAR MATTER, LINE SPECTRA, SPECTROGRAPHS},
	Month = apr,
	Pages = {64-76},
	Title = {{The evolution of C/O in dwarf galaxies from Hubble Space Telescope FOS observations}},
	Volume = 443,
	Year = 1995}

@ARTICLE{Izotov:2008,
   author = {{Izotov}, Y.~I. and {Thuan}, T.~X.},
    title = "{Active Galactic Nuclei in Four Metal-poor Dwarf Emission-Line Galaxies}",
  journal = {\apj},
archivePrefix = "arXiv",
   eprint = {0807.2029},
 keywords = {Galaxies: Abundances, Galaxies: Active, Galaxies: Irregular, Galaxies: ISM, ISM: H II Regions, ISM: Kinematics and Dynamics},
     year = 2008,
    month = nov,
   volume = 687,
    pages = {133-140},
      doi = {10.1086/591660},
   adsurl = {http://adsabs.harvard.edu/abs/2008ApJ...687..133I},
  adsnote = {Provided by the SAO/NASA Astrophysics Data System}
}

@article{Fillippenko:1982,
	Adsnote = {Provided by the SAO/NASA Astrophysics Data System},
	Adsurl = {http://adsabs.harvard.edu/abs/1982PASP...94..715F},
	Author = {{Filippenko}, A.~V.},
	Date-Added = {2008-07-25 14:42:48 +0100},
	Date-Modified = {2008-07-25 14:43:03 +0100},
	Journal = {\pasp},
	Keywords = {ATMOSPHERIC REFRACTION, INSTRUMENT ERRORS, SPECTROPHOTOMETRY, TELESCOPES, AMBIENT TEMPERATURE, ATMOSPHERIC PRESSURE, ERROR ANALYSIS, OPTIMIZATION, PRESSURE EFFECTS, TEMPERATURE EFFECTS},
	Month = aug,
	Pages = {715-721},
	Title = {{The importance of atmospheric differential refraction in spectrophotometry}},
	Volume = 94,
	Year = 1982}

@inproceedings{Walsh:1990,
	Adsnote = {Provided by the SAO/NASA Astrophysics Data System},
	Adsurl = {http://adsabs.harvard.edu/abs/1990daan.work...95W},
	Author = {{Walsh}, J.~R. and {Roy}, J.~R.},
	Booktitle = {ESO Conf. Proc. 34: 2nd ESO/ST-ECF Data Analysis Workshop},
	Date-Added = {2008-07-25 14:41:25 +0100},
	Date-Modified = {2008-07-25 14:41:44 +0100},
	Editor = {{Baade}, D. and {Grosbol}, P.~J.},
	Pages = {95},
	Title = {{Area Spectroscopy and Correction for Differential Atmospheric Refraction}},
	Year = 1990}

@article{Naze:2003,
	Adsnote = {Provided by the SAO/NASA Astrophysics Data System},
	Adsurl = {http://adsabs.harvard.edu/abs/2003A
	Author = {{Naz{\'e}}, Y. and {Rauw}, G. and {Manfroid}, J. and {Chu}, Y.-H. and 
	{Vreux}, J.-M.},
	Date-Added = {2008-07-11 12:05:08 +0100},
	Date-Modified = {2008-07-11 12:05:21 +0100},
	Doi = {10.1051/0004-6361:20030847},
	Eprint = {arXiv:astro-ph/0306084},
	Journal = {\aap},
	Keywords = {Wolf-Rayet, ISM: individual object: LMC N44C, HII regions, ISM: bubbles, ISM: abundances, Magellanic Clouds},
	Month = sep,
	Pages = {171-186},
	Title = {{WR bubbles and He II emission}},
	Volume = 408,
	Year = 2003}

@article{Thuan:2005,
	Adsnote = {Provided by the SAO/NASA Astrophysics Data System},
	Adsurl = {http://adsabs.harvard.edu/abs/2005ApJS..161..240T},
	Author = {{Thuan}, T.~X. and {Izotov}, Y.~I.},
	Date-Added = {2008-07-10 12:11:41 +0100},
	Date-Modified = {2008-07-10 12:11:54 +0100},
	Doi = {10.1086/491657},
	Eprint = {arXiv:astro-ph/0507209},
	Journal = {\apjs},
	Keywords = {Galaxies: Abundances, Galaxies: Evolution, Galaxies: Irregular, Galaxies: ISM},
	Month = dec,
	Pages = {240-270},
	Title = {{High-Ionization Emission in Metal-deficient Blue Compact Dwarf Galaxies}},
	Volume = 161,
	Year = 2005}

@article{Izotov:2007,
	Adsnote = {Provided by the SAO/NASA Astrophysics Data System},
	Adsurl = {http://adsabs.harvard.edu/abs/2007ApJ...671.1297I},
	Author = {{Izotov}, Y.~I. and {Thuan}, T.~X. and {Guseva}, N.~G.},
	Date-Added = {2008-07-10 11:50:19 +0100},
	Date-Modified = {2008-07-10 11:50:37 +0100},
	Doi = {10.1086/522923},
	Eprint = {arXiv:0709.3643},
	Journal = {\apj},
	Keywords = {Galaxies: Abundances, Galaxies: Active, Galaxies: Irregular, Galaxies: ISM, ISM: H II Regions, ISM: Kinematics and Dynamics},
	Month = dec,
	Pages = {1297-1320},
	Title = {{Broad-Line Emission in Low-Metallicity Blue Compact Dwarf Galaxies: Evidence for Stellar Wind, Supernova, and Possible AGN Activity}},
	Volume = 671,
	Year = 2007}

@article{Kingsburgh:1994,
	Adsnote = {Provided by the SAO/NASA Astrophysics Data System},
	Adsurl = {http://adsabs.harvard.edu/abs/1994MNRAS.271..257K},
	Author = {{Kingsburgh}, R.~L. and {Barlow}, M.~J.},
	Date-Added = {2008-07-09 15:41:44 +0100},
	Date-Modified = {2008-07-09 15:41:57 +0100},
	Journal = {\mnras},
	Keywords = {Planetary Nebulae: Element Abundances, Planetary Nebulae: Spectrophotometry, Planetary Nebulae: Electron Temperatures, Planetary Nebulae: Electron Densities},
	Month = nov,
	Pages = {257-299},
	Title = {{Elemental abundances for a sample of southern galctic planetary nebulae.}},
	Volume = 271,
	Year = 1994}

@article{cardelli:1989,
	Adsnote = {Provided by the SAO/NASA Astrophysics Data System},
	Adsurl = {http://adsabs.harvard.edu/abs/1989ApJ...345..245C},
	Author = {{Cardelli}, J.~A. and {Clayton}, G.~C. and {Mathis}, J.~S.},
	Date-Added = {2008-07-09 15:09:00 +0100},
	Date-Modified = {2008-07-09 15:09:10 +0100},
	Doi = {10.1086/167900},
	Journal = {\apj},
	Keywords = {INFRARED SPECTRA, INTERSTELLAR EXTINCTION, ULTRAVIOLET SPECTRA, VISIBLE SPECTRUM, COMPUTATIONAL ASTROPHYSICS, INTERSTELLAR MATTER, IUE},
	Month = oct,
	Pages = {245-256},
	Title = {{The relationship between infrared, optical, and ultraviolet extinction}},
	Volume = 345,
	Year = 1989}

@article{Izotov:2006,
	Adsnote = {Provided by the SAO/NASA Astrophysics Data System},
	Adsurl = {http://adsabs.harvard.edu/abs/2006A
	Author = {{Izotov}, Y.~I. and {Schaerer}, D. and {Blecha}, A. and {Royer}, F. and 
	{Guseva}, N.~G. and {North}, P.},
	Date-Added = {2008-07-09 12:21:45 +0100},
	Date-Modified = {2008-07-09 12:22:05 +0100},
	Doi = {10.1051/0004-6361:20065622},
	Eprint = {arXiv:astro-ph/0608203},
	Journal = {\aap},
	Keywords = {galaxies: fundamental parameters, galaxies: starburst, galaxies: ISM, galaxies: abundances, galaxies: indvidual: SBS 0335-052E},
	Month = nov,
	Pages = {71-84},
	Title = {{VLT/GIRAFFE spectroscopic observations of the metal-poor blue compact dwarf galaxy SBS 0335-052E}},
	Volume = 459,
	Year = 2006}

@article{keenan:2001,
	Adsnote = {Provided by the SAO/NASA Astrophysics Data System},
	Adsurl = {http://adsabs.harvard.edu/abs/2001PNAS...98.9476K},
	Author = {{Keenan}, F.~P. and {Aller}, L.~H. and {Ryans}, R.~S.~I. and 
	{Hyung}, S.},
	Date-Added = {2008-07-09 11:44:08 +0100},
	Date-Modified = {2008-07-09 11:44:23 +0100},
	Doi = {10.1073/pnas.151263098},
	Journal = {Proceedings of the National Academy of Science},
	Pages = {9476-9477},
	Title = {{Theoretical emission line ratios for [Fe III] and [Fe VII] applicable to the optical and infrared spectra of gaseous nebulae}},
	Volume = 98,
	Year = 2001}

@article{Izotov:1994,
	Adsnote = {Provided by the SAO/NASA Astrophysics Data System},
	Adsurl = {http://adsabs.harvard.edu/abs/1994ApJ...435..647I},
	Author = {{Izotov}, Y.~I. and {Thuan}, T.~X. and {Lipovetsky}, V.~A.},
	Date-Added = {2008-07-09 11:34:56 +0100},
	Date-Modified = {2008-07-09 11:35:21 +0100},
	Doi = {10.1086/174843},
	Journal = {\apj},
	Keywords = {EMISSIVITY, GALACTIC EVOLUTION, GALAXIES, NUCLEAR FUSION, SIGNAL TO NOISE RATIOS, SPECTROPHOTOMETRY, ASTRONOMICAL OBSERVATORIES, CHEMICAL COMPOUNDS, COSMOLOGY, MATHEMATICAL MODELS, STELLAR PHYSICS},
	Month = nov,
	Pages = {647-667},
	Title = {{The primordial helium abundance from a new sample of metal-deficient blue compact galaxies}},
	Volume = 435,
	Year = 1994}

@url{vanHoof,
	Author = {{van Hoof}, P.},
	Date-Added = {2008-07-08 14:53:35 +0100},
	Date-Modified = {2008-07-22 10:37:03 +0100},
	Keywords = {line lists},
	Title = {Atomic Line List},
	Url = {http://www.pa.uky.edu/~peter/atomic/},
	Urldate = {2007}}

@article{crowther:2006,
	Adsnote = {Provided by the SAO/NASA Astrophysics Data System},
	Adsurl = {http://adsabs.harvard.edu/abs/2006A
	Author = {{Crowther}, P.~A. and {Hadfield}, L.~J.},
	Date-Added = {2008-07-08 14:37:23 +0100},
	Date-Modified = {2008-07-08 14:37:38 +0100},
	Doi = {10.1051/0004-6361:20054298},
	Eprint = {arXiv:astro-ph/0512183},
	Journal = {\aap},
	Keywords = {stars: Wolf-Rayet, galaxies: stellar content, galaxies:, individual: I Zw 18, stars: atmospheres},
	Month = apr,
	Pages = {711-722},
	Title = {{Reduced Wolf-Rayet line luminosities at low metallicity}},
	Volume = 449,
	Year = 2006}

@article{mould:2000,
	Adsnote = {Provided by the SAO/NASA Astrophysics Data System},
	Adsurl = {http://adsabs.harvard.edu/abs/2000ApJ...529..786M},
	Author = {{Mould}, J.~R. and {Huchra}, J.~P. and {Freedman}, W.~L. and 
	{Kennicutt}, Jr., R.~C. and {Ferrarese}, L. and {Ford}, H.~C. and 
	{Gibson}, B.~K. and {Graham}, J.~A. and {Hughes}, S.~M.~G. and 
	{Illingworth}, G.~D. and {Kelson}, D.~D. and {Macri}, L.~M. and 
	{Madore}, B.~F. and {Sakai}, S. and {Sebo}, K.~M. and {Silbermann}, N.~A. and 
	{Stetson}, P.~B.},
	Date-Added = {2008-07-08 12:06:30 +0100},
	Date-Modified = {2008-07-08 12:06:43 +0100},
	Doi = {10.1086/308304},
	Eprint = {arXiv:astro-ph/9909260},
	Journal = {\apj},
	Keywords = {STARS: VARIABLES: CEPHEIDS, COSMOLOGY: OBSERVATIONS, COSMOLOGY: DISTANCE SCALE, GALAXIES: DISTANCES AND REDSHIFTS},
	Month = feb,
	Pages = {786-794},
	Title = {{The Hubble Space Telescope Key Project on the Extragalactic Distance Scale. XXVIII. Combining the Constraints on the Hubble Constant}},
	Volume = 529,
	Year = 2000}

@article{vacca:1992,
	Adsnote = {Provided by the SAO/NASA Astrophysics Data System},
	Adsurl = {http://adsabs.harvard.edu/abs/1992ApJ...401..543V},
	Author = {{Vacca}, W.~D. and {Conti}, P.~S.},
	Date-Added = {2008-07-08 11:49:47 +0100},
	Date-Modified = {2008-07-08 11:50:02 +0100},
	Doi = {10.1086/172085},
	Journal = {\apj},
	Keywords = {SPECTROPHOTOMETRY, STARBURST GALAXIES, STELLAR COMPOSITION, VISIBLE SPECTRUM, WOLF-RAYET STARS, EMISSION SPECTRA, FORBIDDEN TRANSITIONS, HELIUM, O STARS},
	Month = dec,
	Pages = {543-558},
	Title = {{Optical spectrophotometry of Wolf-Rayet galaxies}},
	Volume = 401,
	Year = 1992}

@article{Thuan:1996,
	Adsnote = {Provided by the SAO/NASA Astrophysics Data System},
	Adsurl = {http://adsabs.harvard.edu/abs/1996ApJ...463..120T},
	Author = {{Thuan}, T.~X. and {Izotov}, Y.~I. and {Lipovetsky}, V.~A.},
	Date-Added = {2008-07-08 10:13:53 +0100},
	Date-Modified = {2008-07-22 10:36:04 +0100},
	Doi = {10.1086/177228},
	Journal = {\apj},
	Keywords = {GALAXIES: ABUNDANCES, GALAXIES: COMPACT, GALAXIES: INDIVIDUAL NAME: MARKARIAN 996, GALAXIES: ISM, GALAXIES: NUCLEI, GALAXIES: STELLAR CONTENT, GALAXIES: STRUCTURE},
	Month = may,
	Pages = {120-+},
	Title = {{Hubble Space Telescope Observations of the Unusual Blue Compact Dwarf Galaxy Markarian 996}},
	Volume = 463,
	Year = 1996}

@article{Westmoquette:2007,
	Adsnote = {Provided by the SAO/NASA Astrophysics Data System},
	Adsurl = {http://adsabs.harvard.edu/abs/2007MNRAS.381..894W},
	Author = {{Westmoquette}, M.~S. and {Exter}, K.~M. and {Smith}, L.~J. and 
	{Gallagher}, J.~S.},
	Date-Added = {2008-07-07 12:06:50 +0100},
	Date-Modified = {2008-07-22 10:35:07 +0100},
	Doi = {10.1111/j.1365-2966.2007.12346.x},
	Eprint = {arXiv:0708.2379},
	Journal = {\mnras},
	Keywords = {ISM: kinematics and dynamics , galaxies: evolution , galaxies: individual: NGC 1569 , galaxies: ISM , galaxies: starburst},
	Month = nov,
	Pages = {894-912},
	Title = {{Gemini GMOS/IFU spectroscopy of NGC 1569 - I. Mapping the properties of a young star cluster and its environment}},
	Volume = 381,
	Year = 2007}

@manual{dimeo:2005,
	Author = {{Dimeo}, R.},
	Date-Added = {2008-07-07 11:57:04 +0100},
	Date-Modified = {2008-08-05 13:42:52 +0100},
	Keywords = {pan, reduction techniques},
	Local-Url = {ftp://ftp.ncnr.nist.gov/pub/staff/dimeo/pandoc.pdf},
	Title = {PAN User Guide},
	Url = {ftp://ftp.ncnr.nist.gov/pub/staff/dimeo/pandoc.pdf},
	Urldate = {2005},
	Year = {2005}}

@article{Zanichelli:2005,
	Adsnote = {Provided by the SAO/NASA Astrophysics Data System},
	Adsurl = {http://adsabs.harvard.edu/abs/2005PASP..117.1271Z},
	Author = {{Zanichelli}, A. and {Garilli}, B. and {Scodeggio}, M. and {Franzetti}, P. and 
	{Rizzo}, D. and {Maccagni}, D. and {Merighi}, R. and {Picat}, J.~P. and 
	{Le F{\`e}vre}, O. and {Foucaud}, S. and {Bottini}, D. and {Le Brun}, V. and 
	{Scaramella}, R. and {Tresse}, L. and {Vettolani}, G. and {Adami}, C. and 
	{Arnaboldi}, M. and {Arnouts}, S. and {Bardelli}, S. and {Bolzonella}, M. and 
	{Cappi}, A. and {Charlot}, S. and {Ciliegi}, P. and {Contini}, T. and 
	{Gavignaud}, I. and {Guzzo}, L. and {Ilbert}, O. and {Iovino}, A. and 
	{McCracken}, H.~J. and {Marano}, B. and {Marinoni}, C. and {Mathez}, G. and 
	{Mazure}, A. and {Meneux}, B. and {Paltani}, S. and {Pell{\`o}}, R. and 
	{Pollo}, A. and {Pozzetti}, L. and {Radovich}, M. and {Zamorani}, G. and 
	{Zucca}, E.},
	Date-Added = {2008-06-27 16:34:36 +0100},
	Date-Modified = {2008-07-22 10:36:31 +0100},
	Doi = {10.1086/496936},
	Eprint = {arXiv:astro-ph/0509454},
	Journal = {\pasp},
	Keywords = {Instrumentation: Spectrographs, Methods: Data Analysis, Techniques: Spectroscopic},
	Month = nov,
	Pages = {1271-1283},
	Title = {{The VIMOS Integral Field Unit: Data-Reduction Methods and Quality Assessment}},
	Volume = 117,
	Year = 2005}

@article{Leitherer:1999,
	Adsnote = {Provided by the Smithsonian/NASA Astrophysics Data System},
	Adsurl = {http://adsabs.harvard.edu/abs/1999ApJS..123....3L},
	Author = {{Leitherer}, C. and {Schaerer}, D. and {Goldader}, J.~D. and 
	{Delgado}, R.~M.~G. and {Robert}, C. and {Kune}, D.~F. and {de Mello}, D.~F. and 
	{Devost}, D. and {Heckman}, T.~M.},
	Date-Added = {2007-06-28 13:31:38 +0100},
	Date-Modified = {2007-06-28 13:31:56 +0100},
	Doi = {10.1086/313233},
	Eprint = {arXiv:astro-ph/9902334},
	Journal = {\apjs},
	Month = jul,
	Pages = {3-40},
	Title = {{Starburst99: Synthesis Models for Galaxies with Active Star Formation}},
	Volume = 123,
	Year = 1999}

@article{ercolano:2005,
	Adsnote = {Provided by the Smithsonian/NASA Astrophysics Data System},
	Adsurl = {http://adsabs.harvard.edu/abs/2005MNRAS.362.1038E},
	Author = {{Ercolano}, B. and {Barlow}, M.~J. and {Storey}, P.~J.},
	Date-Added = {2007-06-26 14:39:48 +0100},
	Date-Modified = {2007-06-26 14:39:58 +0100},
	Doi = {10.1111/j.1365-2966.2005.09381.x},
	Eprint = {arXiv:astro-ph/0507050},
	Journal = {\mnras},
	Month = sep,
	Pages = {1038-1046},
	Title = {{The dusty MOCASSIN: fully self-consistent 3D photoionization and dust radiative transfer models}},
	Volume = 362,
	Year = 2005}

@article{erolano:2004,
	Adsnote = {Provided by the Smithsonian/NASA Astrophysics Data System},
	Adsurl = {http://adsabs.harvard.edu/abs/2004astro.ph..7200E},
	Author = {{Ercolano}, B. and {Barlow}, M.~J. and {Storey}, P.~J. and {Liu}, X.~-.
	},
	Date-Added = {2007-06-26 14:39:08 +0100},
	Date-Modified = {2007-06-26 14:39:16 +0100},
	Eprint = {astro-ph/0407200},
	Journal = {ArXiv Astrophysics e-prints},
	Month = jul,
	Title = {{MOCASSIN: 3D photoionisation and dust radiative transfer modelling of PNe}},
	Year = 2004}

@article{Ercolano:2003a,
	Adsnote = {Provided by the Smithsonian/NASA Astrophysics Data System},
	Adsurl = {http://adsabs.harvard.edu/abs/2003MNRAS.340.1136E},
	Author = {{Ercolano}, B. and {Barlow}, M.~J. and {Storey}, P.~J. and {Liu}, X.-W.
	},
	Date-Added = {2007-06-26 14:37:49 +0100},
	Date-Modified = {2008-10-24 14:25:53 +0100},
	Doi = {10.1046/j.1365-8711.2003.06371.x},
	Eprint = {arXiv:astro-ph/0209378},
	Journal = {\mnras},
	Month = apr,
	Pages = {1136-1152},
	Title = {{MOCASSIN: a fully three-dimensional Monte Carlo photoionization code}},
	Volume = 340,
	Year = 2003}

@inproceedings{Loose:1985,
	Adsnote = {Provided by the Smithsonian/NASA Astrophysics Data System},
	Adsurl = {http://adsabs.harvard.edu/abs/1985sfdg.conf...73L},
	Author = {{Loose}, H.-H. and {Thuan}, T.~X.},
	Booktitle = {Star-Forming Dwarf Galaxies and Related Objects},
	Date-Added = {2007-06-26 12:11:05 +0100},
	Date-Modified = {2007-06-26 12:11:13 +0100},
	Editor = {{Kunth}, D. and {Thuan}, T.~X. and {Tran Thanh van}, J.},
	Pages = {73-+},
	Title = {{The Morphology and Structure of Blue Compact Dwarf Galaxies from CCD Observations}},
	Year = 1985}

@article{papaderos:2001,
	Adsnote = {Provided by the Smithsonian/NASA Astrophysics Data System},
	Adsurl = {http://adsabs.harvard.edu/abs/2001astro.ph.10040P},
	Author = {{Papaderos}, P. and {Izotov}, Y.~I. and {Noeske}, K.~G. and 
	{Thuan}, T.~X. and {Fricke}, K.~J.},
	Date-Added = {2007-06-26 12:09:08 +0100},
	Date-Modified = {2007-06-26 12:09:16 +0100},
	Eprint = {astro-ph/0110040},
	Journal = {ArXiv Astrophysics e-prints},
	Month = oct,
	Title = {{Optical {\amp} NIR Surface Photometry of I Zw 18}},
	Year = 2001}

@article{aloisi:2007,
	Adsnote = {Provided by the Smithsonian/NASA Astrophysics Data System},
	Adsurl = {http://adsabs.harvard.edu/abs/2007astro.ph..2216A},
	Author = {{Aloisi}, A. and {Annibali}, F. and {Mack}, J. and {Tosi}, M. and 
	{van der Marel}, R. and {Clementini}, G. and {Contreras}, R. and 
	{Fiorentino}, G. and {Marconi}, M. and {Musella}, I. and {Saha}, A.
	},
	Date-Added = {2007-06-26 12:08:17 +0100},
	Date-Modified = {2007-06-26 12:08:27 +0100},
	Eprint = {astro-ph/0702216},
	Journal = {ArXiv Astrophysics e-prints},
	Month = feb,
	Title = {{A New Deep HST/ACS CMD of I Zw 18: Evidence for Red Giant Branch Stars}},
	Year = 2007}

@article{corbin:2006,
	Adsnote = {Provided by the Smithsonian/NASA Astrophysics Data System},
	Adsurl = {http://adsabs.harvard.edu/abs/2006ApJ...651..861C},
	Author = {{Corbin}, M.~R. and {Vacca}, W.~D. and {Cid Fernandes}, R. and 
	{Hibbard}, J.~E. and {Somerville}, R.~S. and {Windhorst}, R.~A.
	},
	Date-Added = {2007-06-26 12:07:01 +0100},
	Date-Modified = {2007-06-26 12:07:12 +0100},
	Doi = {10.1086/507575},
	Eprint = {arXiv:astro-ph/0607280},
	Journal = {\apj},
	Month = nov,
	Pages = {861-873},
	Title = {{Ultracompact Blue Dwarf Galaxies: Hubble Space Telescope Imaging and Stellar Population Analysis}},
	Volume = 651,
	Year = 2006}

@article{izotov:2005,
	Adsnote = {Provided by the Smithsonian/NASA Astrophysics Data System},
	Adsurl = {http://adsabs.harvard.edu/abs/2005ApJ...632..210I},
	Author = {{Izotov}, Y.~I. and {Thuan}, T.~X. and {Guseva}, N.~G.},
	Date-Added = {2007-06-12 18:51:38 +0100},
	Date-Modified = {2007-06-12 18:51:48 +0100},
	Doi = {10.1086/432874},
	Eprint = {arXiv:astro-ph/0506498},
	Journal = {\apj},
	Month = oct,
	Pages = {210-216},
	Title = {{SBS 0335-052W: The Lowest Metallicity Star-forming Galaxy Known}},
	Volume = 632,
	Year = 2005}

@inproceedings{dwek:2005,
	Adsnote = {Provided by the Smithsonian/NASA Astrophysics Data System},
	Adsurl = {http://adsabs.harvard.edu/abs/2005AIPC..804..197D},
	Author = {{Dwek}, E.},
	Booktitle = {Planetary Nebulae as Astronomical Tools},
	Date-Added = {2007-06-12 17:24:50 +0100},
	Date-Modified = {2007-06-12 17:24:58 +0100},
	Doi = {10.1063/1.2146272},
	Editor = {{Szczerba}, R. and {Stasinska}, G. and {Gorny}, S.~K.},
	Month = nov,
	Pages = {197-203},
	Series = {American Institute of Physics Conference Series},
	Title = {{The Chemical Evolution of Interstellar Dust}},
	Volume = 804,
	Year = 2005}

@article{mas-hesse:1999,
	Adsnote = {Provided by the Smithsonian/NASA Astrophysics Data System},
	Adsurl = {http://adsabs.harvard.edu/abs/1999A
	Author = {{Mas-Hesse}, J.~M. and {Kunth}, D.},
	Date-Added = {2007-06-12 17:05:28 +0100},
	Date-Modified = {2007-06-12 17:05:49 +0100},
	Eprint = {arXiv:astro-ph/9812072},
	Journal = {\aap},
	Month = sep,
	Pages = {765-795},
	Title = {{A comprehensive study of intense star formation bursts in irregular and compact galaxies}},
	Volume = 349,
	Year = 1999}

@article{galliano:2005,
	Adsnote = {Provided by the Smithsonian/NASA Astrophysics Data System},
	Adsurl = {http://adsabs.harvard.edu/abs/2005A
	Author = {{Galliano}, F. and {Madden}, S.~C. and {Jones}, A.~P. and {Wilson}, C.~D. and 
	{Bernard}, J.-P.},
	Date-Added = {2007-06-12 15:05:48 +0100},
	Date-Modified = {2007-06-12 15:05:58 +0100},
	Doi = {10.1051/0004-6361:20042369},
	Eprint = {arXiv:astro-ph/0501632},
	Journal = {\aap},
	Month = may,
	Pages = {867-885},
	Title = {{ISM properties in low-metallicity environments. III. The spectral energy distributions of II Zw 40, He 2-10 and NGC 1140}},
	Volume = 434,
	Year = 2005}

@article{cohen:2003,
	Adsnote = {Provided by the Smithsonian/NASA Astrophysics Data System},
	Adsurl = {http://adsabs.harvard.edu/abs/2003AJ....125.2645C},
	Author = {{Cohen}, M. and {Megeath}, S.~T. and {Hammersley}, P.~L. and 
	{Mart{\'{\i}}n-Luis}, F. and {Stauffer}, J.},
	Date-Added = {2007-06-11 20:18:54 +0100},
	Date-Modified = {2007-06-11 20:19:05 +0100},
	Doi = {10.1086/374362},
	Eprint = {arXiv:astro-ph/0304349},
	Journal = {\aj},
	Month = may,
	Pages = {2645-2663},
	Title = {{Spectral Irradiance Calibration in the Infrared. XIII. ``Supertemplates'' and On-Orbit Calibrators for the SIRTF Infrared Array Camera}},
	Volume = 125,
	Year = 2003}

@article{higdon:2004,
	Adsnote = {Provided by the Smithsonian/NASA Astrophysics Data System},
	Adsurl = {http://adsabs.harvard.edu/abs/2004PASP..116..975H},
	Author = {{Higdon}, S.~J.~U. and {Devost}, D. and {Higdon}, J.~L. and 
	{Brandl}, B.~R. and {Houck}, J.~R. and {Hall}, P. and {Barry}, D. and 
	{Charmandaris}, V. and {Smith}, J.~D.~T. and {Sloan}, G.~C. and 
	{Green}, J.},
	Date-Added = {2007-06-11 17:53:38 +0100},
	Date-Modified = {2007-06-11 17:53:50 +0100},
	Doi = {10.1086/425083},
	Eprint = {arXiv:astro-ph/0408295},
	Journal = {\pasp},
	Month = oct,
	Pages = {975-984},
	Title = {{The SMART Data Analysis Package for the Infrared Spectrograph on the Spitzer Space Telescope}},
	Volume = 116,
	Year = 2004}

@manual{SOM:2004,
	Address = {http://ssc.spitzer.caltech.edu/documents/SOM/},
	Author = {{Spitzer Science Center}},
	Date-Added = {2007-06-11 17:42:58 +0100},
	Date-Modified = {2007-06-12 18:45:59 +0100},
	Title = {Spitzer Observer's Manual},
	Year = {2004}}

@article{Kunth:2000,
	Adsnote = {Provided by the Smithsonian/NASA Astrophysics Data System},
	Adsurl = {http://adsabs.harvard.edu/abs/2000A
	Author = {{Kunth}, D. and {{\"O}stlin}, G.},
	Date-Added = {2007-06-08 17:04:35 +0100},
	Date-Modified = {2008-10-21 14:44:20 +0100},
	Doi = {10.1007/s001590000005},
	Eprint = {arXiv:astro-ph/9911094},
	Journal = {\aapr},
	Pages = {1-79},
	Title = {{The most metal-poor galaxies}},
	Volume = 10,
	Year = 2000}

@book{osterbrock,
	Author = {{Osterbrock}, D. E.},
	Date-Added = {2007-06-08 16:02:31 +0100},
	Date-Modified = {2007-06-08 16:05:21 +0100},
	Publisher = {University Science Books},
	Title = {Astrophysics of Gaseous Nebulae and Active Galactic Nuclei},
	Year = {2006}}

@article{galliano:2006,
	Adsnote = {Provided by the Smithsonian/NASA Astrophysics Data System},
	Adsurl = {http://adsabs.harvard.edu/abs/2006astro.ph.10852G},
	Author = {{Galliano}, F.},
	Date-Added = {2007-06-08 15:58:09 +0100},
	Date-Modified = {2007-06-08 15:58:24 +0100},
	Eprint = {astro-ph/0610852},
	Journal = {ArXiv Astrophysics e-prints},
	Month = oct,
	Title = {{PAHs in Galaxies: their Properties and Evolution}},
	Year = 2006}

@article{madden:2002,
	Adsnote = {Provided by the Smithsonian/NASA Astrophysics Data System},
	Adsurl = {http://adsabs.harvard.edu/abs/2002Ap
	Author = {{Madden}, S.~C.},
	Date-Added = {2007-06-08 15:56:50 +0100},
	Date-Modified = {2007-06-08 15:57:03 +0100},
	Doi = {10.1023/A:1019510806723},
	Journal = {\apss},
	Month = jul,
	Pages = {247-252},
	Title = {{The Low Metallicity ISM of Dwarf Galaxies}},
	Volume = 281,
	Year = 2002}

@article{wu:2006,
	Adsnote = {Provided by the Smithsonian/NASA Astrophysics Data System},
	Adsurl = {http://adsabs.harvard.edu/abs/2006ApJ...639..157W},
	Author = {{Wu}, Y. and {Charmandaris}, V. and {Hao}, L. and {Brandl}, B.~R. and 
	{Bernard-Salas}, J. and {Spoon}, H.~W.~W. and {Houck}, J.~R.
	},
	Date-Added = {2007-06-08 15:55:47 +0100},
	Date-Modified = {2007-06-08 15:55:55 +0100},
	Doi = {10.1086/499226},
	Eprint = {arXiv:astro-ph/0510856},
	Journal = {\apj},
	Month = mar,
	Pages = {157-172},
	Title = {{Mid-Infrared Properties of Low-Metallicity Blue Compact Dwarf Galaxies from the Spitzer Infrared Spectrograph}},
	Volume = 639,
	Year = 2006}

@article{ohalloran:2006,
	Adsnote = {Provided by the Smithsonian/NASA Astrophysics Data System},
	Adsurl = {http://adsabs.harvard.edu/abs/2006ApJ...641..795O},
	Author = {{O'Halloran}, B. and {Satyapal}, S. and {Dudik}, R.~P.},
	Date-Added = {2007-06-08 15:54:29 +0100},
	Date-Modified = {2007-06-08 15:54:42 +0100},
	Doi = {10.1086/500529},
	Eprint = {arXiv:astro-ph/0512404},
	Journal = {\apj},
	Month = apr,
	Pages = {795-800},
	Title = {{The Polycyclic Aromatic Hydrocarbon Emission Deficit in Low-Metallicity Galaxies-A Spitzer View}},
	Volume = 641,
	Year = 2006}

@article{lebouteiller:2007,
	Adsnote = {Provided by the Smithsonian/NASA Astrophysics Data System},
	Adsurl = {http://adsabs.harvard.edu/abs/2007arXiv0704.2068L},
	Author = {{Lebouteiller}, V. and {Brandl}, B. and {Bernard-Salas}, J. and 
	{Devost}, D. and {Houck}, J.~R.},
	Date-Added = {2007-06-08 15:53:14 +0100},
	Date-Modified = {2007-06-08 15:53:27 +0100},
	Eprint = {0704.2068},
	Journal = {ArXiv e-prints},
	Month = apr,
	Title = {{PAH Strength and the Interstellar Radiation Field around the Massive Young Cluster NGC3603}},
	Volume = 704,
	Year = 2007}

@article{wu:2007,
	Adsnote = {Provided by the Smithsonian/NASA Astrophysics Data System},
	Adsurl = {http://adsabs.harvard.edu/abs/2007astro.ph..3283W},
	Author = {{Wu}, Y. and {Charmandaris}, V. and {Hunt}, L.~K. and {Bernard-Salas}, J. and 
	{Brandl}, B.~R. and {Marshall}, J.~A. and {Lebouteiller}, V. and 
	{Houck}, J.~R.},
	Date-Added = {2007-06-08 15:52:19 +0100},
	Date-Modified = {2007-06-08 15:52:28 +0100},
	Eprint = {astro-ph/0703283},
	Journal = {ArXiv Astrophysics e-prints},
	Month = mar,
	Title = {{Dust in the Extremely Metal-Poor Blue Compact Dwarf Galaxy IZw18: The Spitzer Mid-Infrared View}},
	Year = 2007}

@article{brandl:2006,
	Adsnote = {Provided by the Smithsonian/NASA Astrophysics Data System},
	Adsurl = {http://adsabs.harvard.edu/abs/2006ApJ...653.1129B},
	Author = {{Brandl}, B.~R. and {Bernard-Salas}, J. and {Spoon}, H.~W.~W. and 
	{Devost}, D. and {Sloan}, G.~C. and {Guilles}, S. and {Wu}, Y. and 
	{Houck}, J.~R. and {Weedman}, D.~W. and {Armus}, L. and {Appleton}, P.~N. and 
	{Soifer}, B.~T. and {Charmandaris}, V. and {Hao}, L. and {Higdon}, J.~A.~M.~S.~J. and 
	{Herter}, T.~L.},
	Date-Added = {2007-06-08 15:51:22 +0100},
	Date-Modified = {2007-06-08 15:51:33 +0100},
	Doi = {10.1086/508849},
	Eprint = {arXiv:astro-ph/0609024},
	Journal = {\apj},
	Month = dec,
	Pages = {1129-1144},
	Title = {{The Mid-Infrared Properties of Starburst Galaxies from Spitzer-IRS Spectroscopy}},
	Volume = 653,
	Year = 2006}

@article{thuan:1997,
	Adsnote = {Provided by the Smithsonian/NASA Astrophysics Data System},
	Adsurl = {http://adsabs.harvard.edu/abs/1997ApJ...477..661T},
	Author = {{Thuan}, T.~X. and {Izotov}, Y.~I. and {Lipovetsky}, V.~A.},
	Date-Added = {2007-06-08 15:44:33 +0100},
	Date-Modified = {2007-06-08 15:44:48 +0100},
	Doi = {10.1086/303737},
	Eprint = {arXiv:astro-ph/9607121},
	Journal = {\apj},
	Month = mar,
	Pages = {661-+},
	Title = {{Hubble Space Telescope Observations of the Blue Compact Dwarf SBS 0335-052: A Probable Young Galaxy}},
	Volume = 477,
	Year = 1997}

@article{houck:2004,
	Adsnote = {Provided by the Smithsonian/NASA Astrophysics Data System},
	Adsurl = {http://adsabs.harvard.edu/abs/2004ApJS..154..211H},
	Author = {{Houck}, J.~R. and {Charmandaris}, V. and {Brandl}, B.~R. and 
	{Weedman}, D. and {Herter}, T. and {Armus}, L. and {Soifer}, B.~T. and 
	{Bernard-Salas}, J. and {Spoon}, H.~W.~W. and {Devost}, D. and 
	{Uchida}, K.~I.},
	Date-Added = {2007-06-08 15:43:00 +0100},
	Date-Modified = {2007-06-08 15:43:08 +0100},
	Doi = {10.1086/423137},
	Eprint = {arXiv:astro-ph/0406150},
	Journal = {\apjs},
	Month = sep,
	Pages = {211-214},
	Title = {{The Extraordinary Mid-infrared Spectrum of the Blue Compact Dwarf Galaxy SBS 0335-052}},
	Volume = 154,
	Year = 2004}

@article{engelbracht:2005,
	Adsnote = {Provided by the Smithsonian/NASA Astrophysics Data System},
	Adsurl = {http://adsabs.harvard.edu/abs/2005ApJ...628L..29E},
	Author = {{Engelbracht}, C.~W. and {Gordon}, K.~D. and {Rieke}, G.~H. and 
	{Werner}, M.~W. and {Dale}, D.~A. and {Latter}, W.~B.},
	Date-Added = {2007-06-08 15:41:43 +0100},
	Date-Modified = {2007-06-08 15:41:58 +0100},
	Doi = {10.1086/432613},
	Eprint = {arXiv:astro-ph/0506214},
	Journal = {\apjl},
	Month = jul,
	Pages = {L29-L32},
	Title = {{Metallicity Effects on Mid-Infrared Colors and the 8 {$\mu$}m PAH Emission in Galaxies}},
	Volume = 628,
	Year = 2005}

@article{jackson:2006,
	Adsnote = {Provided by the Smithsonian/NASA Astrophysics Data System},
	Adsurl = {http://adsabs.harvard.edu/abs/2006ApJ...646..192J},
	Author = {{Jackson}, D.~C. and {Cannon}, J.~M. and {Skillman}, E.~D. and 
	{Lee}, H. and {Gehrz}, R.~D. and {Woodward}, C.~E. and {Polomski}, E.
	},
	Date-Added = {2007-06-08 15:40:33 +0100},
	Date-Modified = {2007-06-08 15:40:49 +0100},
	Doi = {10.1086/504707},
	Eprint = {arXiv:astro-ph/0603860},
	Journal = {\apj},
	Month = jul,
	Pages = {192-204},
	Title = {{Hot Dust and Polycyclic Aromatic Hydrocarbon Emission at Low Metallicity: A Spitzer Survey of Local Group and Other Nearby Dwarf Galaxies}},
	Volume = 646,
	Year = 2006}

@article{stasinska:2007,
	Adsnote = {Provided by the Smithsonian/NASA Astrophysics Data System},
	Adsurl = {http://adsabs.harvard.edu/abs/2007arXiv0704.0348S},
	Author = {{Stasinska}, G.},
	Date-Added = {2007-06-08 15:39:10 +0100},
	Date-Modified = {2007-06-08 15:39:30 +0100},
	Eprint = {0704.0348},
	Journal = {ArXiv e-prints},
	Month = apr,
	Title = {{What can emission lines tell us?}},
	Volume = 704,
	Year = 2007}

@article{hunt:2006,
	Adsnote = {Provided by the Smithsonian/NASA Astrophysics Data System},
	Adsurl = {http://adsabs.harvard.edu/abs/2006ApJ...653..222H},
	Author = {{Hunt}, L.~K. and {Thuan}, T.~X. and {Sauvage}, M. and {Izotov}, Y.~I.
	},
	Date-Added = {2007-06-08 15:36:59 +0100},
	Date-Modified = {2007-06-08 15:37:14 +0100},
	Doi = {10.1086/508867},
	Eprint = {arXiv:astro-ph/0610405},
	Journal = {\apj},
	Month = dec,
	Pages = {222-239},
	Title = {{The Spitzer View of Low-Metallicity Star Formation. I. Haro 3}},
	Volume = 653,
	Year = 2006}

@article{rosenberg:2006,
	Adsnote = {Provided by the Smithsonian/NASA Astrophysics Data System},
	Adsurl = {http://adsabs.harvard.edu/abs/2006ApJ...636..742R},
	Author = {{Rosenberg}, J.~L. and {Ashby}, M.~L.~N. and {Salzer}, J.~J. and 
	{Huang}, J.-S.},
	Date-Added = {2007-06-08 15:35:27 +0100},
	Date-Modified = {2007-06-08 15:35:51 +0100},
	Doi = {10.1086/498133},
	Eprint = {arXiv:astro-ph/0509566},
	Journal = {\apj},
	Month = jan,
	Pages = {742-752},
	Title = {{The Diverse Infrared Properties of a Complete Sample of Star-forming Dwarf Galaxies}},
	Volume = 636,
	Year = 2006}

@article{fazio:irac,
	Adsnote = {Provided by the Smithsonian/NASA Astrophysics Data System},
	Adsurl = {http://adsabs.harvard.edu/abs/2004ApJS..154...10F},
	Author = {{Fazio}, G.~G. and others},
	Date-Added = {2007-06-08 14:17:52 +0100},
	Date-Modified = {2007-06-08 14:44:26 +0100},
	Doi = {10.1086/422843},
	Eprint = {arXiv:astro-ph/0405616},
	Journal = {\apjs},
	Month = sep,
	Pages = {10-17},
	Title = {{The Infrared Array Camera (IRAC) for the Spitzer Space Telescope}},
	Volume = 154,
	Year = 2004}

@article{houck:irs,
	Adsnote = {Provided by the Smithsonian/NASA Astrophysics Data System},
	Adsurl = {http://adsabs.harvard.edu/abs/2004ApJS..154...18H},
	Author = {{Houck}, J.~R. and others},
	Date-Added = {2007-06-08 14:16:14 +0100},
	Date-Modified = {2007-06-08 14:45:08 +0100},
	Doi = {10.1086/423134},
	Eprint = {arXiv:astro-ph/0406167},
	Journal = {\apjs},
	Month = sep,
	Pages = {18-24},
	Title = {{The Infrared Spectrograph (IRS) on the Spitzer Space Telescope}},
	Volume = 154,
	Year = 2004}

@article{werner:spitzer,
	Adsnote = {Provided by the Smithsonian/NASA Astrophysics Data System},
	Adsurl = {http://adsabs.harvard.edu/abs/2004ApJS..154....1W},
	Author = {{Werner}, M.~W. and others},
	Date-Added = {2007-06-08 14:14:15 +0100},
	Date-Modified = {2007-06-08 14:29:34 +0100},
	Doi = {10.1086/422992},
	Eprint = {arXiv:astro-ph/0406223},
	Journal = {\apjs},
	Month = sep,
	Pages = {1-9},
	Title = {{The Spitzer Space Telescope Mission}},
	Volume = 154,
	Year = 2004}

@article{rieke:mips,
	Adsnote = {Provided by the Smithsonian/NASA Astrophysics Data System},
	Adsurl = {http://adsabs.harvard.edu/abs/2004ApJS..154...25R},
	Author = {{Rieke}, G.~H. and others},
	Date-Added = {2007-06-08 14:12:02 +0100},
	Date-Modified = {2007-06-08 14:45:47 +0100},
	Doi = {10.1086/422717},
	Journal = {\apjs},
	Month = sep,
	Pages = {25-29},
	Title = {{The Multiband Imaging Photometer for Spitzer (MIPS)}},
	Volume = 154,
	Year = 2004}

@inproceedings{Thuan:1996a,
	Adsnote = {Provided by the Smithsonian/NASA Astrophysics Data System},
	Adsurl = {http://adsabs.harvard.edu/abs/1996ibms.conf..607T},
	Author = {{Thuan}, T.~X. and {Izotov}, Y.~I. and {Lipovetsky}, V.~A.},
	Booktitle = {The Interplay Between Massive Star Formation, the ISM and Galaxy Evolution},
	Date-Added = {2007-06-08 11:47:16 +0100},
	Date-Modified = {2008-07-22 10:36:04 +0100},
	Editor = {{Kunth}, D. and {Guiderdoni}, B. and {Heydari-Malayeri}, M. and {Thuan}, T.~X.},
	Pages = {607-+},
	Title = {{The Blue Compact Galaxy SBS 0335-052: a Young Galaxy.}},
	Year = 1996}
@ARTICLE{Bastian:2006,
   author = {{Bastian}, N. and {Emsellem}, E. and {Kissler-Patig}, M. and 
	{Maraston}, C.},
    title = "{Young star cluster complexes in NGC 4038/39. Integral field spectroscopy using VIMOS-VLT}",
  journal = {\aap},
   eprint = {arXiv:astro-ph/0509249},
 keywords = {galaxies: star clusters, galaxies: interactions, galaxies: individual: NGC 4038},
     year = 2006,
    month = jan,
   volume = 445,
    pages = {471-483},
      doi = {10.1051/0004-6361:20053793},
   adsurl = {http://adsabs.harvard.edu/abs/2006A
  adsnote = {Provided by the SAO/NASA Astrophysics Data System}
}

@ARTICLE{Takase:1986,
   author = {{Takase}, B. and {Miyauchi-Isobe}, N.},
    title = "{KISO survey for ultraviolet-excess galaxies. IV}",
  journal = {Annals of the Tokyo Astronomical Observatory},
 keywords = {ASTRONOMICAL CATALOGS, GALACTIC STRUCTURE, ULTRAVIOLET SPECTRA, MAGNITUDE, SPIRAL GALAXIES},
     year = 1986,
   volume = 21,
    pages = {127-180},
   adsurl = {http://adsabs.harvard.edu/abs/1986AnTok..21..127T},
  adsnote = {Provided by the SAO/NASA Astrophysics Data System}
}
@ARTICLE{Taylor:1995,
   author = {{Taylor}, C.~L. and {Brinks}, E. and {Grashuis}, R.~M. and {Skillman}, E.~D.
	},
    title = "{An H i/Optical Atlas of H II Galaxies and Their Companions}",
  journal = {\apjs},
 keywords = {ATLASES, GALAXIES: FUNDAMENTAL PARAMETERS, ISM: H II REGIONS, RADIO LINES: GALAXIES},
     year = 1995,
    month = aug,
   volume = 99,
    pages = {427-+},
      doi = {10.1086/192193},
   adsurl = {http://adsabs.harvard.edu/abs/1995ApJS...99..427T},
  adsnote = {Provided by the SAO/NASA Astrophysics Data System}
}
@ARTICLE{Telles:1995,
   author = {{Telles}, E. and {Terlevich}, R.},
    title = "{The environment of HII galaxies}",
  journal = {\mnras},
   eprint = {arXiv:astro-ph/9501084},
 keywords = {HII REGIONS, GALAXIES: INTERACTIONS, GALAXIES: STARBURST},
     year = 1995,
    month = jul,
   volume = 275,
    pages = {1-8},
   adsurl = {http://adsabs.harvard.edu/abs/1995MNRAS.275....1T},
  adsnote = {Provided by the SAO/NASA Astrophysics Data System}
}
@PHDTHESIS{Telles:1995thesis,
   author = {{Telles}, J.~E.},
   school = {, Univ.~Cambridge, (1995)},
     year = 1995,
    month = jan,
   adsurl = {http://adsabs.harvard.edu/abs/1995PhDT........99T},
  adsnote = {Provided by the SAO/NASA Astrophysics Data System}
}
@ARTICLE{Izotov:1998,
   author = {{Izotov}, Y.~I. and {Thuan}, T.~X.},
    title = "{The Primordial Abundance of 4He Revisited}",
  journal = {\apj},
 keywords = {GALAXIES: ABUNDANCES, GALAXIES: IRREGULAR, GALAXIES: ISM, ISM: H II REGIONS, ISM: ABUNDANCES},
     year = 1998,
    month = jun,
   volume = 500,
    pages = {188-+},
      doi = {10.1086/305698},
   adsurl = {http://adsabs.harvard.edu/abs/1998ApJ...500..188I},
  adsnote = {Provided by the SAO/NASA Astrophysics Data System}
}
@ARTICLE{Lopez-Sanchez:2008,
   author = {{L{\'o}pez-S{\'a}nchez}, {\'A}.~R. and {Esteban}, C.},
    title = "{Massive star formation in Wolf-Rayet galaxies. I. Optical and NIR photometric results}",
  journal = {\aap},
archivePrefix = "arXiv",
   eprint = {0811.0202},
 keywords = {galaxies: dwarf, galaxies: starburst, galaxies: photometry, galaxies: interactions, stars: Wolf-Rayet},
     year = 2008,
    month = nov,
   volume = 491,
    pages = {131-156},
      doi = {10.1051/0004-6361:200809409},
   adsurl = {http://adsabs.harvard.edu/abs/2008A
  adsnote = {Provided by the SAO/NASA Astrophysics Data System}
}
@ARTICLE{Macalpine:1978,
   author = {{MacAlpine}, G.~M. and {Lewis}, D.~W.},
    title = "{Curtis Schmidt-thin prism survey for extragalactic emission-line objects University of Michigan list IV}",
  journal = {\apjs},
 keywords = {ASTRONOMICAL CATALOGS, EMISSION SPECTRA, GALAXIES, QUASARS, RED SHIFT, SCHMIDT TELESCOPES, SEYFERT GALAXIES, SKY SURVEYS (ASTRONOMY)},
     year = 1978,
    month = apr,
   volume = 36,
    pages = {587-593},
      doi = {10.1086/190512},
   adsurl = {http://adsabs.harvard.edu/abs/1978ApJS...36..587M},
  adsnote = {Provided by the SAO/NASA Astrophysics Data System}
}

@ARTICLE{James:2009,
   author = {{James}, B.~L. and {Tsamis}, Y.~G. and {Barlow}, M.~J. and {Westmoquette}, M.~S. and 
	{Walsh}, J.~R. and {Cuisinier}, F. and {Exter}, K.~M.},
    title = "{A VLT VIMOS study of the anomalous BCD Mrk996: mapping the ionized gas kinematics and abundances}",
  journal = {\mnras},
 keywords = {stars: Wolf-Rayet , galaxies: abundances , galaxies: dwarf , galaxies: individual: Markarian996 , galaxies: kinematics and dynamics , galaxies: starburst},
     year = 2009,
    month = sep,
   volume = 398,
    pages = {2-22},
      doi = {10.1111/j.1365-2966.2009.15172.x},
   adsurl = {http://adsabs.harvard.edu/abs/2009MNRAS.398....2J},
  adsnote = {Provided by the SAO/NASA Astrophysics Data System}
}

@ARTICLE{Cairos:2001,
   author = {{Cair{\'o}s}, L.~M. and {V{\'{\i}}lchez}, J.~M. and {Gonz{\'a}lez P{\'e}rez}, J.~N. and 
	{Iglesias-P{\'a}ramo}, J. and {Caon}, N.},
    title = "{Multiband Analysis of a Sample of Blue Compact Dwarf Galaxies. I. Surface Brightness Distribution, Morphology, and Structural Parameters}",
  journal = {\apjs},
 keywords = {Galaxies: Dwarf, Galaxies: Starburst, Galaxies: Stellar Content, Galaxies: Structure},
     year = 2001,
    month = apr,
   volume = 133,
    pages = {321-343},
      doi = {10.1086/320350},
   adsurl = {http://adsabs.harvard.edu/abs/2001ApJS..133..321C},
  adsnote = {Provided by the SAO/NASA Astrophysics Data System}
}
@ARTICLE{Schaerer:1999,
   author = {{Schaerer}, D. and {Contini}, T. and {Pindao}, M.},
    title = "{New catalogue of Wolf-Rayet galaxies and high-excitation extra-galactic HII regions}",
  journal = {\aaps},
   eprint = {arXiv:astro-ph/9812347},
 keywords = {GALAXIES: STARBURST, GALAXIES: STELLAR CONTENT, GALAXIES: ISM, STARS: WOLF-RAYET},
     year = 1999,
    month = apr,
   volume = 136,
    pages = {35-52},
      doi = {10.1051/aas:1999197},
   adsurl = {http://adsabs.harvard.edu/abs/1999A
  adsnote = {Provided by the SAO/NASA Astrophysics Data System}
}
@ARTICLE{Guseva:2000,
   author = {{Guseva}, N.~G. and {Izotov}, Y.~I. and {Thuan}, T.~X.},
    title = "{A Spectroscopic Study of a Large Sample Of Wolf-Rayet Galaxies}",
  journal = {\apj},
   eprint = {arXiv:astro-ph/9910432},
 keywords = {GALAXIES: ABUNDANCES, GALAXIES: STARBURST, GALAXIES: STELLAR CONTENT, ISM: H II REGIONS, STARS: WOLF-RAYET},
     year = 2000,
    month = mar,
   volume = 531,
    pages = {776-803},
      doi = {10.1086/308489},
   adsurl = {http://adsabs.harvard.edu/abs/2000ApJ...531..776G},
  adsnote = {Provided by the SAO/NASA Astrophysics Data System}
}
@ARTICLE{Mendez:2000,
   author = {{M{\'e}ndez}, D.~I. and {Esteban}, C.},
    title = "{Deep optical imaging and spectroscopy of a sample of Wolf-Rayet galaxies}",
  journal = {\aap},
 keywords = {GALAXIES: INTERACTIONS, GALAXIES: KINEMATICS AND DYNAMICS, GALAXIES: STARBURST, STARS: WOLF-RAYET},
     year = 2000,
    month = jul,
   volume = 359,
    pages = {493-508},
   adsurl = {http://adsabs.harvard.edu/abs/2000A
  adsnote = {Provided by the SAO/NASA Astrophysics Data System}
}
@ARTICLE{Iglesias:2001,
   author = {{Iglesias-P{\'a}ramo}, J. and {V{\'{\i}}lchez}, J.~M.},
    title = "{Star-forming Objects in the Tidal Tails of Compact Groups}",
  journal = {\apj},
   eprint = {arXiv:astro-ph/0011115},
 keywords = {Galaxies: Evolution, Galaxies: Interactions, Galaxies: Starburst, ISM: H II Regions},
     year = 2001,
    month = mar,
   volume = 550,
    pages = {204-211},
      doi = {10.1086/319710},
   adsurl = {http://adsabs.harvard.edu/abs/2001ApJ...550..204I},
  adsnote = {Provided by the SAO/NASA Astrophysics Data System}
}

@ARTICLE{Verdes:2002,
   author = {{Verdes-Montenegro}, L. and {Del Olmo}, A. and {Iglesias-P{\'a}ramo}, J.~I. and 
	{Perea}, J. and {V{\'{\i}}lchez}, J.~M. and {Yun}, M.~S. and 
	{Huchtmeier}, W.~K.},
    title = "{Ripples and tails in the compact group of galaxies Hickson 54}",
  journal = {\aap},
   eprint = {arXiv:astro-ph/0210162},
 keywords = {galaxies: interactions, galaxies: kinematics and dynamics, galaxies: evolution, galaxies: structure, radio lines: galaxies, galaxies: individual: HCG 54},
     year = 2002,
    month = dec,
   volume = 396,
    pages = {815-832},
      doi = {10.1051/0004-6361:20021420},
   adsurl = {http://adsabs.harvard.edu/abs/2002A
  adsnote = {Provided by the SAO/NASA Astrophysics Data System}
}
@ARTICLE{Tran:2003,
   author = {{Tran}, H.~D. and {Sirianni}, M. and {Ford}, H.~C. and {Illingworth}, G.~D. and 
	{Clampin}, M. and {Hartig}, G. and {Becker}, R.~H. and {White}, R.~L. and 
	{Bartko}, F. and {Ben{\'{\i}}tez}, N. and {Blakeslee}, J.~P. and 
	{Bouwens}, R. and {Broadhurst}, T.~J. and {Brown}, R. and {Burrows}, C. and 
	{Cheng}, E. and {Cross}, N. and {Feldman}, P.~D. and {Franx}, M. and 
	{Golimowski}, D.~A. and {Gronwall}, C. and {Infante}, L. and 
	{Kimble}, R.~A. and {Krist}, J. and {Lesser}, M. and {Magee}, D. and 
	{Martel}, A.~R. and {McCann}, W.~J. and {Meurer}, G.~R. and 
	{Miley}, G. and {Postman}, M. and {Rosati}, P. and {Sparks}, W.~B. and 
	{Tsvetanov}, Z.},
    title = "{Advanced Camera for Surveys Observations of Young Star Clusters in the Interacting Galaxy UGC 10214}",
  journal = {\apj},
   eprint = {arXiv:astro-ph/0211371},
 keywords = {Galaxies: Individual: Name: Arp 188, Galaxies: Individual: Alphanumeric: UGC 10214, Galaxies: Individual: Alphanumeric: VV 29, Galaxies: Star Clusters},
     year = 2003,
    month = mar,
   volume = 585,
    pages = {750-755},
      doi = {10.1086/346125},
   adsurl = {http://adsabs.harvard.edu/abs/2003ApJ...585..750T},
  adsnote = {Provided by the SAO/NASA Astrophysics Data System}
}
@ARTICLE{Fanelli:1988,
   author = {{Fanelli}, M.~N. and {O'Connell}, R.~W. and {Thuan}, T.~X.},
    title = "{Spectral synthesis in the ultraviolet. II - Stellar populations and star formation in blue compact galaxies}",
  journal = {\apj},
 keywords = {COMPACT GALAXIES, STAR DISTRIBUTION, STAR FORMATION, STELLAR SPECTRA, ULTRAVIOLET SPECTRA, ABSORPTION SPECTRA, DWARF GALAXIES, MAIN SEQUENCE STARS, STELLAR LUMINOSITY, SUPERGIANT STARS},
     year = 1988,
    month = nov,
   volume = 334,
    pages = {665-687},
      doi = {10.1086/166869},
   adsurl = {http://adsabs.harvard.edu/abs/1988ApJ...334..665F},
  adsnote = {Provided by the SAO/NASA Astrophysics Data System}
}
@ARTICLE{VanZee:1998,
   author = {{van Zee}, L. and {Skillman}, E.~D. and {Salzer}, J.~J.},
    title = "{Neutral Gas Distributions and Kinematics of Five Blue Compact Dwarf Galaxies}",
  journal = {\aj},
   eprint = {arXiv:astro-ph/9806246},
 keywords = {GALAXIES: COMPACT, GALAXIES: DWARF, GALAXIES: INDIVIDUAL: ALPHANUMERIC: II ZW 40, GALAXIES: INDIVIDUAL: ALPHANUMERIC: UGC 4483, GALAXIES: INDIVIDUAL: ALPHANUMERIC: UM 439, GALAXIES: INDIVIDUAL: ALPHANUMERIC: UM 461, GALAXIES: INDIVIDUAL: ALPHANUMERIC: UM 462, GALAXIES: KINEMATICS AND DYNAMICS},
     year = 1998,
    month = sep,
   volume = 116,
    pages = {1186-1204},
      doi = {10.1086/300510},
   adsurl = {http://adsabs.harvard.edu/abs/1998AJ....116.1186V},
  adsnote = {Provided by the SAO/NASA Astrophysics Data System}
}
@ARTICLE{Wesson:2008,
   author = {{Wesson}, R. and {Barlow}, M.~J. and {Liu}, X.-W. and {Storey}, P.~J. and 
	{Ercolano}, B. and {de Marco}, O.},
    title = "{The hydrogen-deficient knot of the `born-again' planetary nebula Abell 58 (V605 Aql)}",
  journal = {\mnras},
archivePrefix = "arXiv",
   eprint = {0711.1139},
 keywords = {ISM: abundances , planetary nebulae: individual: Abell 58},
     year = 2008,
    month = feb,
   volume = 383,
    pages = {1639-1648},
      doi = {10.1111/j.1365-2966.2007.12683.x},
   adsurl = {http://adsabs.harvard.edu/abs/2008MNRAS.383.1639W},
  adsnote = {Provided by the SAO/NASA Astrophysics Data System}
}
@ARTICLE{Smits:1996,
   author = {{Smits}, D.~P.},
    title = "{Theoretical HeI line intensities in low-density plasmas}",
  journal = {\mnras},
 keywords = {ATOMIC DATA, ATOMIC PROCESSES, LINE: FORMATION},
     year = 1996,
    month = feb,
   volume = 278,
    pages = {683-687},
   adsurl = {http://adsabs.harvard.edu/abs/1996MNRAS.278..683S},
  adsnote = {Provided by the SAO/NASA Astrophysics Data System}
}
@ARTICLE{Benjamin:1999,
   author = {{Benjamin}, R.~A. and {Skillman}, E.~D. and {Smits}, D.~P.},
    title = "{Improving Predictions for Helium Emission Lines}",
  journal = {\apj},
   eprint = {arXiv:astro-ph/9810087},
 keywords = {ATOMIC DATA, ISM: GENERAL},
     year = 1999,
    month = mar,
   volume = 514,
    pages = {307-324},
      doi = {10.1086/306923},
   adsurl = {http://adsabs.harvard.edu/abs/1999ApJ...514..307B},
  adsnote = {Provided by the SAO/NASA Astrophysics Data System}
}
@ARTICLE{Izotov:2006a,
   author = {{Izotov}, Y.~I. and {Stasi{\'n}ska}, G. and {Meynet}, G. and 
	{Guseva}, N.~G. and {Thuan}, T.~X.},
    title = "{The chemical composition of metal-poor emission-line galaxies in the Data Release 3 of the Sloan Digital Sky Survey}",
  journal = {\aap},
   eprint = {arXiv:astro-ph/0511644},
 keywords = {galaxies: ISM, galaxies: starburst, galaxies: abundances},
     year = 2006,
    month = mar,
   volume = 448,
    pages = {955-970},
      doi = {10.1051/0004-6361:20053763},
   adsurl = {http://adsabs.harvard.edu/abs/2006A
  adsnote = {Provided by the SAO/NASA Astrophysics Data System}
}
@ARTICLE{Hopkins:2002,
   author = {{Hopkins}, A.~M. and {Schulte-Ladbeck}, R.~E. and {Drozdovsky}, I.~O.
	},
    title = "{Star Formation Rates of Local Blue Compact Dwarf Galaxies. I. 1.4 GHz and 60 Micron Luminosities}",
  journal = {\aj},
   eprint = {arXiv:astro-ph/0204528},
 keywords = {Galaxies: Dwarf, Galaxies: Evolution, Galaxies: Starburst},
     year = 2002,
    month = aug,
   volume = 124,
    pages = {862-876},
      doi = {10.1086/341584},
   adsurl = {http://adsabs.harvard.edu/abs/2002AJ....124..862H},
  adsnote = {Provided by the SAO/NASA Astrophysics Data System}
}
@ARTICLE{Sage:1992,
   author = {{Sage}, L.~J. and {Salzer}, J.~J. and {Loose}, H.-H. and {Henkel}, C.
	},
    title = "{Star formation and molecular clouds in blue compact galaxies}",
  journal = {\aap},
 keywords = {COMPACT GALAXIES, INTERSTELLAR GAS, MOLECULAR CLOUDS, STAR FORMATION, CARBON DIOXIDE, DWARF GALAXIES, ELLIPTICAL GALAXIES, RADIATIVE TRANSFER},
     year = 1992,
    month = nov,
   volume = 265,
    pages = {19-31},
   adsurl = {http://adsabs.harvard.edu/abs/1992A
  adsnote = {Provided by the SAO/NASA Astrophysics Data System}
}
@article{Kennicutt:1998,
	Adsnote = {Provided by the SAO/NASA Astrophysics Data System},
	Adsurl = {http://adsabs.harvard.edu/abs/1998ARA
	Author = {{Kennicutt}, Jr., R.~C.},
	Date-Added = {2008-11-30 22:32:21 +0000},
	Date-Modified = {2008-11-30 22:32:30 +0000},
	Doi = {10.1146/annurev.astro.36.1.189},
	Eprint = {arXiv:astro-ph/9807187},
	Journal = {\araa},
	Pages = {189-232},
	Title = {{Star Formation in Galaxies Along the Hubble Sequence}},
	Volume = 36,
	Year = 1998}
@article{Lee:2002,
	Adsnote = {Provided by the SAO/NASA Astrophysics Data System},
	Adsurl = {http://adsabs.harvard.edu/abs/2002AJ....124.3088L},
	Author = {{Lee}, J.~C. and {Salzer}, J.~J. and {Impey}, C. and {Thuan}, T.~X. and {Gronwall}, C.},
	Date-Added = {2008-11-30 22:41:04 +0000},
	Date-Modified = {2008-11-30 22:41:10 +0000},
	Doi = {10.1086/344309},
	Eprint = {arXiv:astro-ph/0208482},
	Journal = {\aj},
	Keywords = {Galaxies: Dwarf, Galaxies: Irregular, Galaxies: ISM, Galaxies: Luminosity Function, Mass Function, Galaxies: Starburst, ISM: H I},
	Month = dec,
	Pages = {3088-3117},
	Title = {{H I Properties of Low-Luminosity Star-forming Galaxies in the KPNO International Spectroscopic Survey}},
	Volume = 124,
	Year = 2002}
@ARTICLE{Lennon:1994,
   author = {{Lennon}, D.~J. and {Burke}, V.~M.},
    title = "{Atomic data from the IRON project. II. Effective collision strength S for infrared transitions in carbon-like ions}",
  journal = {\aaps},
     year = 1994,
    month = feb,
   volume = 103,
    pages = {273-277},
   adsurl = {http://adsabs.harvard.edu/abs/1994A
  adsnote = {Provided by the SAO/NASA Astrophysics Data System}
}
@BOOK{Wiese:1996,
   author = {{Wiese}, W.~L. and {Fuhr}, J.~R. and {Deters}, T.~M.},
    title = "{Atomic transition probabilities of carbon, nitrogen, and oxygen : a critical data compilation}",
booktitle = {Atomic transition probabilities of carbon, nitrogen, and oxygen : a critical data compilation.~ Edited by W.L.~Wiese, J.R.~Fuhr, and T.M.~Deters.~Washington, DC :  American Chemical Society ...~for the National Institute of Standards and Technology (NIST) c1996.~QC 453 .W53 1996.},
     year = 1996,
   adsurl = {http://adsabs.harvard.edu/abs/1996atpc.book.....W},
  adsnote = {Provided by the SAO/NASA Astrophysics Data System}
}
@ARTICLE{Scott:2009,
   author = {{Scott}, P. and {Asplund}, M. and {Grevesse}, N. and {Sauval}, A.~J.
	},
    title = "{On the Solar Nickel and Oxygen Abundances}",
  journal = {\apjl},
archivePrefix = "arXiv",
   eprint = {0811.0815},
 keywords = {line: formation, line: profiles, Sun: abundances, Sun: atmosphere, Sun: photosphere, techniques: polarimetric},
     year = 2009,
    month = feb,
   volume = 691,
    pages = {L119-L122},
      doi = {10.1088/0004-637X/691/2/L119},
   adsurl = {http://adsabs.harvard.edu/abs/2009ApJ...691L.119S},
  adsnote = {Provided by the SAO/NASA Astrophysics Data System}
}
@ARTICLE{Salpeter:1955,
   author = {{Salpeter}, E.~E.},
    title = "{The Luminosity Function and Stellar Evolution.}",
  journal = {\apj},
     year = 1955,
    month = jan,
   volume = 121,
    pages = {161-+},
      doi = {10.1086/145971},
   adsurl = {http://adsabs.harvard.edu/abs/1955ApJ...121..161S},
  adsnote = {Provided by the SAO/NASA Astrophysics Data System}
}
@ARTICLE{Tsamis:2003,
   author = {{Tsamis}, Y.~G. and {Barlow}, M.~J. and {Liu}, X.-W. and {Danziger}, I.~J. and 
	{Storey}, P.~J.},
    title = "{Heavy elements in Galactic and Magellanic Cloud HII regions: recombination-line versus forbidden-line abundances}",
  journal = {\mnras},
   eprint = {arXiv:astro-ph/0209534},
 keywords = {stars: individual: R139, stars: individual: R140, stars: individual: P3157, ISM: abundances, HII regions},
     year = 2003,
    month = jan,
   volume = 338,
    pages = {687-710},
      doi = {10.1046/j.1365-8711.2003.06081.x},
   adsurl = {http://cdsads.u-strasbg.fr/abs/2003MNRAS.338..687T},
  adsnote = {Provided by the SAO/NASA Astrophysics Data System}
}
@ARTICLE{Thuan:1981,
   author = {{Thuan}, T.~X. and {Martin}, G.~E.},
    title = "{Blue compact dwarf galaxies. I - Neutral hydrogen observations of 115 galaxies}",
  journal = {\apj},
 keywords = {COMPACT GALAXIES, GALACTIC STRUCTURE, INTERSTELLAR GAS, RADIO SOURCES (ASTRONOMY), STELLAR EVOLUTION, ASTRONOMICAL CATALOGS, CENTIMETER WAVES, H II REGIONS, NEUTRAL GASES, STOCHASTIC PROCESSES},
     year = 1981,
    month = aug,
   volume = 247,
    pages = {823-848},
      doi = {10.1086/159094},
   adsurl = {http://cdsads.u-strasbg.fr/abs/1981ApJ...247..823T},
  adsnote = {Provided by the SAO/NASA Astrophysics Data System}
}
@INBOOK{Brinks:1990,
   author = {{Brinks}, E.},
    title = "{II Zwicky 33: star formation induced by a recent interaction.}",
 keywords = {Interacting Galaxies: Star Formation, Interacting Galaxies: H I Clouds},
booktitle = {Dynamics and Interactions of Galaxies},
     year = 1990,
   editor = {{Wielen}, R.},
    pages = {146-149},
   adsurl = {http://cdsads.u-strasbg.fr/abs/1990dig..book..146B},
  adsnote = {Provided by the SAO/NASA Astrophysics Data System}
}


\begin{thebibliography}{49}
\expandafter\ifx\csname natexlab\endcsname\relax\def\natexlab#1{#1}\fi

\bibitem[{{Baldwin} {et~al.}(1981){Baldwin}, {Phillips}, \&
  {Terlevich}}]{Baldwin:1981}
{Baldwin} J.~A., {Phillips} M.~M., {Terlevich} R., 1981, \pasp, 93, 5

\bibitem[{{Bastian} {et~al.}(2006){Bastian}, {Emsellem}, {Kissler-Patig}, \&
  {Maraston}}]{Bastian:2006}
{Bastian} N., {Emsellem} E., {Kissler-Patig} M., {Maraston} C., 2006, \aap,
  445, 471

\bibitem[{{Benjamin} {et~al.}(1999){Benjamin}, {Skillman}, \&
  {Smits}}]{Benjamin:1999}
{Benjamin} R.~A., {Skillman} E.~D., {Smits} D.~P., 1999, \apj, 514, 307

\bibitem[{{Brinks}(1990)}]{Brinks:1990}
{Brinks} E., 1990, {II Zwicky 33: star formation induced by a recent
  interaction.}, {Wielen} R., ed., pp. 146--149

\bibitem[{{Cair{\'o}s} {et~al.}(2001){Cair{\'o}s}, {V{\'{\i}}lchez},
  {Gonz{\'a}lez P{\'e}rez}, {Iglesias-P{\'a}ramo}, \& {Caon}}]{Cairos:2001}
{Cair{\'o}s} L.~M., {V{\'{\i}}lchez} J.~M., {Gonz{\'a}lez P{\'e}rez} J.~N.,
  {Iglesias-P{\'a}ramo} J., {Caon} N., 2001, \apjs, 133, 321

\bibitem[{{DeBernardis} {et~al.}(2008){DeBernardis}, {Melchiorri}, {Verde}, \&
  {Jimenez}}]{Bernardis:2008}
{DeBernardis} F., {Melchiorri} A., {Verde} L., {Jimenez} R., 2008, Journal of
  Cosmology and Astro-Particle Physics, 3, 20

\bibitem[{{Dimeo}(2005)}]{Dimeo:2005}
{Dimeo} R., 2005, PAN User Guide

\bibitem[{{Dopita} {et~al.}(2001){Dopita}, {Kewley}, {Sutherland}, \&
  {Heisler}}]{Dopita:2001}
{Dopita} M.~A., {Kewley} L.~J., {Sutherland} R.~S., {Heisler} C.~A., 2001, in
  Starburst Galaxies: Near and Far, {Tacconi} L., {Lutz} D., eds., pp. 225--+

\bibitem[{{Fanelli} {et~al.}(1988){Fanelli}, {O'Connell}, \&
  {Thuan}}]{Fanelli:1988}
{Fanelli} M.~N., {O'Connell} R.~W., {Thuan} T.~X., 1988, \apj, 334, 665

\bibitem[{{Guseva} {et~al.}(2000){Guseva}, {Izotov}, \& {Thuan}}]{Guseva:2000}
{Guseva} N.~G., {Izotov} Y.~I., {Thuan} T.~X., 2000, \apj, 531, 776

\bibitem[{{Hopkins} {et~al.}(2002){Hopkins}, {Schulte-Ladbeck}, \&
  {Drozdovsky}}]{Hopkins:2002}
{Hopkins} A.~M., {Schulte-Ladbeck} R.~E., {Drozdovsky} I.~O., 2002, \aj, 124,
  862

\bibitem[{{Howarth}(1983)}]{Howarth:1983}
{Howarth} I.~D., 1983, \mnras, 203, 301

\bibitem[{{Hummer} \& {Storey}(1987)}]{Hummer:1987}
{Hummer} D.~G., {Storey} P.~J., 1987, \mnras, 224, 801

\bibitem[{{Iglesias-P{\'a}ramo} \& {V{\'{\i}}lchez}(2001)}]{Iglesias:2001}
{Iglesias-P{\'a}ramo} J., {V{\'{\i}}lchez} J.~M., 2001, \apj, 550, 204

\bibitem[{{Izotov} {et~al.}(2006){Izotov}, {Stasi{\'n}ska}, {Meynet}, {Guseva},
  \& {Thuan}}]{Izotov:2006a}
{Izotov} Y.~I., {Stasi{\'n}ska} G., {Meynet} G., {Guseva} N.~G., {Thuan} T.~X.,
  2006, \aap, 448, 955

\bibitem[{{Izotov} \& {Thuan}(1998)}]{Izotov:1998}
{Izotov} Y.~I., {Thuan} T.~X., 1998, \apj, 500, 188

\bibitem[{{James} {et~al.}(2009){James}, {Tsamis}, {Barlow}, {Westmoquette},
  {Walsh}, {Cuisinier}, \& {Exter}}]{James:2009}
{James} B.~L., {Tsamis} Y.~G., {Barlow} M.~J., {Westmoquette} M.~S., {Walsh}
  J.~R., {Cuisinier} F., {Exter} K.~M., 2009, \mnras, 398, 2

\bibitem[{{Kennicutt}(1998)}]{Kennicutt:1998}
{Kennicutt} Jr. R.~C., 1998, \araa, 36, 189

\bibitem[{{Kewley} {et~al.}(2001){Kewley}, {Dopita}, {Sutherland}, {Heisler},
  \& {Trevena}}]{Kewley:2001}
{Kewley} L.~J., {Dopita} M.~A., {Sutherland} R.~S., {Heisler} C.~A., {Trevena}
  J., 2001, \apj, 556, 121

\bibitem[{{Kingsburgh} \& {Barlow}(1994)}]{Kingsburgh:1994}
{Kingsburgh} R.~L., {Barlow} M.~J., 1994, \mnras, 271, 257

\bibitem[{{Kunth} \& {{\"O}stlin}(2000)}]{Kunth:2000}
{Kunth} D., {{\"O}stlin} G., 2000, \aapr, 10, 1

\bibitem[{{Lee} {et~al.}(2002){Lee}, {Salzer}, {Impey}, {Thuan}, \&
  {Gronwall}}]{Lee:2002}
{Lee} J.~C., {Salzer} J.~J., {Impey} C., {Thuan} T.~X., {Gronwall} C., 2002,
  \aj, 124, 3088

\bibitem[{{Leitherer} {et~al.}(1999){Leitherer}, {Schaerer}, {Goldader},
  {Delgado}, {Robert}, {Kune}, {de Mello}, {Devost}, \&
  {Heckman}}]{Leitherer:1999}
{Leitherer} C., {Schaerer} D., {Goldader} J.~D., {Delgado} R.~M.~G., {Robert}
  C., {Kune} D.~F., {de Mello} D.~F., {Devost} D., {Heckman} T.~M., 1999,
  \apjs, 123, 3

\bibitem[{{Lennon} \& {Burke}(1994)}]{Lennon:1994}
{Lennon} D.~J., {Burke} V.~M., 1994, \aaps, 103, 273

\bibitem[{{Loose} \& {Thuan}(1985)}]{Loose:1985}
{Loose} H.-H., {Thuan} T.~X., 1985, in Star-Forming Dwarf Galaxies and Related
  Objects, {Kunth} D., {Thuan} T.~X., {Tran Thanh van} J., eds., pp. 73--+

\bibitem[{{L{\'o}pez-S{\'a}nchez} \& {Esteban}(2008)}]{Lopez-Sanchez:2008}
{L{\'o}pez-S{\'a}nchez} {\'A}.~R., {Esteban} C., 2008, \aap, 491, 131

\bibitem[{{MacAlpine} \& {Lewis}(1978)}]{Macalpine:1978}
{MacAlpine} G.~M., {Lewis} D.~W., 1978, \apjs, 36, 587

\bibitem[{{M{\'e}ndez} \& {Esteban}(2000)}]{Mendez:2000}
{M{\'e}ndez} D.~I., {Esteban} C., 2000, \aap, 359, 493

\bibitem[{{Pustilnik} {et~al.}(2004){Pustilnik}, {Kniazev}, {Pramskij},
  {Izotov}, {Foltz}, {Brosch}, {Martin}, \& {Ugryumov}}]{Pustilnik:2004}
{Pustilnik} S., {Kniazev} A., {Pramskij} A., {Izotov} Y., {Foltz} C., {Brosch}
  N., {Martin} J.-M., {Ugryumov} A., 2004, \aap, 419, 469

\bibitem[{{Sage} {et~al.}(1992){Sage}, {Salzer}, {Loose}, \&
  {Henkel}}]{Sage:1992}
{Sage} L.~J., {Salzer} J.~J., {Loose} H.-H., {Henkel} C., 1992, \aap, 265, 19

\bibitem[{{Salpeter}(1955)}]{Salpeter:1955}
{Salpeter} E.~E., 1955, \apj, 121, 161

\bibitem[{{Schaerer} {et~al.}(1999){Schaerer}, {Contini}, \&
  {Pindao}}]{Schaerer:1999}
{Schaerer} D., {Contini} T., {Pindao} M., 1999, \aaps, 136, 35

\bibitem[{{Schaerer} \& {Vacca}(1998)}]{Schaerer:1998}
{Schaerer} D., {Vacca} W.~D., 1998, \apj, 497, 618

\bibitem[{{Schlegel} {et~al.}(1998){Schlegel}, {Finkbeiner}, \&
  {Davis}}]{Schlegel:1998}
{Schlegel} D.~J., {Finkbeiner} D.~P., {Davis} M., 1998, \apj, 500, 525

\bibitem[{{Scott} {et~al.}(2009){Scott}, {Asplund}, {Grevesse}, \&
  {Sauval}}]{Scott:2009}
{Scott} P., {Asplund} M., {Grevesse} N., {Sauval} A.~J., 2009, \apjl, 691, L119

\bibitem[{{Smits}(1996)}]{Smits:1996}
{Smits} D.~P., 1996, \mnras, 278, 683

\bibitem[{{Takase} \& {Miyauchi-Isobe}(1986)}]{Takase:1986}
{Takase} B., {Miyauchi-Isobe} N., 1986, Annals of the Tokyo Astronomical
  Observatory, 21, 127

\bibitem[{{Taylor} {et~al.}(1995){Taylor}, {Brinks}, {Grashuis}, \&
  {Skillman}}]{Taylor:1995}
{Taylor} C.~L., {Brinks} E., {Grashuis} R.~M., {Skillman} E.~D., 1995, \apjs,
  99, 427

\bibitem[{{Telles} \& {Terlevich}(1995)}]{Telles:1995}
{Telles} E., {Terlevich} R., 1995, \mnras, 275, 1

\bibitem[{{Telles}(1995)}]{Telles:1995thesis}
{Telles} J.~E., 1995, PhD thesis, , Univ.~Cambridge, (1995)

\bibitem[{{Thuan} \& {Martin}(1981)}]{Thuan:1981}
{Thuan} T.~X., {Martin} G.~E., 1981, \apj, 247, 823

\bibitem[{{Tran} {et~al.}(2003){Tran}, {Sirianni}, {Ford}, {Illingworth},
  {Clampin}, {Hartig}, {Becker}, {White}, {Bartko}, {Ben{\'{\i}}tez},
  {Blakeslee}, {Bouwens}, {Broadhurst}, {Brown}, {Burrows}, {Cheng}, {Cross},
  {Feldman}, {Franx}, {Golimowski}, {Gronwall}, {Infante}, {Kimble}, {Krist},
  {Lesser}, {Magee}, {Martel}, {McCann}, {Meurer}, {Miley}, {Postman},
  {Rosati}, {Sparks}, \& {Tsvetanov}}]{Tran:2003}
{Tran} H.~D., {Sirianni} M., {Ford} H.~C., {Illingworth} G.~D., {Clampin} M.,
  {Hartig} G., {Becker} R.~H., {White} R.~L., {Bartko} F., {Ben{\'{\i}}tez} N.,
  {Blakeslee} J.~P., {Bouwens} R., {Broadhurst} T.~J., {Brown} R., {Burrows}
  C., {Cheng} E., {Cross} N., {Feldman} P.~D., {Franx} M., {Golimowski} D.~A.,
  {Gronwall} C., {Infante} L., {Kimble} R.~A., {Krist} J., {Lesser} M., {Magee}
  D., {Martel} A.~R., {McCann} W.~J., {Meurer} G.~R., {Miley} G., {Postman} M.,
  {Rosati} P., {Sparks} W.~B., {Tsvetanov} Z., 2003, \apj, 585, 750

\bibitem[{{Tsamis} {et~al.}(2003){Tsamis}, {Barlow}, {Liu}, {Danziger}, \&
  {Storey}}]{Tsamis:2003}
{Tsamis} Y.~G., {Barlow} M.~J., {Liu} X.-W., {Danziger} I.~J., {Storey} P.~J.,
  2003, \mnras, 338, 687

\bibitem[{{Verdes-Montenegro} {et~al.}(2002){Verdes-Montenegro}, {Del Olmo},
  {Iglesias-P{\'a}ramo}, {Perea}, {V{\'{\i}}lchez}, {Yun}, \&
  {Huchtmeier}}]{Verdes:2002}
{Verdes-Montenegro} L., {Del Olmo} A., {Iglesias-P{\'a}ramo} J.~I., {Perea} J.,
  {V{\'{\i}}lchez} J.~M., {Yun} M.~S., {Huchtmeier} W.~K., 2002, \aap, 396, 815

\bibitem[{{Walsh} \& {Roy}(1990)}]{Walsh:1990}
{Walsh} J.~R., {Roy} J.~R., 1990, in ESO Conf. Proc. 34: 2nd ESO/ST-ECF Data
  Analysis Workshop, {Baade} D., {Grosbol} P.~J., eds., p.~95

\bibitem[{{Wesson} {et~al.}(2008){Wesson}, {Barlow}, {Liu}, {Storey},
  {Ercolano}, \& {de Marco}}]{Wesson:2008}
{Wesson} R., {Barlow} M.~J., {Liu} X.-W., {Storey} P.~J., {Ercolano} B., {de
  Marco} O., 2008, \mnras, 383, 1639

\bibitem[{{Westmoquette} {et~al.}(2007){Westmoquette}, {Exter}, {Smith}, \&
  {Gallagher}}]{Westmoquette:2007}
{Westmoquette} M.~S., {Exter} K.~M., {Smith} L.~J., {Gallagher} J.~S., 2007,
  \mnras, 381, 894

\bibitem[{{Wiese} {et~al.}(1996){Wiese}, {Fuhr}, \& {Deters}}]{Wiese:1996}
{Wiese} W.~L., {Fuhr} J.~R., {Deters} T.~M., 1996, {Atomic transition
  probabilities of carbon, nitrogen, and oxygen : a critical data compilation}

\bibitem[{{Zanichelli} {et~al.}(2005){Zanichelli}, {Garilli}, {Scodeggio},
  {Franzetti}, {Rizzo}, {Maccagni}, {Merighi}, {Picat}, {Le F{\`e}vre},
  {Foucaud}, {Bottini}, {Le Brun}, {Scaramella}, {Tresse}, {Vettolani},
  {Adami}, {Arnaboldi}, {Arnouts}, {Bardelli}, {Bolzonella}, {Cappi},
  {Charlot}, {Ciliegi}, {Contini}, {Gavignaud}, {Guzzo}, {Ilbert}, {Iovino},
  {McCracken}, {Marano}, {Marinoni}, {Mathez}, {Mazure}, {Meneux}, {Paltani},
  {Pell{\`o}}, {Pollo}, {Pozzetti}, {Radovich}, {Zamorani}, \&
  {Zucca}}]{Zanichelli:2005}
{Zanichelli} A., {Garilli} B., {Scodeggio} M., {Franzetti} P., {Rizzo} D.,
  {Maccagni} D., {Merighi} R., {Picat} J.~P., {Le F{\`e}vre} O., {Foucaud} S.,
  {Bottini} D., {Le Brun} V., {Scaramella} R., {Tresse} L., {Vettolani} G.,
  {Adami} C., {Arnaboldi} M., {Arnouts} S., {Bardelli} S., {Bolzonella} M.,
  {Cappi} A., {Charlot} S., {Ciliegi} P., {Contini} T., {Gavignaud} I., {Guzzo}
  L., {Ilbert} O., {Iovino} A., {McCracken} H.~J., {Marano} B., {Marinoni} C.,
  {Mathez} G., {Mazure} A., {Meneux} B., {Paltani} S., {Pell{\`o}} R., {Pollo}
  A., {Pozzetti} L., {Radovich} M., {Zamorani} G., {Zucca} E., 2005, \pasp,
  117, 1271

\end{thebibliography}

\appendix
\bsp
\label{lastpage}
\end{document}